\documentclass[12pt]{article}

\usepackage{amsfonts,amssymb,amstext}

\usepackage[normal]{subfigure}
\usepackage{epsfig}
\usepackage{graphicx}
\usepackage{enumerate}
\usepackage{mathtools}
\usepackage{color}
\usepackage[normalem]{ulem}
\usepackage{float}
\usepackage{amsmath}
\usepackage{amsthm}

\usepackage{ulem}

\usepackage[titletoc,title]{appendix}

\allowdisplaybreaks

\input epsf.tex

\newcommand{\wb}[1]{\overline{#1}}

\newcommand{\abs}[1]{\left\vert #1 \right\vert}

\def\Q{  {{\mathcal U}_{\mathcal Z}}}
\def\QH{  { {\mathcal U}_{\mathcal Z}^H}}
\def\QV{  { {\mathcal U}_{\mathcal Z}^V}}

\def\R{\mathbb{R}}

\def\vu{\boldsymbol{u}}

\def\vn{\boldsymbol{n}}

\def\a{\alpha}
\def\b{\beta}
\def\r{\rho}

\def\T{\tau}
\def\G{\Gamma}

\def\p{\partial}

\def\bS{\boldsymbol{\sigma}}
\def\bT{\boldsymbol{\tau}}

\def\dint{\displaystyle\int}
\def\dsum{\displaystyle\sum}

\def\dfrac{\displaystyle\frac}

\def\bd{\boldsymbol}

\def\wtd{\widetilde}

\def\mb{\mbox}
\def\B{\mathcal{B}}
\def\F{\mathcal{F}}

\title{2D granular flows with the $\mu(I)$ rheology and side walls friction: a well balanced multilayer discretization}
\author{E.D. Fern\'andez-Nieto \thanks{Dpto. Matem\'atica Aplicada I. ETS Arquitectura - Universidad de Sevilla.
  Avda. Reina Mercedes S/N, 41012-Sevilla, Spain, (edofer@us.es, gnarbona@us.es)}\ , J. Garres-D{\'i}az \thanks{IMUS {\&} Dpto. Matem{\'a}tica Aplicada I. ETS Arquitectura - Universidad de Sevilla.
  Avda. Reina Mercedes S/N, 41012-Sevilla, Spain, (jgarres@us.es)}\ , A. Mangeney
 \thanks{Institut de Physique du Globe de Paris, Seismology team, University Paris-Diderot, Sorbonne Paris Cit\'e, 75238, Paris, France, (mangeney@ipgp.fr)}\ \thanks{ANGE team, CEREMA, INRIA, Lab. J. Louis Lions, 75252, Paris, France}\ , G. Narbona-Reina $^*$}

\begin{document}
\date{}
\maketitle

\abstract{
We present here numerical modelling of granular flows with the $\mu(I)$ rheology in confined channels. The contribution is twofold: (i) a model to approximate the Navier-Stokes equations with the $\mu(I)$ rheology through an asymptotic analysis. Under the hypothesis of a one-dimensional flow, this model takes into account side walls friction; (ii) a multilayer discretization following Fern\'andez-Nieto \textit{et al}. (\textit{J. Fluid Mech.}, vol. 798, 2016, pp. 643-681). In this new numerical scheme, we propose an appropriate treatment of the rheological terms through a hydrostatic reconstruction which allows this scheme to be well-balanced and therefore to deal with dry areas. Based on academic tests, we first evaluate the influence of the width of the channel on the normal profiles of the downslope velocity thanks to the multilayer approach that is intrinsically able to describe changes from Bagnold to S-shaped (and vice versa) velocity profiles. We also check the well balance property of the proposed numerical scheme. We show that approximating side walls friction using single-layer models may lead to strong errors. Secondly, we compare the numerical results with experimental data on granular collapses. We show that the proposed scheme allows us to qualitatively reproduce the deposit in the case of a rigid bed (i. e. dry area) and that the error made by replacing the dry area by a small layer of material may be large if this layer is not thin enough. The proposed model is also able to reproduce the time evolution of the free surface and of the flow/no-flow interface. In addition, it reproduces the effect of erosion for granular flows over initially static material lying on the bed. This is possible when using a variable friction coefficient $\mu(I)$ but not with a constant friction coefficient.
}

\bigskip



\frenchspacing   
\newpage
\section{Introduction}
Granular flows have been widely studied in recent years owing to their importance in industrial processes and geophysical flows such as avalanches, debris flows, etc. In particular, numerical models provide a unique tool to study the dynamics of these very complex flows and to predict their behaviour in natural environment (see e. g. Delannay et al. \cite{delannay:2017} for a review). Defining an appropriate rheological law to describe these flows is still a challenge. Currently, the most accepted rheological law is the viscoplastic so-called $\mu(I)$ rheology introduced by Jop et al. \cite{jop:2006}. It considers a Drucker-Prager type model with the friction coefficient
$$\mu(I) = \mu_s + \dfrac{\mu_2 - \mu_s}{I_0 + I}\,I,$$
 where $I_0$ and $\mu_2>\mu_s$ are constant parameters depending on the material properties and $I$ is the inertial number defined by
$$ I = \frac{2d_s\|D(\vu)\|}{\sqrt{p/\r_{s}}},$$
with $\vu\in\mathbb{R}^3$ the velocity field. $D(\vu)=\frac 1 2 (\nabla \vu +(\nabla \vu)')$ is the strain rate tensor and $\|D\| = \sqrt{0.5\;D:D}$. As usual, $p\in\mathbb{R}$ denotes the pressure, $d_s$ the particle diameter and $\rho_s$ the particle density. Lagr\'ee et al. \cite{lagree:2011} defined a viscosity with a regularization of the $\mu(I)$ rheology, in order to model granular flows using the full Navier-Stokes solver \textit{Gerris}. Applications of this model are presented in  \cite{staron:2012,staron:2014}. Following these works the $\mu(I)$-viscosity can be defined as
\begin{equation}
\label{eq:viscosity_prev}
\eta = \dfrac{\mu(I)p}{\sqrt{\|D(\vu)\|^2 + \delta^2}},
\end{equation}
where $\delta$ is a regularization parameter  (see e.g. \cite{engelman:1980,lusso:2017a}). Then, the total stress tensor is written $\bS= -p\bd{\mathcal{I}} + \bd{\tau}$,
where $\bd{\mathcal{I}}$ is the 3D identity tensor and $\bd{\bT}$ the deviatoric stress tensor given by $\bT =\eta D(\vu)$.

By using a finite elements discretization and an augmented Lagrangian formulation, Ionescu et al. \cite{ionescu:2015} and Martin et al. \cite{martin:2017} showed that the $\mu(I)$ rheology reproduces laboratory experiments of granular collapses on horizontal and inclined planes. Using an Arbitrary Lagrangian Eulerian (ALE) formulation for the displacement of the domain, Lusso et al. \cite{lusso:2017a} showed that similar results where obtained when using either a regularization method or an augmented Lagrangian formulation in the case of the collapse and spreading of a granular column. These studies showed the difficulty of the ALE method to deal with detailed description of the front propagation due to the deformation of the mesh and possible overturning of the elements at the front, in particular when trying to simulate granular flows over an initially layer of material lying on the bed. The $\mu(I)$ rheology has also been implemented in a three-dimensional numerical model by Chauchat \& M\'edale \cite{chauchat:2014}, where they used a finite element method combined with the Newton-Raphson algorithm.\\

Nevertheless, 3D Navier-Stokes solvers have a high associated computational cost. In order to avoid solving these equations, granular flows has been studied through depth-averaged models (e. g. \cite{savage:1989, iverson:1997,mangeney:2003,mangeney:2005,mangeney:2007}), in  particular for application on natural geophysical flows on Earth and on other planets (e. g. \cite{mcdougall:2004,lucas:2007,pirulli:2008,favreau:2010,mangold:2010,lucas:2014}). Recently, Gray and Edwards \cite{gray:2014} introduced a depth-averaged model with the $\mu(I)$ rheology by prescribing the well known Bagnold profile that is used in \cite{edwards:2015} to reproduce erosion-deposition waves. However, depth-averaged models do not describe the change in time of velocity profiles. Indeed, a given velocity profile or an assymptotic argument is assumed during the derivation of the equations. \\

To go beyond this limitation, Fernandez-Nieto et al. \cite{fernandezNieto:2016} derived a multilayer shallow model with the $\mu(I)$ rheology making it possible to recover the vertical structure of the velocity without prescribing a typical vertical profile (e.g. Bagnold or S-shaped profile). In this approach the vertical direction is solved but the flow is still assumed to be shallow as was also done in Lusso et al. \cite{lusso:2017b,bouchut:2016}. Interestingly, these authors showed using analytical solutions that the flow/no-flow interface evolution is related to the normal gradient of the velocity at this interface. As a result, describing erosion/deposition processes related to static/flowing transition requires a model that is able to recover the time and space variation of the velocity profile. However, the model proposed by Lusso et al. is restricted to unform flows in the downslope direction and does not take into account side wall friction.

We present here an extension of the multilayer shallow model \cite{fernandezNieto:2016} that describes granular flows in a rectangular channel by including Coulomb friction at the lateral walls. This model is obtained by a dimensional analysis and the integration along the transversal direction of the channel.
Taberley et al. \cite{taberlet:2003} and Jop et al. \cite{jop:2005} showed the importance of side walls friction for uniform flows in inclined channels. They proposed to model this effect by adding an extra term to $\mu(I)$ for the case of uniform flows. Jop et al. \cite{jop:2007} used this additional term to simulate the transient normal profile of velocity in narrow channels and compared their simulation with laboratory experiments.
Recently, Baker et al. \cite{baker:2016} extended the depth-averaged model introduced in \cite{gray:2014} to the two horizontal dimensions case for steady uniform flows between parallel plates. They included a new viscous term for the side walls friction and studied the normal profiles of velocity in narrow and wide channels, where these profiles are reconstructed by assuming a Bagnold profile. They compared the full and the depth-averaged $\mu(I)$ rheology and conclude that they cannot reproduce the different profiles of the velocity observed in transient flows because of the prescribed vertical profile, in particular close to the lateral walls of narrow channels. On the contrary, Capart et al. \cite{capart:2015} prescribed a typical S-shaped profile for the downslope velocity, so that they were able to reproduce velocity profiles when flow was decelerating but not the Bagnold profile observed in other regimes.

The numerical solution of the new multilayer shallow model is compared here to laboratory experiments \cite{mangeney:2010, jop:2007} and analytical solutions, showing that it appropriately reproduces the evolution of the shape of the normal velocity profile for uniform flows. The other strong advantage of this multilayer shallow models is the low cost associated with the numerical treatment of the free surface, and the exact conservation of mass, see \cite{audusse:2011,audusse:2011b,fernandezNieto:2014,sainteMarie:2011}. Furthermore, contrary to ALE formulation, the description of the front could be very precise because the number of layers in the direction normal to the slope does not depend on the thickness of the flow and there is no deformation of the mesh. However, multilayer models could obviously not describe overturning of the front that may occur in some specific situations \cite{lusso:2017a}. Finally, multilayer discretization is well adapted to describe erosion processes in a thin layer of erodible material because, again, the vertical discretization does not depend on the material thickness and the numerical cost is quite low.\\

From a numerical point of view, many efforts have been devoted to the development of numerical schemes for depth-averaged models verifying the well-balance property and dealing properly with wet/dry fronts. For example, the hydrostatic reconstruction method is a technique that allows to recover the well-balanced property for depth-averaged models of avalanches by including wet/dry fronts (see e.g. \cite{audusse:2004,bouchut:2004,castro:2007}).  Another numerical treatment to deal with dry areas was introduced by Castro et al. \cite{castro:2005}, which was improved in \cite{castro:2006}. Par\'es \& Castro \cite{pares:2004} investigated the well-balancing of the Roe's method for non-conservative hyperbolic systems.   \\

In this paper we present a well-balanced discretization of the proposed multilayer shallow model with $\mu(I)$ rheology and lateral Coulomb friction. This discretization combines a particular hydrostatic reconstruction (see \cite{bouchut:2004}) and a specific treatment of wet/dry fronts for the elliptic part of the multilayer system. We present here several tests with dry areas, by including a comparison with the laboratory data of \cite{mangeney:2010} that includes wet/dry fronts. \\

The paper is organised as follows. Section \ref{se:finalModel} is devoted to the derivation of the 2D-model with the new approach to account for the side walls friction. In this section we also present the multilayer discretization of the proposed model (see \cite{fernandezNieto:2016}), and  add the discretization of the side walls friction term. In Section \ref{se:numerico} we propose a numerical scheme for the multilayer system, for which the well-balance property is achieved as consequence of the particular treatment of the rheological and friction terms. In section \ref{se:numericalTest} we present numerical tests, including the comparison with laboratory experiments. The influence of the side walls effect on the velocity profile as a function of the width of the channel for a uniform flow is also shown. Finally, some conclusions are presented in Section \ref{se:conclusions}.

\section{A 2D-model including lateral walls friction}\label{se:finalModel}
 \begin{figure}
\begin{center}
\includegraphics[width=0.4\textwidth]{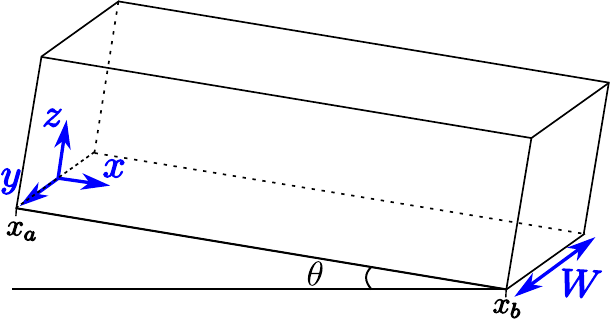}
\end{center}
 \caption{\label{fig:dominio} \it{Sketch of the rectangular domain}}
 \end{figure}
Let us consider tilted coordinates $(x,y,z)\in [x_a,x_b]\times[-W/2,W/2]\times\mathbb{R}$, with a constant slope $\theta$. Here, $W$ denotes the channel width, and the channel length is $x_b-x_a$, see Figure \ref{fig:dominio}. Fernandez-Nieto et al. \cite{fernandezNieto:2016} showed that the multilayer shallow model with the $\mu(I)$ rheology is able to reproduce typical velocity profiles of granular flows in the presence of lateral walls. In order to approximate lateral wall friction, they follow \cite{jop:2005} where the flow is assumed to be one-dimensional and uniform in the downslope direction $x$. In that case, side wall friction is introduced by adding a second term in the definition of the $\mu(I)$ law:
\begin{equation}\label{eq:frictioneffect}
\wtd{\mu(I)} =\mu(I) + \mu_w\dfrac{z_b+h-z}{W},
\end{equation}
where  $\mu_w$ is the constant friction coefficient at the lateral walls, and $z_b+h$ is the level of the free surface. The coefficient $\mu_w$ is usually different and lower than the coefficient used to model friction at the bottom (see e.g. \cite{baker:2016,capart:2015,ionescu:2015,martin:2017,jop:2006,jop:2007}). Let us look for a one-dimensional model for non-uniform flow in the $x$-direction that takes into account the friction with the lateral walls.\\

The three-dimensional Navier-Stokes system can be written as
\begin{equation}\label{eq:NS_3D}
\left\{
\begin{array}{l}
\p_{x} u+\p_y v + \p_{z}w= 0,\\[3mm]
\r\big(\p_{t}u+u\,\p_{x}u + v\,\p_{y}u+w\,\p_{z}u\big)+\p_{x}p = - \r g\sin\theta+\p_{x}\T_{xx}+\p_{y}\T_{xy}+\p_{z}\T_{xz},\\[3mm]
\r\big(\p_{t}v+u\,\p_{x}v + v\,\p_{y}v+w\,\p_{z}v\big)+\p_{y}p = \p_{x}\T_{yx}+\p_{y}\T_{yy}+\p_{z}\T_{yz},\\[3mm]
\r\big(\p_{t}w+ u\,\p_{x}w + v\,\p_{y}w + w\,\p_{z}w\big) +\p_{z}p=-\r g\cos\;\theta+\p_{x} \T_{zx} + \p_{y} \T_{zy} +\p_{z}\T_{zz},
\end{array}
\right.
\end{equation}
where $\vu = (u,v,w)$ is the velocity vector and $\rho=\varphi_s \rho_s$, $\varphi_s$ being the solid volume fraction assumed to be constant. At the free surface, we set the usual kinematic condition and we assume that the pressure vanishes. At the bottom, either the no-slip condition or Coulomb type friction can be considered. Moreover,  we consider a Coulomb type friction at the lateral boundaries, described as follows (see \cite{martin:2017}):
\begin{equation}\label{eq:BC_wall}
\bS\ \vn^{w} - \left(\left(\bS\ \vn^{w}\right)\cdot\vn^{w}\right)\vn^{w} = \left(\begin{matrix}
-\mu_w p\dfrac{u}{\abs{u}}, \ 0, \ 0\
\end{matrix}\right)',
\end{equation}
being $\vn^{w} = (0,\pm 1,0)'$ the normal vector at $y=\pm W/2$, respectively.\\

To derive a multilayer shallow model from dimensional analysis (see  \cite{fernandezNieto:2016}), we assume that the aspect ratio between the characteristic height $(H)$ and length $(L)$,
$$
\varepsilon = \dfrac{H}{L},
$$
is small.  Note that the influence of the lateral walls on the friction coefficient \eqref{eq:frictioneffect} is measured by the term $\mu_w(z_b+h-z)/W$. The dimension of this term is $H/L_y$, where $L_y$ is the characteristic width of the channel. Therefore, the lateral walls have a higher influence on the flow when the characteristic width $(L_y)$ of the channel is small in comparison with its characteristic height $(H)$.  We would like to study the influence of this scale into the system, then, in order to take it into account in the model we perform a dimensional analysis by also introducing the parameter
$$
\lambda = \dfrac{H}{L_y}.
$$
Notice that the higher the value of $\lambda$, the more important the lateral friction becomes.
Denoting the dimensionless variables with the tilde symbol ($\tilde{.}$), we define
\begin{equation*}
\label{nondim_var}
\begin{array}{c}(x,y,z,t) = (L\wtd{x},L_y\wtd{y},H\wtd{z},(L/U)\wtd{t}),
\quad W= L_y \wtd{W}, \\ \\
(u,v,w) =  (U\wtd{u},{\dfrac{\varepsilon}{\lambda }}U \wtd{v}, \varepsilon U\wtd{w}),\\
\\
h = H\wtd{h},
\quad
\r = \r_{0}\wtd{\r},\\
\\
p = \r_{0}U^{2}\wtd{p},
\quad
\eta= \r_{0}UH\wtd{\eta},\quad\eta_M = \r_{0}UH\wtd{\eta_M},\\
\end{array}
\end{equation*}

\begin{equation} \label{eq_adim_tau}
\left(\T_{xx},\T_{xy},\T_{yy},\T_{xz},\T_{yz},\T_{zz}\right) = \r_{0}U^2\left(\varepsilon\wtd{\T_{xx}},\wtd{\T_{xy}},\varepsilon\wtd{\T_{yy}},\wtd{\T_{xz}},\varepsilon\wtd{\T_{yz}},\varepsilon\wtd{\T_{zz}}\right).
\end{equation}

\noindent with $\r_0$ the characteristic density. Note that since

$$
D(\bd{\vu}) = \dfrac{U}{H}\ \frac 1 2 \left( \begin{matrix}
2\varepsilon \p_{\wtd{x}} \wtd{u} & {{\lambda}\p_{\wtd{y}} \wtd{u} + \frac{\varepsilon^2}{ \lambda}\p_{\wtd{x}} \wtd{v}} &         \p_{\wtd{z}}\wtd{u}+\varepsilon^2\p_{\wtd{x}}\wtd{w}\\
\\
{{\lambda}\p_{\wtd{y}} \wtd{u} + \frac{\varepsilon^2}{ \lambda}\p_{\wtd{x}} \wtd{v}} & 2\varepsilon \p_{\wtd{y}} \wtd{v} & { \varepsilon {\lambda}\p_{\wtd{y}} \wtd{w} +\frac{\varepsilon}{\lambda}\p_{\wtd{z}} \wtd{v}}\\
 \\
\p_{\wtd{z}}\wtd{u}+\varepsilon^2\p_{\wtd{x}}\wtd{w} &  { \varepsilon {\lambda}\p_{\wtd{y}} \wtd{w} +\frac{\varepsilon}{\lambda}\p_{\wtd{z}} \wtd{v}} & 2\varepsilon\p_{\wtd{z}}\wtd{w}
\end{matrix}\right),
$$
we obtain by (\ref{eq_adim_tau})
$$
\begin{array}{lll}
\wtd{\T_{xx}}= \wtd{\eta} \p_{\wtd{x}} \wtd{u}, &  \quad
\wtd{\T_{xy}}= \frac{\wtd{\eta}}{2} \left( {{\lambda}}\p_{\wtd{y}} \wtd{u} + {\frac{\varepsilon^2}{ \lambda}}\p_{\wtd{x}} \wtd{v} \right), & \quad
\wtd{\T_{xz}}= \frac{\wtd{\eta}}{2} \left( \p_{\wtd{z}}\wtd{u}+\varepsilon^2\p_{\wtd{x}}\wtd{w} \right),\\  \\
\wtd{\T_{yy}}= \wtd{\eta}  \p_{\wtd{y}} \wtd{v}, & \quad
\wtd{\T_{yz}}= \frac{\wtd{\eta}}{2} \left( {{\lambda}}\p_{\wtd{y}} \wtd{w} +{\frac{1}{\lambda}}\p_{\wtd{z}} \wtd{v} \right),  &
\quad
\wtd{\T_{zz}}= \wtd{\eta} \p_{\wtd{z}}\wtd{w}.
\end{array}
$$

\noindent Then, the system of equations (\ref{eq:NS_3D}) can be rewritten using the non-dimensional variables as (tildes have been dropped for simplicity):
\begin{equation}\label{eq:nondim_NS_3D}
\left\{
\begin{array}{l}
\p_{x} u+\p_y v +\p_{z}w= 0,\\[3mm]
\r\big(\p_{t}u+u\p_{x} u+ v\p_y u + w\p_{z}u\big)+\p_{x}p  =-\dfrac{1}{\varepsilon}\r \dfrac{1}{Fr^2}\tan\;\theta+\varepsilon\p_{x}\T_{xx}+{\frac{\lambda}{\varepsilon}}\p_{y}\T_{xy} +\dfrac{1}{\varepsilon}\p_{z}\T_{xz}, \\[3mm]
\r\big(\p_{t}v+u\,\p_{x}v + v\,\p_{y}v+w\,\p_{z}v\big)+{\dfrac{\lambda^2}{\varepsilon^2}}\p_{y}p = {\dfrac{\lambda}{\varepsilon}}\p_{x}\T_{yx}+{\dfrac{\lambda^2}{\varepsilon}}\p_{y}\T_{yy}+{\dfrac{\lambda}{\varepsilon}}\p_{z}\T_{yz},\\[3mm]
\r\varepsilon^2\big(\p_{t}w+ u\,\p_{x}w + v\,\p_{y}w + w\,\p_{z}w\big) +\p_{z}p=-\r \dfrac{1}{Fr^2}+\varepsilon\p_{x} \T_{zx} + {\varepsilon{\lambda}}\p_{y} \T_{zy} +\varepsilon\p_{z}\T_{zz},
\end{array}
\right.
\end{equation}
where $Fr$ denotes the Froude number,
\[Fr = \frac{U}{\sqrt{gH\cos\theta }}.\]

We now assume that the flow is one dimensional (i.e. $v=0$) and keeping all the terms involving $\lambda$, the previous system reads
\begin{equation}\label{eq:nondim_NS_1D}
\left\{
\begin{array}{l}
\p_{x} u +\p_{z}w= 0, \\[3mm]
\r\big(\p_{t}u+u\p_{x} u + w\p_{z}u\big)+\p_{x}p  =-\dfrac{1}{\varepsilon}\r \dfrac{1}{Fr^2}\tan\;\theta+{\frac{\lambda^2}{2\varepsilon}}\p_{y}\left(\eta \p_y u\right) +\dfrac{1}{2 \varepsilon}\p_{z}\left(\eta \p_z u \right) + {\mathcal O}(\varepsilon), \\[3mm]
\p_{y}p ={\mathcal O}(\varepsilon), \\[3mm]
\p_{z}p=-\r \dfrac{1}{Fr^2} + {\dfrac{\varepsilon\lambda^2}{2}\p_{y}\left(\eta \p_y w\right)}+\,{\mathcal O}(\varepsilon).
\end{array}
\right.
\end{equation}

\noindent Note that the term
\begin{equation}\label{eq:dyu}
{\frac{\lambda^2}{2\varepsilon}}\p_{y}\left(\eta \p_y u\right)
\end{equation}
is of the main order $1/\epsilon$ and 
collects the lateral friction effect on the momentum equation. Lateral walls friction has then a high influence on the flow, both on the norm of the maximum velocity and on its normal velocity profile.

Moreover, this is the term that allows us to introduce the lateral Coulomb friction in the model, by integrating in the horizontal transversal direction. To this aim, we define
 $$
 \begin{array}{l}
 \wb{u} = \dfrac{1}{W} \dint_{- W/2}^{W/2} u\,dy,  \quad \wb{w} = \dfrac{1}{W} \dint_{-W/2}^{W/2} w\,dy, \quad
 \wb{p} = \dfrac{1}{W} \dint_{-W/2}^{W/2} p\,dy,\quad \wb{\eta} = \dfrac{1}{W} \dint_{-W/2}^{W/2} \eta\,dy.
 \end{array}
 $$

We also assume that the perturbation with respect to the transversal averages are small, therefore we can approximate $\wb{fg}$ by $\bar{f}\,\bar{g}$, for any two variables $f$, $g$.
By integrating system \eqref{eq:nondim_NS_1D} with respect to the transversal direction between $-W/2$ and $W/2$ we obtain
$$
\left\{
\begin{array}{l}
\p_{x} \wb{u} +\p_{z}\wb{w}= 0, \\[2mm]
\r\big(\p_{t}\wb{u}+\wb{u}\p_{x} \wb{u} + \wb{w}\p_{z}\wb{u}\big)+\p_{x}\wb{p}  =-\dfrac{1}{\varepsilon}\r \dfrac{1}{Fr^2}\tan\;\theta +\dfrac{1}{\varepsilon}\p_{z}\left(\wb{\eta}\dfrac{\p_z \wb{u}}{2}\right)+
\,{\dfrac{\lambda^2}{W \varepsilon}} (\eta \p_y u)_{ |_{W/2}} + {{\mathcal O}(\varepsilon)}, \\[4mm]
p_{|_{W/2}} =p_{|_{-W/2}}+ {\mathcal O}(\varepsilon),  \\[3mm]
\p_{z}\wb{p}=-\r \dfrac{1}{Fr^2} +\dfrac{\varepsilon\lambda^2}{2 W }\left(\left(\eta\p_y w\right)_{|_{W/2}} - \left(\eta\p_y w\right)_{|_{-W/2}}\right)+{\mathcal O}(\varepsilon). \\[3mm]
\end{array}
\right.
$$
In the previous equation we have supposed a symmetric profile of $\eta\p_y u$, i.e., we assume that $\eta\p_y u_{ |_{-W/2}} = -\eta\p_y u_{ |_{W/2}}$. Moreover, from lateral friction condition (\ref{eq:BC_wall}) it follows that
\begin{equation}
{\lambda}\left(\dfrac{\eta}{2}{\p_y u}\right)_{|_{W/2}} = -\,\mu_w {p}_{|_{W/2}} \left(\dfrac{u}{\abs{u}}\right)_{|_{W/2}}
\qquad\text{and}\qquad \left(\eta\p_y w\right)_{|_{W/2}} = \left(\eta\p_y w\right)_{|_{-W/2}} = 0.
\end{equation}
Therefore, to obtain the final model, we neglect terms of order $\varepsilon$, leading to the first order model approximation
\begin{equation}\label{eq:nondim_NS_1D_2}
\left\{
\begin{array}{l}
\p_{x} \wb{u} +\p_{z}\wb{w}= 0, \\[2mm]
\r\big(\p_{t}\wb{u}+\wb{u}\p_{x} \wb{u} + \wb{w}\p_{z}\wb{u}\big)+\p_{x}\wb{p}  =-\dfrac{1}{\varepsilon}\r \dfrac{1}{Fr^2}\tan\;\theta +\dfrac{1}{\varepsilon}\p_{z}\left(\wb{\eta}\dfrac{\p_z \wb{u}}{2}\right)
-\,{\dfrac{2\lambda}{W \varepsilon}}\,\mu_w\, \wb{p} \,\dfrac{u}{\abs{u}}, \\[4mm]
\p_{z}\wb{p}=-\r \dfrac{1}{Fr^2}. \\[3mm]
\end{array}
\right.
\end{equation}

\noindent Going back to the dimensional variables, we get
\begin{equation}\label{eq:dim_NS_1D_2}
\left\{
\begin{array}{l}
\p_{x} \wb{u} +\p_{z}\wb{w}= 0, \\[5mm]
\r\big(\p_{t}\wb{u}+\wb{u}\p_{x} \wb{u} + \wb{w}\p_{z}\wb{u}\big)+\p_{x}\wb{p}  =-\r\,g\,\sin\theta +\p_{z}\left(\wb{\eta}\dfrac{\p_z \wb{u}}{2}\right)
-\,\dfrac{2}{W}\,\mu_w\, \wb{p}\,\dfrac{u}{\abs{u}}\, ,
\\[5mm]
\p_{z}\wb{p}=-\r \,g\cos\theta.
\end{array}
\right.
\end{equation}
\\
Hereafter bars are dropped by simplicity. We can find some similarities to previous models presented in the literature. In order to  make the comparison we must take into account that terms of order $\epsilon$ have been neglected to obtain the proposed model (\ref{eq:dim_NS_1D_2}). Thus, since
  $\|D(\vu)\| = \abs{\p_z u}/2+{\mathcal O}(\varepsilon)$, we obtain that model (\ref{eq:dim_NS_1D_2}) matches with the one introduced in \cite{martin:2017} for hydrostatic pressure and neglecting also here terms in $\varepsilon$.

  In addition, if the horizontal velocity verifies that $sign(u)=sign(\p_z u)$, which is usually the case, then it also coincides with the model proposed in \cite{jop:2005}.
In \cite{jop:2005} the lateral friction effect is defined for  uniform flows, by adding an extra term to the definition of $\mu(I)$ (equation \eqref{eq:frictioneffect}). Let us see that in fact it is equivalent up to first order in $\epsilon$ to model (\ref{eq:dim_NS_1D_2}), for flows verifying $sign(u)=sign(\p_z u)$ and not only for uniform flows.

We use \eqref{eq:viscosity_prev} with $\delta = 0$ and the previous approximation of $\|D(\vu)\|=\abs{\p_z u}/2+{\mathcal O}(\varepsilon)$, then the viscous term in \eqref{eq:dim_NS_1D_2} neglecting terms of order $\epsilon$ reads
\begin{equation}\label{eq:comp_jop}
\begin{array}{l}
\p_z\left( \dfrac{\mu(I)\,p(z)}{\abs{\p_z u}}\p_z u\right) -\,\dfrac{2}{W}\,\mu_w\,p(z)\,\dfrac{u}{\abs{u}}.
\end{array}
\end{equation}
Since we have a hydrostatic pressure, $p(z) = \r\,g\,(z_b+h-z)$, the second term in the previous equation can be rewritten as
$$
\begin{array}{c}
 -\,\dfrac{2}{W}\,\mu_w\,\r\,g\,\cos\theta\,(z_b+h-z)\,sign(u) \,=  \p_z\left(\dfrac{1}{W}\,\mu_w\,\r\,g\,\cos\theta\,(z_b+h-z)^2 sign\left(u\right)\right) = \\[4mm]
 = \, \p_z\left(\mu_w\dfrac{z_b+h-z}{W}\,p(z)\,sign\left(u\right)\right).
\end{array}$$
Therefore \eqref{eq:comp_jop} yields
$$
\begin{array}{l}
\p_z\left( \mu(I)\,p(z)\,sign\left(\p_z u\right) \,+ \,\mu_w\dfrac{z_b+h-z}{W}\,p(z)\,sign\left(u\right)\right).
\end{array}
$$
Now, if $sign(\p_z u) = sign(u) $,  we obtain
\begin{equation} \label{eq:defmudzu}
\p_z\left( \left(\mu(I) + \mu_w\dfrac{z_b+h-z}{W}\right)\,p(z)\,sign(\p_z u)\right) = \p_z\left( \wtd{\mu(I)}\,p(z)sign(\p_z u)\right),
\end{equation}
which is the viscous term resulting of considering the modified friction coefficient proposed in \cite{jop:2005}. Then, we obtain that the model (\ref{eq:dim_NS_1D_2}) and the one proposed in \cite{jop:2005} match in this case.

Note also that the term on the right hand side of equation (\ref{eq:defmudzu}) is  an approximation at order $\varepsilon$ of $div(\wtd{\mu(I)} p\frac{D(u)}{\|D(u)\|})$.  For example,  these terms are equal in the case of a uniform flow. However, $div(\wtd{\mu(I)} p\frac{D(u)}{\|D(u)\|})$
 cannot be rewritten as
$$
div({\mu(I)} p\frac{D(u)}{\|D(u)\|}) -\,\dfrac{2}{W}\,\mu_w\,p(z)\,\dfrac{u}{\abs{u}},
$$
 which is the term that appears in the full model (see \cite{ionescu:2015}). As a result, using $\wtd{\mu(I)}$ to describe side walls friction is not correct in general in the full 3D model. It can be justified if  we consider a model at first order in $\varepsilon$,  with hydrostatic pressure, and  $sign(\p_z u) = sign(u)$.

\bigskip

\noindent In the next subsection we present a multilayer discretization of the model (\ref{eq:dim_NS_1D_2}).

\subsection{A multilayer discretization}\label{subse:multilayer}
In this section we briefly describe  the multilayer approach for the system \eqref{eq:dim_NS_1D_2} (see \cite{fernandezNieto:2016,fernandezNieto:2014} for more details).
\subsubsection{General description}

 \begin{figure}
\begin{center}
\includegraphics[width=0.5\textwidth]{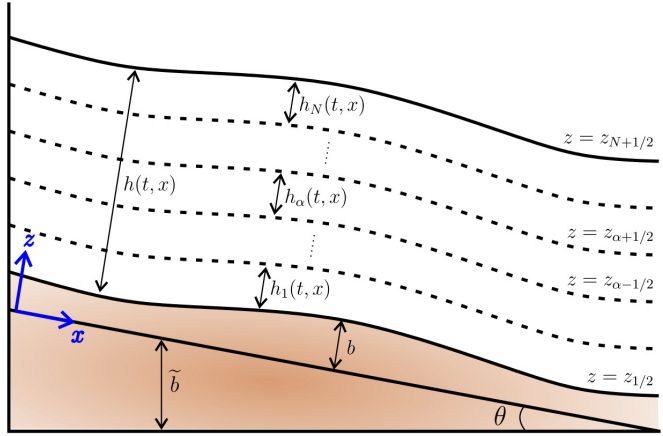}
\end{center}
 \caption{\label{fig:Multilayers} \it{Sketch of the multilayer division of the domain}}
 \end{figure}

 We consider a vertical partition of the domain in $N\in\mathbb{N}^*$ layers of thicknesses $h_\a(t,x)$ for $\alpha=1,...,N$, (see figure \ref{fig:Multilayers}), and therefore $\sum^N_{\a=1}h_\a = h$. In practice, we define the coefficients  $l_{\a}>0$ such that
$$
h_{\a} = l_{\a}h \quad\text{ for } \a = 1,...,N; \quad\quad
\dsum_{\a=1}^N l_{\a} = 1.
$$
These layers are separated by $N+1$ smooth interfaces $\G_{\a+\frac12}(t)$, whose equations are $z = z_{\a+\frac12}(t,x)$ for $\a = 0,1,..,N$. Note that the fixed bottom and the free surface are respectively the first and last interfaces $b = z_{\frac12}$ and $b+h = z_{N+\frac12}$. Note that $z_{\a+\frac12} = b + \sum_{\b=1}^{\a}h_{\b}$ and $h_\a = z_{\a+\frac{1}{2}} - z_{\a-\frac{1}{2}}$, for $\a = 1,...,N$. We consider the approximation of an arbitrary function $f$ at the interface $\G_{\a+\frac12}$ as $f_{\a+\frac{1}{2}}$.  Finally, $u_\a$ denotes the velocity in the layer $\a$, that is,
$$
u_{\a}(x) := \dfrac{1}{h_\a}\dint_{z_{\a-1/2}}^{z_{\a+1/2}}\bar{u}(x,z) dz \,.
$$
\noindent Now, to be consistent with our model \eqref{eq:dim_NS_1D_2}, we need to approximate the viscosity at the interface $\eta_{\a+\frac12}$ neglecting the terms of order $\varepsilon$. With this purpose, we consider as before $\|D(\vu)\|=\abs{\p_z u}/2+{\mathcal O}(\varepsilon)$, so the approximation at $z=z_{\a+\frac{1}{2}}$ is given by
\begin{equation} \label{eq:approx_norm_Du_o1}
\|D(\vu)\|_{{\a+\frac{1}{2}}} \approx \abs{ \QH_{,\a+\frac{1}{2}} }/2,
\end{equation}
where $\QH$ is introduced in order to approximate $\partial_z u$ in the multilayer framework, that is,
the possible discontinuity in the vertical profile of $\vu$. Then, $\Q$ satisfies
\begin{equation} \label{def_q}
\Q-\partial_z \vu=0, \quad \mbox{with} \quad \Q=(\QH,\QV).
\end{equation}
Firstly, we approximate $\vu$ by $\widetilde{\vu}$, a $\mathbb P_1(z)$ interpolation such that $\widetilde{\vu}_{|z=\frac{1}{2}(z_{\a-\frac{1}{2}}+z_{\a+\frac{1}{2}})}={\vu_{\alpha}}$. Thus, $\Q_{\a+\frac{1}{2}} = \left(\QH_{,\a+\frac{1}{2}},\QV_{,\a+\frac{1}{2}}\right)$ is an approximation of  $\Q(\widetilde{\vu})$ at $\Gamma_{\a+\frac{1}{2}}$. We choose
\begin{equation}\label{eq:dzu}
\QH_{,\a+\frac{1}{2}}= \frac{u_{\a+1}-u_{\a}}{h_{\a+\frac{1}{2}}}, \quad \quad \mbox{for} \quad \alpha=1, \dots,N-1, \quad
\end{equation}
with $h_{\a+\frac{1}{2}}$ the distance between the midpoints of layers $\a$ and $\a+1$. Note that $\QH_{,\frac12}$ and $\QH_{,N+\frac12}$ must be defined by the boundary condition at the bottom and free surface, respectively.\\\\

\noindent Therefore, the viscosity coefficient at the interface $\Gamma_{\a+\frac12}$ reads
\begin{equation} \label{eq:approx_eta_interfaz}
\eta_{\a+\frac{1}{2}} =\eta_{\a+\frac{1}{2}}(\QH_{,\a+\frac{1}{2}}) = \dfrac{\mu(I_{\a+\frac{1}{2}}) p_{\a+\frac{1}{2}}}{\sqrt{ \abs{ \QH_{,\a+\frac{1}{2}} }^2/4 + \delta^2}}\, ,
\end{equation}
for $\a=0,\dots,N-1$, and $\eta_{N+1/2}=0$ since we fix the atmospheric pressure, $p_S = 0$. In \eqref{eq:approx_eta_interfaz} the pressure is assumed hydrostatic, then
\begin{equation} \label{eq:def_I_interfaz}
p_{\a+\frac{1}{2}} = \r g\,cos\theta\,\sum_{\beta=\a+1}^N h_{\beta}\, , \quad\quad
I_{\a+\frac{1}{2}} = \frac{d_s \abs{ \QH_{,\a+\frac{1}{2}}}}{\sqrt{p_{\a+\frac{1}{2}}/ \r_{s}}}= \dfrac{d_s |u_{\a+1}-u_{\a}|}{h_{\a+\frac{1}{2}} \sqrt{ \varphi_s
g\,cos\theta\,\sum_{\beta=\a+1}^N h_{\beta}}} ,
\end{equation}
for $\a=0,\dots,N-1$. The definition of the viscosity at the bottom $\eta_{1/2}$ is particularly interesting. It will depend on the considered boundary condition, either no-slip or a Coulomb type friction. This will be discussed later.\\

Following the procedure presented in  \cite{fernandezNieto:2016}, the final $\mu(I)$ rheology multilayer model at first order in $\varepsilon$, including the lateral wall friction, reads, for $\smash\a = 1,...,N$,

\begin{align}
\label{eq:FinalModel}
\left\{
\begin{array}{l}
l_{\a} \bigg( \p_{t}h + \p_{x}(h u_{\a}) \bigg) = G_{\a+\frac{1}{2}} - G_{\a-\frac{1}{2}}, \\
\\
l_{\a} \bigg( \r\p_{t}\left(h u_{\a}\right) \;+\; \r\p_{x}\left(h u_{\a}^2\right) \, + \r g\;\cos\theta \,h\,\p_{x} \left(z_b+h\right)  \bigg) \, = K_{\a-\frac{1}{2}} - K_{\a+\frac{1}{2}}\, +\\[4mm]
\qquad \qquad \qquad +\, \dfrac{1}{2} \r G_{\a+\frac{1}{2}}\left(u_{\a+1} + u_{\a}\right) \;-\; \dfrac{1}{2} \r G_{\a-\frac{1}{2}}\left(u_{\a} + u_{\a-1}\right) \,+\, M_{\a,W},
\end{array}
\right.
\end{align}
\\
where $$z_b = b + \wtd{b}; \qquad \wtd{b} = -x\,\tan\theta$$ is the bottom topography. $G_{\a+\frac{1}{2}}$ is the mass transference between the layers $\a$ and $\a+1$, written as
$$
G_{\a+\frac{1}{2}} = \p_{t}z_{\a+\frac{1}{2}} + \frac{u_{\a}+ u_{\a+1}}{2}\p_{x}z_{\a+\frac{1}{2}} - w_{\a+\frac{1}{2}},
\quad \mbox{where} \quad w_{\a+\frac{1}{2}}= \frac{w_{\a+\frac{1}{2}}^+ + w_{\a+\frac{1}{2}}^-}{2}.
$$
The vertical velocity is a piecewise linear function defined through its upper and lower limits at the interfaces. The limit at the interface $z=z_{\a+\frac{1}{2}}$ verifies the jump condition
$$
w_{\a+\frac{1}{2}}^+ = w_{\a+\frac{1}{2}}^- + (u_{\a+1}-u_{\a}) \partial_x z_{\a+\frac{1}{2}},
$$
and by the linear profile of the vertical velocity inside layer $\a$ and the incompressibility condition we have
$$
w_{\a+\frac{1}{2}}^- = w_{\a-\frac{1}{2}}^+ - h_{\a} \partial_x u_{\a}.
$$

The side walls friction is taking into account through the term $M_{\a,W}$. Following the multilayer procedure we obtain that
$$
\begin{array}{l}
M_{\a,W} = -\dint_{z_{\a-1/2}}^{z_{\a+1/2}} \dfrac{2}{W}\mu_w\dfrac{u_\a}{\abs{u_\a}}\,\r\,g\,\cos\theta\,(z_b+h-z)\,dz = \dfrac{2}{W}\mu_w\dfrac{u_\a}{\abs{u_\a}}\,\r\,g\,\cos\theta\,\left.\dfrac{(z_b+h-z)^2}{2}\right]_{z_{\a-1/2}}^{z_{\a+1/2}}
\end{array}
$$
After some algebra we get
$$
\left.\dfrac{(z_b+h-z)^2}{2}\right]_{z_{\a-1/2}}^{z_{\a+1/2}} = -h_\a\left(\dsum_{\b=\a+1}^N h_\b + \dfrac{h_\a}{2} \right) =-h_\a\left(z_b + h -\left(z_b+\dsum_{\b=1}^{\a-1}h_\b + \dfrac{h_\a}{2} \right) \right)
$$
Therefore, denoting $p_\a= \r \,g\,\cos\theta\left(z_b + h -\left(z_b+\dsum_{\b=1}^{\a-1}h_\b + \dfrac{h_\a}{2} \right) \right) $, the pressure in the midpoint of layer $\a$, the lateral walls friction term is written
\begin{equation}\label{eq:friccionlateral}
M_{\a,W} = -l_\a\,h\,\dfrac{2}{W}\mu_w\,p_\a\,\dfrac{u_\a}{\abs{u_\a}}.
\end{equation}

\noindent Finally, the viscous term $K_{\a+\frac12}$ is defined by
\begin{equation}\label{eq:FinalModel2}
K_{\a+\frac{1}{2}} = - \dfrac12\eta_{\a+\frac{1}{2}}(\QH_{,\a+\frac{1}{2}})\, \QH_{,\a+\frac{1}{2}}, \quad \a=1, \dots, N-1
\end{equation}
for $\eta_{\a+\frac{1}{2}}$ defined in (\ref{eq:approx_eta_interfaz})-(\ref{eq:def_I_interfaz}). The terms $K_{\frac{1}{2}}$ and $K_{N+\frac{1}{2}}$ are defined by the boundary conditions at the bottom and the free surface, respectively (see section \ref{se:bc}).
\\

\noindent Model \eqref{eq:FinalModel} has $2N$ equations and unknowns, however the continuity equations can be combined (see \cite{fernandezNieto:2014}) to achieve a system with $N+1$ equations and unknowns: the total height and the discharge of each layer, i.e., $(h, q_{1},\dots,q_N)$, where $q_\a = h\,u_\a$, for $\a = 1,\dots,N$. By defining the auxiliary coefficients
\begin{align*}
\xi_{\a,\gamma} =
\begin{cases}
\bigl(  1 - (l_{1} + \dots + l_{\alpha}) \bigr)l_{\gamma}, & \text{if $\gamma \leq \a$} ,\\
\\
- (l_{1} + \dots + l_{\alpha}) l_{\gamma}, &  \text{otherwise},
\end{cases}
\end{align*}
for $\alpha,\;\gamma \in \{1,\dots,N\}$, the  system \eqref{eq:FinalModel}-\eqref{eq:FinalModel2} is rewritten as
\begin{equation}
\label{eq:swHydro_syst_num}
\left\{
\begin{array}{l}
\p_{t}h  + \p_{x}\,\Biggl( \dsum_{\b=1}^N l_{\b}q_{\b}\Biggr) =  0, \\
\\
\p_{t}q_{\a} + \p_{x}\left(\dfrac{q_{\a}^2}{h} \;+ \, g\,\cos\theta\dfrac{h^2}{2}  \right)\,+\, g\,\cos\theta\,h\,\p_{x}z_b \, +\\[4mm]
\quad+ \dsum_{\gamma=1}^N\dfrac{1}{2hl_{\a}}\bigg(  \left(q_{\a} + q_{\a-1}\right)\xi_{\a-1,\gamma} - \left(q_{\a+1} + q_{\a}\right)\xi_{\a,\gamma}\bigg)\,\p_{x}q_{\gamma} \ = \\[4mm]
\quad =  \dfrac{1}{\r l_{\a}} \Big(K_{\a-\frac{1}{2}} - K_{\a+\frac{1}{2}} \,+\,M_{\a,W}\bigg) \hfill \a=1,\dots,N.
\end{array}
\right.
\end{equation}
\bigskip

\subsubsection{Boundary conditions}\label{se:bc}
The boundary condition at the free surface is simply defined by taking into account that the atmospheric pressure is neglected ($p_S = 0$), therefore $K_{N+\frac{1}{2}} = 0$.
 \begin{figure}[!h]
\begin{center}
\includegraphics[width=0.62\textwidth]{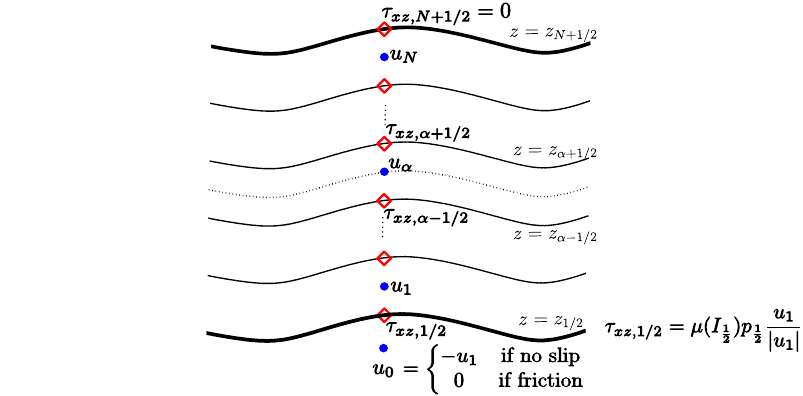}
\end{center}
 \caption{\label{fig:grid} \it{Sketch of the placing of the variables in the multilayer domain}}
 \end{figure}
The term $K_{1/2}$ is defined by the boundary condition at the bottom. A difficult task is to strongly impose the no-slip or Coulomb type friction boundary condition at the bottom in multilayer models. A good way to impose strongly the no slip condition at the bottom would be to calculate the velocities at the vertical interfaces $\Gamma_{\alpha+1/2}$. On the contrary, multilayer models calculate averaged velocities within the layer, which in turn is a second order approximation of the velocity at the middle of the layer (see figure \ref{fig:grid}). As an example, in the first layer we have $u_1 = u(h_1/2) + {\mathcal O}(h_1^2)$. As a result, we can only impose the boundary conditions in a weak sense and the no slip condition is not exactly achieved, as it can be observed when looking in details the numerical results (see section \ref{se:test_unif}).\\ 
Furthermore, we can not impose strongly a Coulomb type boundary condition since the unknowns of the system are the velocities and not the stresses, contrary to Lagrangian Augmented method (see \cite{ionescu:2015}), for example.

Let us propose simple ways to weakly impose no-slip and Coulomb type boundary condition. The key point is to approximate $\p_z u$ at the bottom. The value of $\QH$ as an approximation of $\partial_z u$ at the bottom $z=1/2$ depends on the velocity in the first layer $u_1$. In general we assume the following approximation, $\QH_{,\frac{1}{2}}=\frac{u_1-u_{0}}{h_1}$ where $u_{0}$ represents the velocity in a fictitious layer under the bottom level.\\
If we consider a Coulomb friction law, the stress tensor must verify the condition
\begin{equation}\label{eq:BC_fondo}
\bS\ \vn^{b} - \left(\left(\bS\ \vn^{b}\right)\cdot\vn^{b}\right)\vn^{b} = \left(\begin{matrix}
\mu(I_{\frac12})p_{\frac12}\dfrac{u_1}{\abs{u_1}}, \ 0\
\end{matrix}\right)',
\end{equation}
where $\vn^{b}$ is the downward unit normal vector to the bottom. We can consider either a friction law with a constant parameter ($\mu(I_{\frac12})=\mu_s$) as in \cite{martin:2017}, or given by the expression of the $\mu(I)$ rheology. In this case, we consider the approximation of $\p_zu$ at the bottom considering that the velocity $u_{0}=0$,
$$
\QH_{,\frac{1}{2}}= \frac{u_1}{h_1}.
$$
This makes it possible to obtain a non-zero velocity at the bottom.

\medskip

\noindent Then, the term $K_{\frac{1}{2}}$ is given by condition \eqref{eq:BC_fondo},
\begin{equation}\label{eq:friccion}
K_{\frac{1}{2}} = - \r\,g\,h\,cos\theta\,\mu(I_{\frac12})\dfrac{u_1}{\abs{u_1}}, \quad\mbox{where}\quad I_{\frac12} = \frac{d_s \abs{u_1/h_1}}{\sqrt{\varphi_s \,g\,h\,cos\theta}}.
\end{equation}
If no-slip condition is considered then we must change the approximation $\QH_{,\frac12}$ because now the velocity must vanish at $z=1/2$, so we introduce $u_{0}=-u_1$. Hence we consider the approximation
$$
\QH_{,\frac{1}{2}}= \frac{2 \, u_1}{h_1}.
$$
Then $K_{1/2}$ is given by
\begin{equation}\label{eq:noslip}
K_{\frac{1}{2}} = - \r\,g\,h\,cos\theta\,\mu(I_{\frac12})\dfrac{u_1}{\abs{u_1}}, \quad\mbox{where}\quad I_{\frac12} = \frac{d_s \abs{2\,u_1/h_1}}{\sqrt{\varphi_s \,g\,h\,cos\theta}}.
\end{equation}
As conclusion, the viscous term at the bottom ($K_{1/2}$) defined from a no-slip condition only differs from the one when considering a Coulomb friction law in the inertial number:
\begin{equation} \label{eq:noslip2}
I_{\frac12,\,\text{No slip}} = 2\,I_{\frac12,\,\text{Coulomb}}\ .
\end{equation}

\bigskip

\noindent In the next section we detail the numerical discretization of the proposed multilayer system \eqref{eq:swHydro_syst_num}.
\section{Numerical approximation}\label{se:numerico}
In the literature, multilayer systems have been discretized by combining a usual finite volume method with a splitting procedure \cite{audusse:2005,audusse:2014,audusse:2011,audusse:2008,audusse:2011b,fernandezNieto:2014}. Thus, authors usually separate the viscous terms, which are treated with a semi-implicit scheme. We follow this structure in a particular way. One of the main contribution of our previous work \cite{fernandezNieto:2016} was to introduce a multilayer system with non-constant viscosity. Nevertheless, its numerical approximation was not explained. To our knowledge it's the first time that a numerical scheme for a multilayer system with non-constant viscosity is exposed. 
These rheological terms add new difficulties, namely when looking for a well-balanced scheme that was not achieved in \cite{fernandezNieto:2016}. Here we consider a hydrostatic reconstruction in a finite volume method in order to ensure the well-balance property.\\

Firstly, we can write the system \eqref{eq:swHydro_syst_num} in matrix notation as
\begin{align}
\label{eq:compact_form}
\p_{t}\bd{w} + \p_{x}\bd{F}(\bd{w}) +  \bd{S}(\bd{w})\p_{x} z_{b}+ \bd{B}(\bd{w})\p_{x}\bd{w} = \bd{E}(\bd{w})
\end{align}
where $\bd{w} = \left(h, q_{1}, q_{2}, ... , q_{N} \right)^{'} \in \Omega\subset\R^{N+1}$ is the unknown vector, $\bd{F}(\bd{w}) = (\bd{F}_{\a}(\bd{w}))_{\a=0,1,...,N}$ is a regular function from $\R^{N+1}$ to $\R^{N+1}$, $\,\bd{B}(\bd{w}) = (\bd{B}_{\a,\b}(\bd{w}))_{\a,\b=0,1,...,N}\,$ is a regular matrix function from $\R^{N+1}$ to $\mathcal{M}_{N+1}(\R)$, $\bd{S}(\bd{w}) = (\bd{S}_{\a}(\bd{w}))_{\a=0,1,...,N}$, and $\bd{E}(\bd{w}) = (\bd{E}_{\a}(\bd{w}))_{\a=0,1,...,N}\,$ are vectorial functions from $\R^{N+1}$ to $\R^{N+1}$.

$\bd{F}_{\a}(\bd{w})$ and $\bd{S}_{\a}(\bd{w})$ are defined by the convective and pressure terms, respectively,
$$
\bd{F}_{\a}(\bd{w}) =
\left\{
\begin{array}{l}
\dsum_{\b=1}^N l_{\b}q_{\b},  \quad \mb{if\;\;} \a = 0, \\
\\
\dfrac{q_\a^2}{h} +g \cos \theta \dfrac{h^2}{2}, \quad \mb{if\;\;} \a = 1, ... , N;\\
\end{array}
\right. \qquad \qquad
\bd{S}_{\a}(\bd{w}) =
\left\{
\begin{array}{l}
0,  \quad \mb{if\;\;} \a = 0, \\
\\
 g \cos \theta h, \quad \mb{if\;\;} \a = 1, ... , N.
\end{array}
\right.
$$
Note that the addition of convective and pressure terms can be written as
$$
\p_{x}\bd{F}(\bd{w}) +  \bd{S}(\bd{w})\p_{x} z_{b}= \p_{x}\bd{F_c}(\bd{w}) +  \bd{S}(\bd{w})\p_{x} (z_{b}+h),
\quad
\mbox{with}
\quad
\bd{F_c}(\bd{w}) =\bd{F}(\bd{w}) - \frac{h}{2}  \bd{S}(\bd{w}).
$$
Then, $\bd{F_c}(\bd{w})$ contains the convective term and pressure terms are defined by $\bd{S}(\bd{w})\p_{x} (z_{b}+h)$. On the other hand, $\bd{B}_{\a, \b}(\bd{w})$ is defined in terms of the momentum transference terms,
$$
\bd{B}_{\a, \b}(\bd{w}) =
\left\{
\begin{array}{l}
0 , \quad \mb{if\;\;} (\a, \b) \in \{0\}\times\{0,1,...,N\}\, \cup \,\{1,...,N\}\times\{0\},\\
\\
\dfrac{1}{2hl_{\a}} \left(q_{\a} + q_{\a-1}\right)\xi_{\a-1,\b} - \dfrac{1}{2hl_{\a}} \left(q_{\a+1} + q_{\a}\right)\xi_{\a,\b}, \quad \mb{if\;\;} \a,\b = 1, ... , N.\\
\\
\end{array}
\right.
$$
The viscous terms are included in the definition of $\bd{E}_{\a}(\bd{w})$:\\
$$
\bd{E}_{\a}(\bd{w}) =
\left\{
\begin{array}{l}
0,  \quad \mb{if\;\;} \a = 0, \\
\\
\dfrac{1}{\r l_{\a}}\Big(K_{\a-\frac{1}{2}} - K_{\a+\frac{1}{2}} + M_{\a,W}\Big), \quad \mb{if\;\;} \a = 1, ... , N.\\
\end{array}
\right.
$$
\\
Next, we detail the two step of the splitting procedure. In the first step we consider the hyperbolic system with the non-conservative products, corresponding to the momentum transference terms between the vertical layers. In the second step we deal with the viscous terms.

Before describing these processes, let us focus on the treatment of the bottom condition because it plays a crucial role in order to achieve the well-balance property.
The numerical discretization must solve two different difficulties related to the well-balance property. The first one is physical, i.e.,  the Coulomb friction at the bottom and the walls  must behave as a force which opposes the movement of the granular flow. When the total friction is greater than the sum of the other forces acting on the system, then we should obtain $u_{\a} = 0$ for $\a=1,\dots,N$. This effect is achieved in the second step, through the discretization of the bottom friction term $K_{1/2}$. The second difficulty is a numerical issue. We use a Riemann solver in order to solve the hyperbolic part of the system, which introduces numerical diffusion. This artificial diffusion must be zero in order to ensure $\p_t h = 0$ when the granular flow has stopped, i.e., when $u_\a = 0$ for $\a=1,\dots,N$. Next, we describe the two steps of the numerical approximation:

\bigskip

\noindent{\bf Step 1}: Firstly, we do not consider viscous effects, that is, $\bd{E}(\bd{w}) = \bd{0}$. Then, we consider a finite volume solver to discretize system \eqref{eq:compact_form}. Namely, we consider a HLL type method defined as follows:
\begin{equation} \label{eq:step1}
\begin{array}{l}
\bd{w}_{i}^{n+1/2}=\bd{w}_i^n +\dfrac{\Delta t}{\Delta x} \left( \F_{c,i-1/2}^n-\F_{c,i+1/2}^n \,
+ \,\dfrac{1}{2} \left(\B_{i+1/2}^n+\B_{i-1/2}^n + \mathcal{S}_{i+1/2}^n+\mathcal{S}_{i-1/2}^n \right)\right),
\end{array}
\end{equation}
with
\begin{equation*}
\B_{i+1/2}^n  =\dfrac{1}{2} (\bd{B}(w_{i+1}^n)+\bd{B}(w_{i}^n)) \bigl(\bd{w}_{i+1}^n-\bd{w}_i^n\bigr) ,
\end{equation*}
and
\begin{equation*}
\mathcal{S}_{i+1/2}^n =  \dfrac{1}{2}(\bd{S}(w_{i+1}^n)+\bd{S}(w_{i}^n)) \bigl(h_{i+1/2+}^n - h_{i+1/2-}^n  \bigr),
\end{equation*}
where $h_{i+1/2\pm}$ is defined by the hydrostatic reconstruction introduced in \cite{audusse:2004}:
\begin{equation}\label{eq:recfondo}
\begin{array}{c}
h_{i+1/2-} = \mbox{max}(0,h_i - (\Delta Z_{i+1/2})_+);\\[4mm]
h_{i+1/2+} = \mbox{max}(0,h_{i+1} - (-\Delta Z_{i+1/2})_+),
\end{array}
\end{equation}
where
\begin{equation}\label{eq:recfriccion1a}
 (\Delta Z_{i+1/2})_+ = \mbox{max}(0,z_{b,i+1}-z_{b,i}).
\end{equation}

The numerical flux associated to the convective terms, $\F_{c,i+1/2}^n$, is
\begin{equation*}
\begin{array}{l}
\F_{c,i+1/2}^n = \dfrac{1}{2} \left(\bd{F_c} (\bd{w}_i^n)+\bd{F_c}(\bd{w}_{i+1}^n)\right) \,
 - \,\dfrac{1}{2}\mathcal{D}_{i+1/2}^n, \\
\end{array}
\end{equation*}
where $\mathcal{D}_{i+1/2}^n$ is the numerical diffusion of the scheme. Let us remark that this method can be seen as a path-conservative method with a second order approximation of the Roe matrix by setting the paths as segments (see \cite{pares:2004}).\\
Thus, in order to define the numerical diffusion, we consider the HLL extension proposed in \cite{castro:2012}. In this paper authors proposed a general formulation of numerical methods where the numerical viscosity matrix is defined in terms of the evaluation of a polynomial on the Roe matrix. In our case, taking into account that we use a second order approximation of Roe matrix by segments and the fact that we introduce a well-balanced correction associated to the Coulomb friction term, the numerical diffusion is defined as follows:
\begin{equation}\label{eq:difusion}
\mathcal{D}_{i+1/2} = a_0\,\left(\bd{\widehat{w}}_{i+1}^n-\bd{\widehat{w}}_i^n \right) \,+\, a_1\,
 \left(\bd{F_c} (\bd{w}_{i+1}^n)-\bd{F_c}(\bd{w}_{i}^n) +  \B_{i+1/2}^n + \mathcal{S}_{i+1/2}^n\right),
 \end{equation}
 with
 \[a_0 = \dfrac{S_R|S_L|-S_L|S_R|}{S_R-S_L},\hspace{0.2cm}a_1 = \dfrac{|S_R|-|S_L|}{S_R-S_L},\]
being $S_L$ and $S_R$ approximations of the minimum and maximum wave speed. In practice, to defined $S_L$ and $S_R$ we consider a baroclinic approximation,
$$
S_L= \min \left(\dsum_{\a=1}^N l_{\a} u_{\a,i}^n- \sqrt{g \cos \theta h_i^n}, \, \dsum_{\a=1}^N l_{\a} u_{\a,i+1/2}^n -\sqrt{g \cos \theta h_{i+1/2}^n} \right),
$$
$$
S_R= \max \left(\dsum_{\a=1}^N l_{\a} u_{\a,i+1}^n+ \sqrt{g \cos \theta h_{i+1}^n}, \, \dsum_{\a=1}^N l_{\a} u_{\a,i+1/2}^n +\sqrt{g \cos \theta h_{i+1/2}^n} \right).
$$
In \eqref{eq:difusion} we use the reconstructed states
$$
\bd{\widehat{w}}_i^n = (\hat{h}_{i+1/2-},q_{1,i},\dots,q_{N,i}),
\qquad
\bd{\widehat{w}}_{i+1}^n = (\hat{h}_{i+1/2+},q_{1,i+1},\dots,q_{N,i+1}),
$$
where $\hat{h}_{i+1/2 \pm}$ is defined by \eqref{eq:recfondo} taking in this case
\begin{equation}\label{eq:recfriccion1}
(\Delta Z_{i+1/2})_+ = \mbox{max}(0,z_{b,i+1}-z_{b,i} + \Delta {\mathcal C}_{i+1/2}),
\end{equation}
with $\Delta {\mathcal C}_{i+1/2} = -f_{i+1/2}\Delta x_{i+1/2}$ defined by considering the Coulomb (or no slip) friction term. Several definitions of $f_{i+1/2}$ can be given (see \cite{bouchut:2004}), in this work we set

\begin{equation}\label{eq:recfriccion2}
f_{i+1/2} = -\mathop{proj}\limits_{g\mu_{\beta}}\left(\dfrac{-g(h_{i+1}  + z_{b,i+1} -h_i-z_{b,i})}{\Delta x} - \dfrac{u_{\b,i+1/2}}{\Delta t}\right),
\end{equation}
where
$$
\mu_{\beta}= \mu_s + \frac{2}{W} \mu_w\left(\dsum_{\gamma=\b+1}^{N}h_{\gamma} + \dfrac{h_\b}{2} \right)
$$
being $\b$  the lowest layer that is moving, i.e., $\abs{u_\b} > 0$. If all the layers are at rest then $\beta=N$. Moreover,
\begin{equation}\label{eq:recfriccion3}
\mathop{proj}\limits_{g\mu_{\beta}}(X) = \left\{\begin{array}{lll}
X & \mbox{if} & |X| \leq g\mu_{\beta};\\
g\mu_{\beta} \dfrac{X}{|X|} & \mbox{if} &  |X| > g\mu_{\beta},
\end{array}\right.
\end{equation}
and $u_{\b,i+1/2}$ is an average state of the velocity at layer $\b$. For example we can set the Roe average state
$$
u_{\b,i+1/2} = \dfrac{\sqrt{h_i}\,u_{\b,i}+\sqrt{h_{i+1}}\,u_{\b,i+1}}{\sqrt{h_i}+\sqrt{h_{i+1}}}.
$$
In practice, this term is important when the granular flow is stopping. In general, upper layers are the last ones that stop in granular flows, then we can also consider $u_{\b,i+1/2} = u_{N,i+1/2}$. Note that the first condition in \eqref{eq:recfriccion3} gives the well-balance property by ensuring that the numerical diffusion is zero when the velocity is also zero.

\bigskip
\noindent{\bf Step 2}: Now, we must add the contribution of $\bd{E}(\bd{w})$. With this purpose, a semi-implicit discretization is considered:
 \begin{equation}\label{eq:step2}
\bd{w}^{n+1}_i = \bd{w}_{i}^{n+1/2} + \Delta t\ \bd{E}(\bd{w}^{n}_i,\bd{w}^{n+1}_i),
\end{equation}
where $\bd{w}^{n+1/2}_i$ is the approximation \eqref{eq:step1}. Note that the first component of $\bd{E}(\bd{w})$ is 0, therefore we clearly obtain $h^{n+1} = h^{n+1/2}$.

We get $\bd{q}^{n+1}_i = \left(q_1^{n+1},\dots,q_N^{n+1}\right)_i$ as solution of the $N\,\times\,N$ tridiagonal system
\begin{equation}\label{eq:semiq}
q^{n+1}_{\a,i}  = q_{\a,i}^{n+1/2} + \dfrac{\Delta t}{\r\ l_\a}\left(  \eta_{\a+\frac{1}{2}}(\bd{w}^{n}_i)\dfrac{u_{\a+1,i}^{n+1}-u_{\a,i}^{n+1}}{l_{\a+\frac12}h_{i}^{n+1}} - \eta_{\a-\frac{1}{2}}(\bd{w}^{n}_i)\dfrac{u_{\a,i}^{n+1}-u_{\a-1,i}^{n+1}}{l_{\a-\frac12}h_{i}^{n+1}}   +M_{\a,W}^{n,n+1} \right),
\end{equation}
 for $\a = 2,...,N-1$, where $\eta_{\a+\frac12}$ is defined by \eqref{eq:approx_eta_interfaz}. The lateral side walls friction terms approximation is
 \begin{equation}\label{eq:friccionlateral2}
M_{\a,W, i}^{n,n+1} = -l_\a\,h^n_i\,\dfrac{2}{W}\mu_w\,p_{\a,i}^n\,\dfrac{u^{n+1}_{\a,i}}{\sqrt{\abs{u^n_{\a,i}}^2 + \delta^2}}\, ,
\end{equation}
with
$$
p_{\a,i}^n= \r\,g\cos\theta\,\left(z_{b,i} + h_i^n -\left(z_{b,i}+\dsum_{\b=1}^{\a-1}h_{\b,i}^n + \dfrac{h_{\a,i}^n}{2} \right) \right) .
$$

The equations for $\a = 1$ and $\a = N$ can be analogously obtained taking into account that $\eta_{N+1/2} = 0$ and the definition of $K_{1/2}$. For the first layer we obtain
\begin{equation}\label{eq:semiq_1}
q^{n+1}_{1,i}  = q_{1,i}^{n+1/2} + \dfrac{\Delta t}{\r\ l_1}\left(  \eta_{\frac{3}{2}}(\bd{w}^{n}_i)\dfrac{u_{2,i}^{n+1}-u_{1,i}^{n+1}}{l_{\frac32}h_{i}^{n+1}} - \r\,g\,h\,cos\theta\,\mu(I_{\frac12})\dfrac{u_1}{\abs{u_1}}   +M_{1,W}^{n,n+1} \right),
\end{equation}
and for the last one
\begin{equation}\label{eq:semiq_N}
q^{n+1}_{N,i}  = q_{N,i}^{n+1/2} + \dfrac{\Delta t}{\r\ l_N}\left(  M_{N,W}^{n,n+1}  - \eta_{N-\frac{1}{2}}(\bd{w}^{n}_i)\dfrac{u_{N,i}^{n+1}-u_{N-1,i}^{n+1}}{l_{N-\frac12}h_{i}^{n+1}} \right),
\end{equation}

Note that the symmetric matrix associated to this linear system is a strictly diagonally dominant matrix, therefore the system is well-conditioned. Finally, we use the Thomas algorithm to solve each tridiagonal system.

Let us remark that the friction conditions at the bottom and lateral walls are considered directly in the definition of the linear system. With this purpose we consider two different hydrostatic reconstruction in the first step, which are defined by (\ref{eq:recfondo})-(\ref{eq:recfriccion1a}) and (\ref{eq:recfondo})-(\ref{eq:recfriccion1}). In the first one we deal with the change on the topography but the friction at the bottom is not taken into account. In the second one, the friction condition is managed in order to achieve a well balanced scheme. An important remark is that the numerical treatment of the friction condition would not be consistent if we include the friction condition in the first reconstruction. This is because in that case the friction at the bottom would be added twice in a time step.

The last consideration that we do is related with solving the linear systems. It corresponds with solving a vertical diffusion in each cell. We only solve the linear system in the cell $I_i$ if the total height $h_i^{n+1}$ is larger than $\varepsilon_S$. Otherwise, the friction law at the bottom together with the lateral walls friction are considered as in the case of a single-layer model. The friction term (bottom and side walls) is applied to the first layer and we neglect the vertical variations in those cell, i.e., we set $q_{\a,i}^{n+1} = q_{1,i}^{n+1}$ for $\a = 2,...,N$.

Therefore, in those cells for which $0< h_i^{n+1} < \varepsilon_S$ (or if we consider the single-layer model $N=1$),  we define
\begin{equation}\label{eq:coulomb}
q_{\a,i}^{n+1}=\left\{ \begin{array}{ll}
q_{1,i}^{n+1/2}- \Delta t  \r\,g\,\,cos\theta\, h_i^{n+1} \mu_{1,i}^n\dfrac{q_{1,i}^{n+1}}{\abs{q_{1,i}^n}}, & \mbox{if }  |q_{1,i}^{n+1/2}| > \Delta t  \, \sigma_{c,i}^{n,n+1}\\
 0, & \mbox{otherwise,}
 \end{array} \right. \quad  \, \alpha=1, \dots, N,
\end{equation}
being $\sigma_{c,i}^{n,n+1} = \r\,g\,h_i^{n+1} \,cos\theta\, \mu_{1,i}^n$, with $\mu_{1,i}^n = \mu(I_{\frac12,i}^n) + \dfrac{h^n_i}{W} \mu_w$, where $I_{\frac12,i}^n$ is defined by (\ref{eq:friccion}) if we consider a Coulomb friction law, or by (\ref{eq:noslip}) for a no slip condition.
\bigskip

\section{Numerical tests}\label{se:numericalTest}
In this section we show four numerical tests in order to validate the model and the numerical scheme presented in previous sections. Firstly, in Subsection \ref{se:test_unif}, we consider a uniform flow and investigate the influence of the lateral wall friction on the vertical profile of velocity. We also study the evolution of two critical values of the channel width for different bottom slopes: (i) $W_c$ that is the first value for which all the granular mass is moving, i.e., if $W\geq W_c$ there is no flow/no-flow interface; (ii) $W_b$ that is the first value for which the downslope velocity along the normal direction has a Bagnold profile and not a S-shaped profile. Secondly, in Subsection \ref{se:WB}, we perform a test focused on the well-balance property of the scheme, combined with the treatment of the wet/dry front. Third, in Subsection \ref{se:h_est}, we show that approximating the side walls friction through
a single-layer model could lead to non-physical solutions that strongly differ from those computed with a multilayer model. Finally, in Subsection \ref{se:labexp}, we compare the results obtained with our model to laboratory experiments of granular column collapse.\\
\medskip

All the tests  are computed over a reference inclined plane of angle $\theta$ (titled coordinates), specified for each test. Note that for the tests where a solution at rest is expected, we cannot obtain exactly $u = 0$ m.s$^{-1}$ because of the regularization method. However, we get velocities of order $10^{-7}$ m.s$^{-1}$, which can be considered as zero without meaning a loss of accuracy in the results.

\subsection{Uniform flow: influence of the channel width}\label{se:test_unif}
In this test we consider a uniform flow of granular material, whose height is $h=50\,d_s \approx 2.65$ cm, which flows within a narrow channel of width $W $ and slope $\theta$. The grain diameter is $d_s = 0.53$ mm and the volume fraction is $\varphi_s = 0.6$. The rheological parameters are $\mu_s = \tan(20.9^o)$, $\mu_2 = \tan(32.76^o)$, $\mu_w = \tan(13.1^o)$ and $I_0 = 0.279$, which are typical values for experiments with glass beads.

For the simulations, we impose no slip condition at the bottom and zero velocity at the initial time. The material starts to flow because of the gravitational force. We use 50 layers in the multilayer discretization and consider the regularization parameter $\delta = 10^{-5}$ s$^{-1}$ in equation \eqref{eq:friccionlateral2}. We let the material flow until the uniform steady state is reached, then all the results will refer to this steady state.

\begin{figure}[!h]
 \begin{center}
 \includegraphics[scale = 0.45]{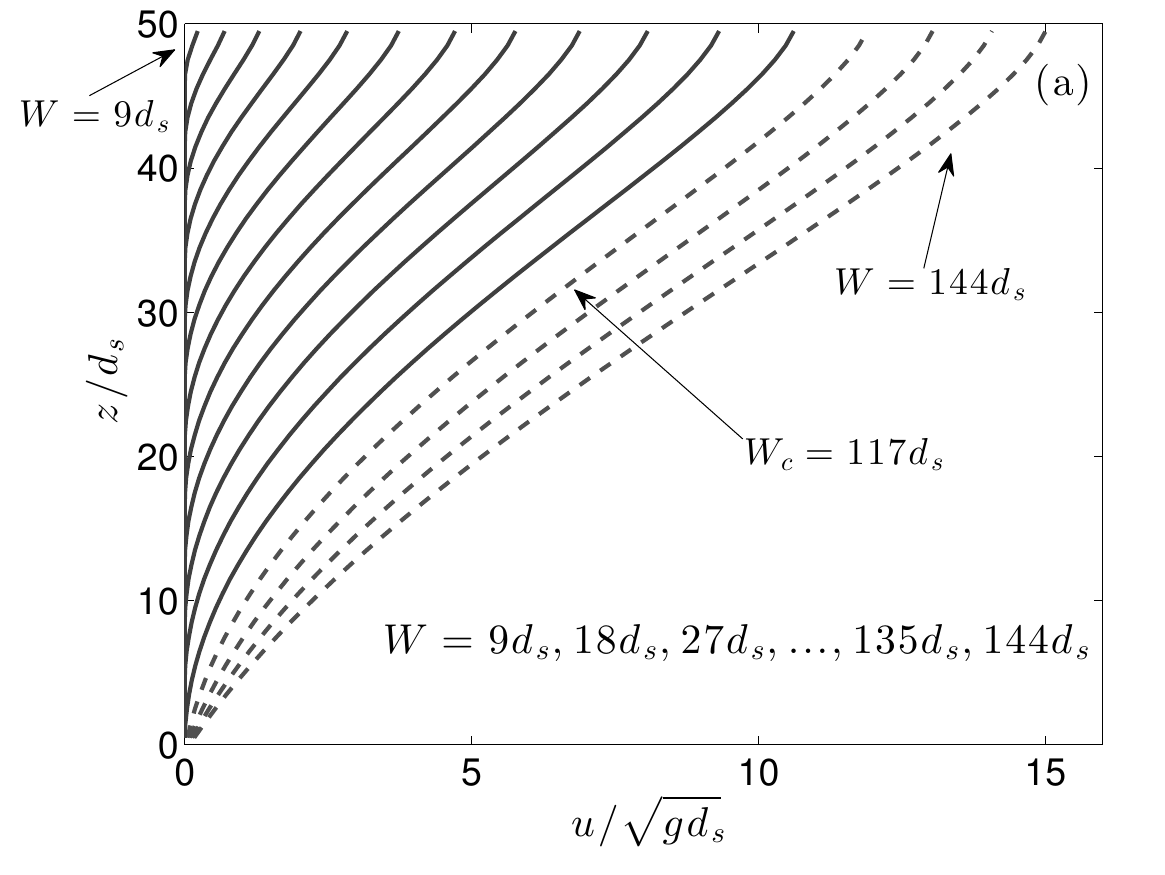}
  \includegraphics[scale = 0.45]{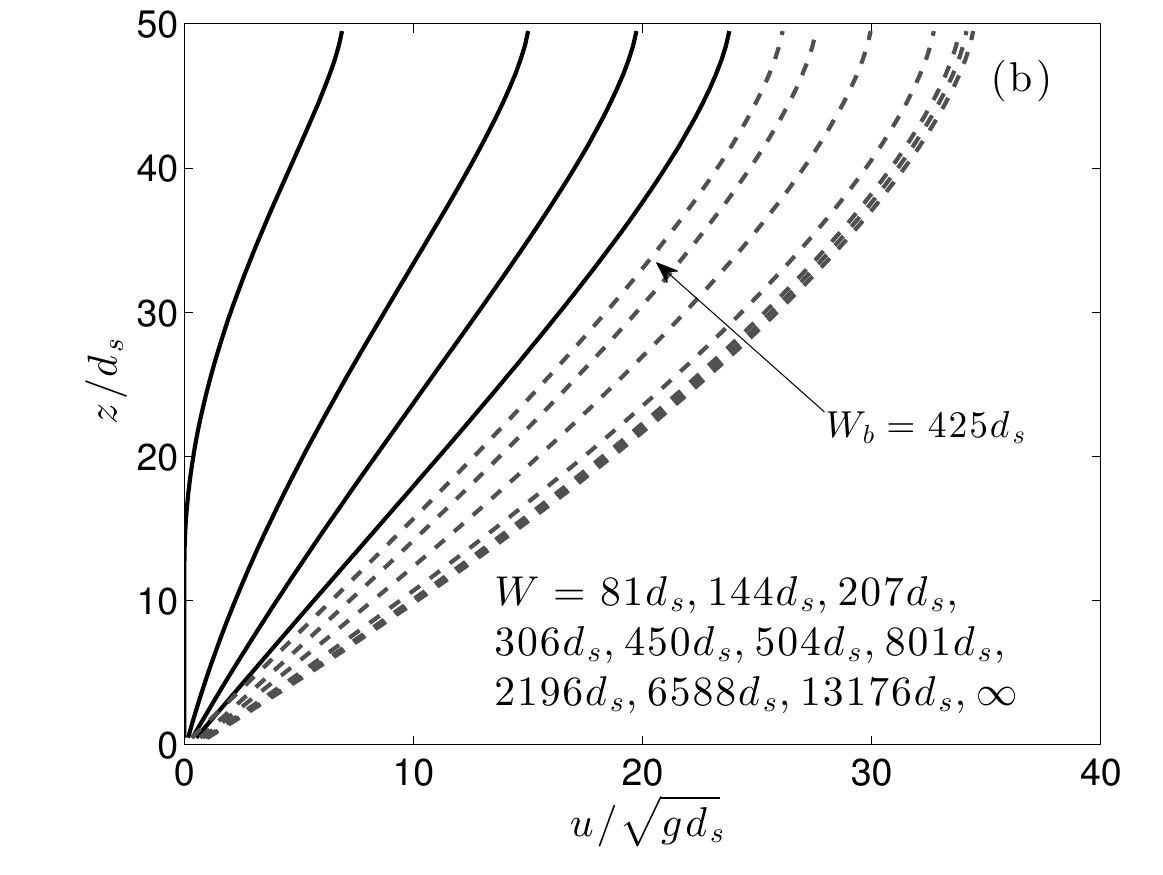}
 \caption{\label{fig:anchos26_h50} \footnotesize \textit{Normalized normal profiles of the downslope velocity $(u)$ on a slope $\theta = 26.1^\circ$ and for different channel widths $(W)$, increasing from left to right lines. Solid lines correspond to the cases when (a) the flow/no-flow interface exists, (b) the velocity profile is still S-shaped; and dashed lines are the cases when (a) the flow/no-flow interface does not exist, (b) the velocity follows a Bagnold type profile.}}
 \end{center}
 \end{figure}

First, we focus on the velocity profiles in the direction normal to the slope when going from a narrow to a wider channel. Figure \ref{fig:anchos26_h50} shows these profiles at a slope $\theta = 26.1^\circ$ for increasing channel widths $W =9\,d_s, 18\,d_s, ... , 13176 \,d_s$. Note that in this test the parameter $\lambda = H/L_y$ take values in a range from $3\cdot10^{-3}$ to $5.5$. In particular, the influence of the channel width on the position of the flow/no-flow interface is shown. The thickness of the flowing layer increases as the width does so. Moreover, all the granular layer flows when the thickness $W\geq W_c = 117\,d_s \approx 6.2$ cm (see figure \ref{fig:anchos26_h50}a). Figure \ref{fig:anchos26_h50}b shows an S-shaped velocity profile until $W\geq W_b = 425\,d_s \approx 22.5$ cm where the flow then exhibits a Bagnold profile. The critical value $W_b$ is measured by approximating the second derivative of the downslope velocity along the normal direction. Then $W_b$ is the first value of $W$ for which the second derivative changes its sign. Interestingly, an asymptotic velocity profile is reached for values of $W$ greater than 3 meters approximately ($W = 6588\,d_s$). Then, the velocity profile is independent of the channel width. Note that these values are related to the chosen thickness $h=50\,d_s$ and slope angle $\theta = 26.1^\circ$.

\begin{figure}[!h]
 \begin{center}
 \includegraphics[scale = 0.45]{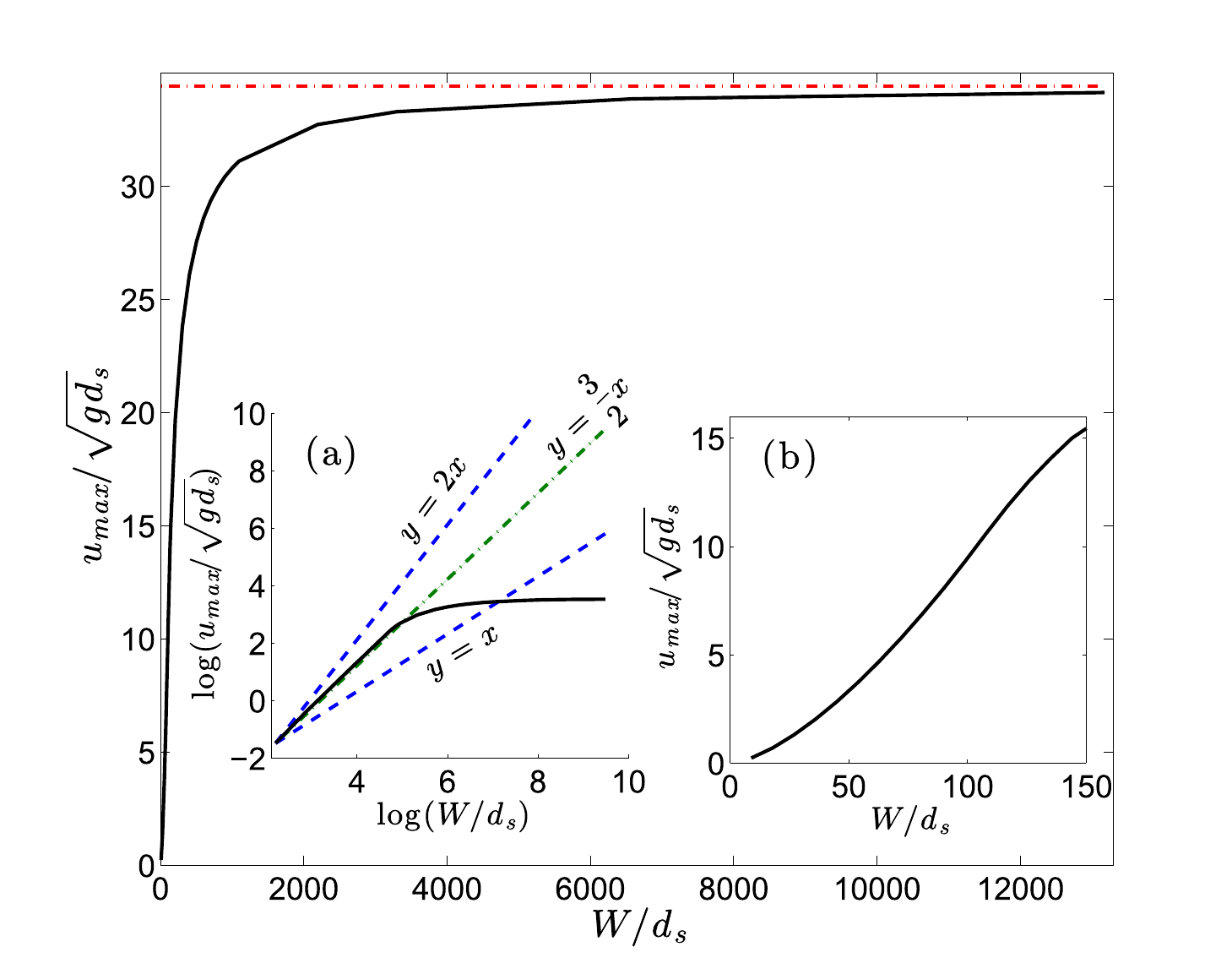}
 \caption{\label{fig:anchos26_h50_V} \footnotesize \textit{Normalized velocity (solid black lines) at the free surface as a function of the channel width $W$. The inner figures are (a) the logarithmic scale; (b) zoom of main figure for short widths $W$. Dashed-dot red line is the velocity at the free surface without side walls effect. }}
 \end{center}
 \end{figure}

 Figure \ref{fig:anchos26_h50_V} shows the influence of the channel width on the maximum velocity (i. e. the velocity at the free surface). We can observe in this figure a nonlinear behavior of the maximum velocity in terms of the channel width. For small values of the width $u_{max}$ scales approximately as $(W/d_s)^{3/2}$ (see inset (a)). When the width increases the maximum velocity tends to the velocity reached when the lateral friction term is not considered (i.e. $W = \infty$). For $W_b=425 d_s \approx 22.52$ cm, the maximum velocity is still 1.3 times lower than $W = \infty$.

   \begin{figure}[!h]
 \begin{center}
 \includegraphics[scale = 0.45]{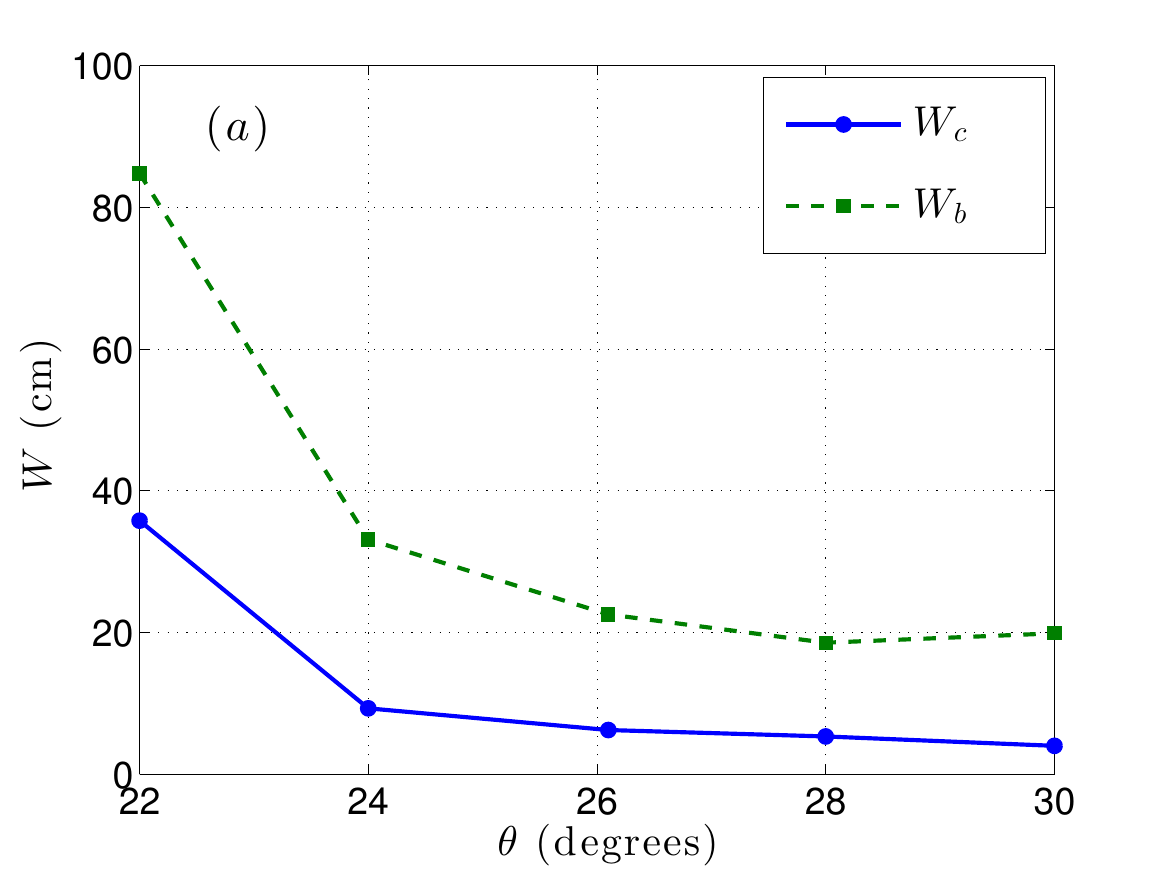}
  \includegraphics[scale = 0.45]{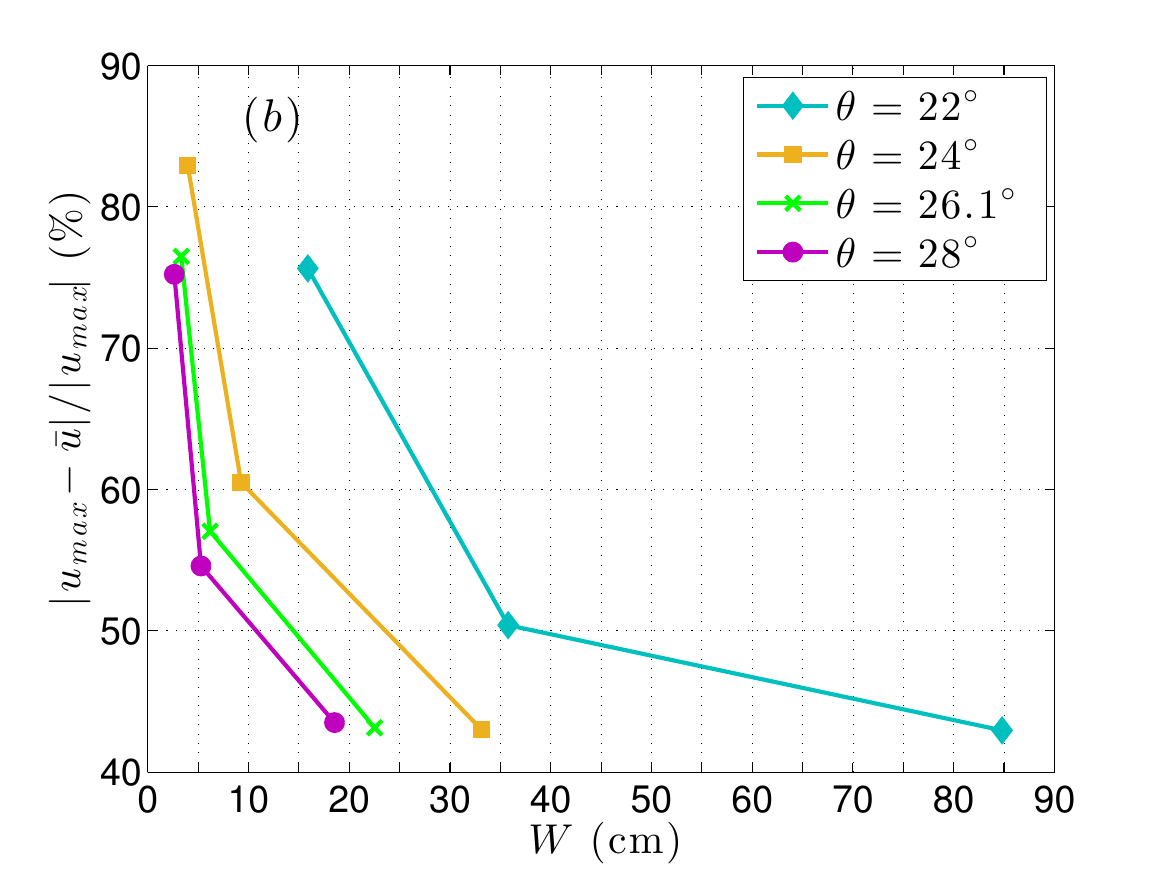}
 \caption{\label{fig:WcWb} \footnotesize \textit{(a) Evolution of the threshold widths $W_c$ (solid-circles blue line) and $W_b$ (cross-dashed green line) as a function of the the slope angle $\theta$. (b) Relative difference between the maximum and the averaged velocity for different slopes (represented by different colors) measured for widths $W_0 < W_c$, $W_c$ and $W_b$ in table \ref{tablew}.
 }}
 \end{center}
 \end{figure}

 \medskip
Let us investigate how $W_c$ (minimum width for all the granular mass to flow) and $W_b$ (minimum width for the flow to exhibit a Bagnold profile) vary with the slope angle for a given flow thickness $h=50\,d_s \approx 2.65$ cm (Figure \ref{fig:WcWb}a). $W_c$ and $W_b$ are calculated by increasing the width $W$ in steps of $25\,d_s \approx 1.33$ cm for fixed slopes $\theta = 22^\circ,\ 24^\circ,\ 26.1^\circ,\ 28^\circ$ and $30^\circ$. For small slopes, high values of $W$ should be reached to get fully flowing materials with Bagnold velocity profile (i. e. $W_c \approx 35.77$ cm and $W_b \approx 84.8$ cm at $\theta=22^o$). The values of $W_c$ and $W_b$ rapidly decrease with increasing slope angles and reach almost constant values $W_c \approx 4$ cm and $W_b \approx 20$ cm. For example for a slope $\theta=24^\circ$, we see that all the material flows for $W>9.28$ cm and that the hypothesis of Bagnold profile is valid only when $W>33.12$ cm. These results could help choosing the good dimensions of the channel in laboratory experiments.

\begin{table}
 \begin{center}
 \begin{tabular}{c|c|c|c}
$\theta$ ($^\circ$) & $W_0$ (cm) & $W_c$ (cm) & $W_b$ (cm)\\
\hline & & & \\
22 & 16 & 35.77 & 84.8\\
24 & 4 & 9.28 & 33.12 \\
26.1 & 3.3 & 6.2 & 22.52\\
28 & 2.6 & 5.3 & 18.55
 \end{tabular}
 \caption{\footnotesize \it{Values of the channel width: $W_c$ (minimum width for which all the granular layer is moving), $W_b$ (minimum width for which the velocity follows a Bagnold profile) and $W_0 < W_c$ as represented in figure \ref{fig:WcWb}.
 }}
\label{tablew}
  \end{center}
 \end{table}

\begin{figure}[!h]
 \begin{center}
 \includegraphics[scale = 0.45]{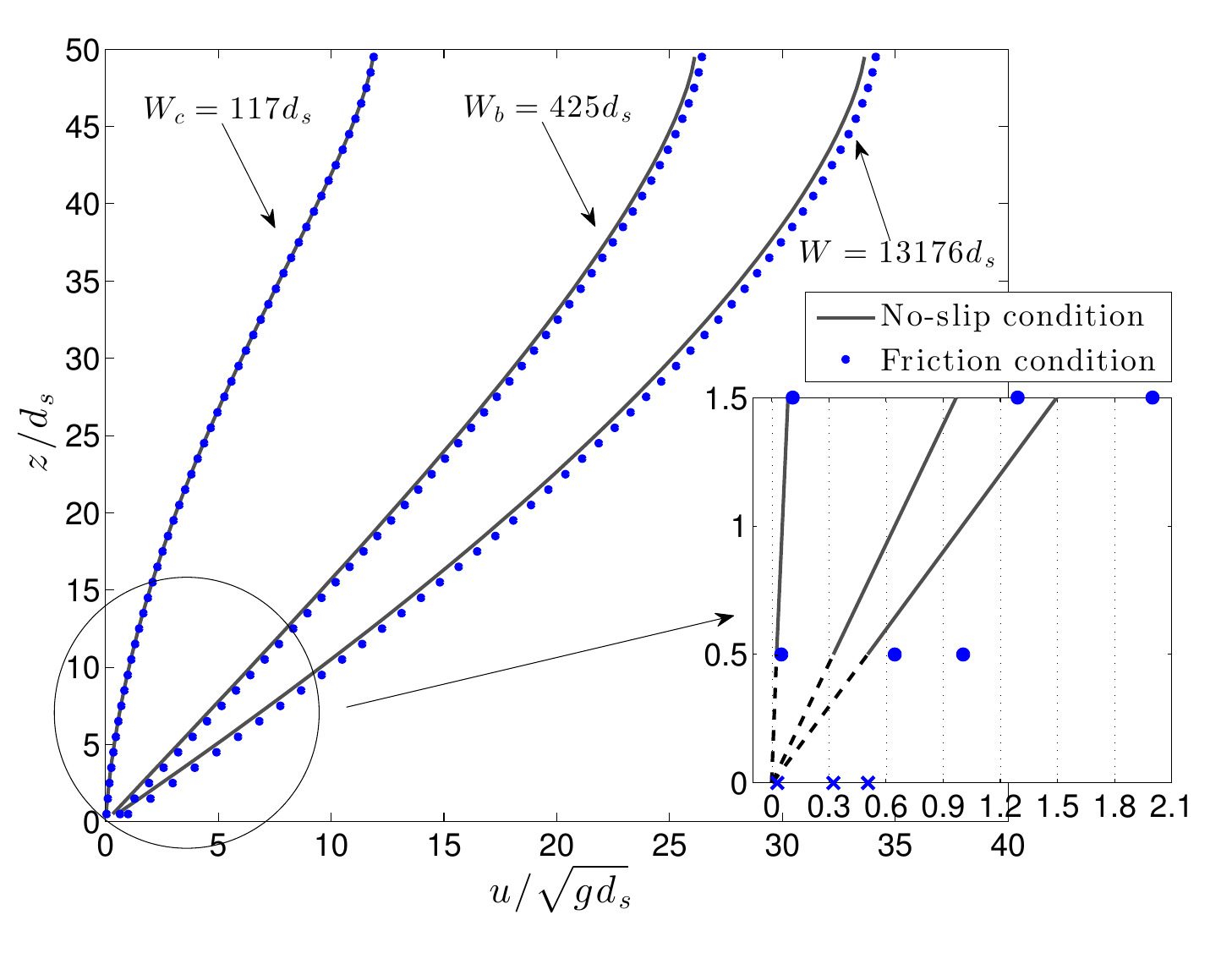}
 \caption{\label{fig:anchos26_friccion} \footnotesize \textit{Normalized vertical profiles of the downslope velocity $(u)$ for different widths $(W_c = 117\,d_s,\,W_b = 425\,d_s$ and $W= 13176\,d_s$) of the channel and $\theta = 26.1^\circ$. Solid lines (respectively symbols) correspond to the simulations with no slip (respectively friction) condition at the bottom. Dashed black lines and blue crosses are the extrapolated velocities at the bottom.}}
 \end{center}
 \end{figure}

Another key issue in shallow depth-averaged models is how to relate the depth-averaged velocity calculated with these models to the free surface velocity that is generally the one measured in laboratory experiments. Figure \ref{fig:WcWb}b shows the difference between the maximum velocity and the velocity averaged along the normal direction, normalized by this maximum velocity, for three different values of the width channel: $W_0 < W_c$, $W_c$ and $W_b$ (see table \ref{tablew}). We see that this difference is huge (greater than $75\,\%$ of the maximum velocity) in the case of small widths. It is because in that case only the layers close to the free surface are moving. This difference decreases for larger widths since all the granular layer is moving. Note that for $W =W_b$, the difference is almost constant for all the slopes and still of about $43\,\%$ of the maximum velocity.

In these tests a no-slip condition was considered at the bottom. Figure \ref{fig:anchos26_friccion} shows the velocity profiles with both no slip and friction condition, for a slope $\theta = 26.1^\circ$ and three widths: $W_c$, $W_b$ and $W \approx 7$ m for which the influence of the side walls is almost insignificant. We see that there is no difference between no-slip and basal friction conditions for the small width $W_c$, and a slight increase of the velocity obtained with basal friction condition for larger widths. We have also checked that the value of $W_c$ and $W_b$ are almost the same in both conditions, for the slope $\theta = 26.1^\circ$.

\subsection{Well-Balanced test including dry areas: granular collapse over an arbitrary bottom} \label{se:WB}

In this test we consider a granular collapse over an arbitrary topography. We show that the hydrostatic reconstruction \eqref{eq:recfondo}-\eqref{eq:recfriccion3} is the key point making it possible to obtain the well-balance property. By comparing the normal profiles of velocity at different times/points, we also show that our model produces results similar to the model considered in Jop et al. \cite{jop:2005}.

 \begin{figure}[!h]
 \begin{center}
 \includegraphics[width=0.9\textwidth]{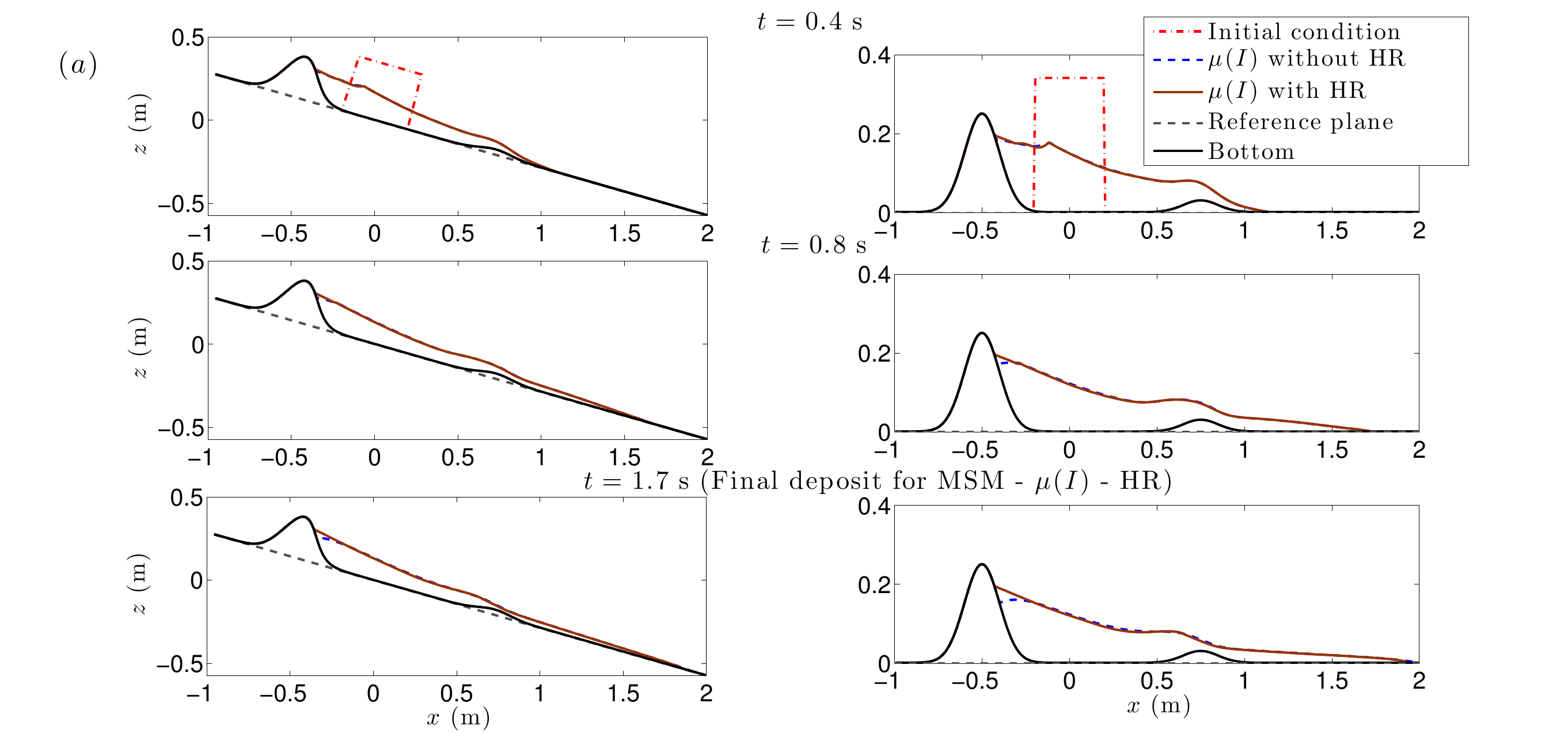}
  \includegraphics[width=0.9\textwidth]{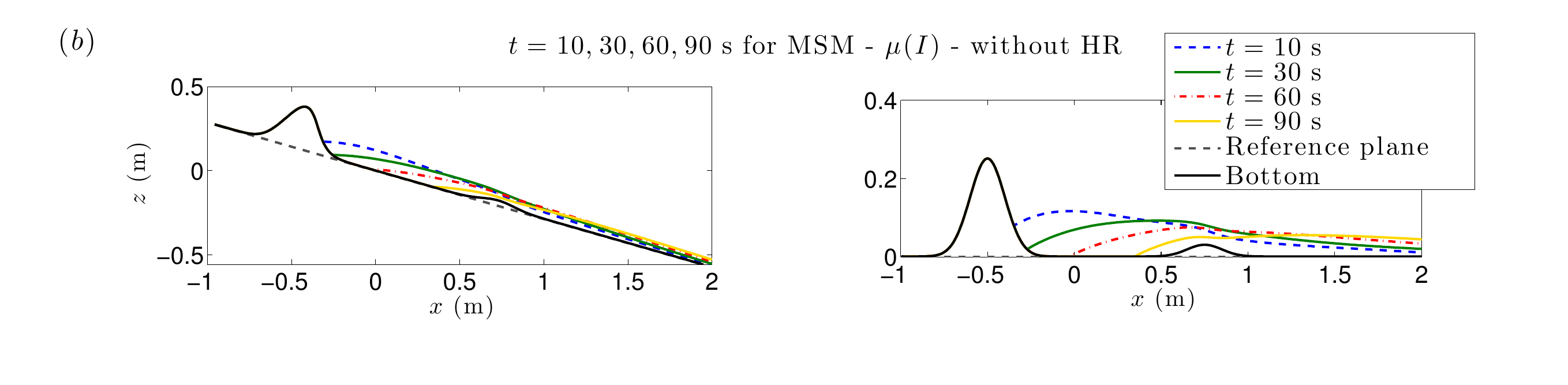}
 \caption{\label{fig:WB_16} \footnotesize \textit{Left: representation in cartesian coordinates. Right: representation in local coordinates on the reference plane.  (a) Free surface evolution during a granular collapse at different times, computed with the multilayer model, taking into account the hydrostatic reconstruction for the Coulomb friction (solid brown line) and without the hydrostatic reconstruction for the Coulomb friction (dashed blue line). (b) The free surface without taking into account the hydrostatic reconstruction at larger times.}}
 \end{center}
 \end{figure}
We take the grain diameter $d_s = 0.7$ mm and the solid volume fraction $\varphi_s = 0.62$, leading to an apparent flow density $\r = 1550$ kg m$^{-3}$. The friction coefficients are $\mu_s = \tan(25.5^\circ)$, $\mu_2 = \tan (36.5^\circ)$ and $I_0 = 0.279$. We also consider the following topography (in m) over  an inclined plane with slope $\theta = 16^{\circ}$ (see Figure \ref{fig:Multilayers}),
$$
b(x) = 0.25 e^{-50(x+0.5)^2} + 0.03 e^{-50(x-0.75)^2}.
$$
The initial condition is given by $q = 0$ m$^2$ s$^{-1}$ and
$$
h(x) = \left\{\begin{array}{ll}
0.34-b(x) & \mbox{if }  |x| \leq 0.2;\\
0 & \mbox{otherwise}.
\end{array}\right.
$$

The channel width is $W=10$ cm and the side walls friction is included through the proposed model with the friction coefficient $\mu_w = \tan (10.5^\circ)$. We use 50 layers in the multilayer system.   Figure \ref{fig:WB_16} shows the evolution of the computed free surface with the multilayer model with and without the hydrostatic reconstruction.  The results are shown  in cartesian (left) and local (right) coordinates. We obtain similar profiles of the flowing mass in both cases at the first times. Nevertheless, when using the hydrostatic reconstruction \eqref{eq:recfondo}-\eqref{eq:recfriccion3} the mass stops at the final time ($t = 1.7$ s), whereas it never stops if the hydrostatic reconstruction for the Coulomb friction is not taken into account (see Figure \ref{fig:WB_16}b for longer times). The hydrostatic reconstruction \eqref{eq:recfondo}-\eqref{eq:recfriccion3} is thus a key ingredient of the well-balance property of the scheme, since it allows to cancel the numerical diffusion \eqref{eq:difusion} when the velocities are close to zero ($u = 0$ is not exactly achieved due to the regularization method). In the following, we will always use the hydrostatic reconstruction.\\

\begin{figure}[!h]
 \begin{center}
 \includegraphics[width=0.9\textwidth]{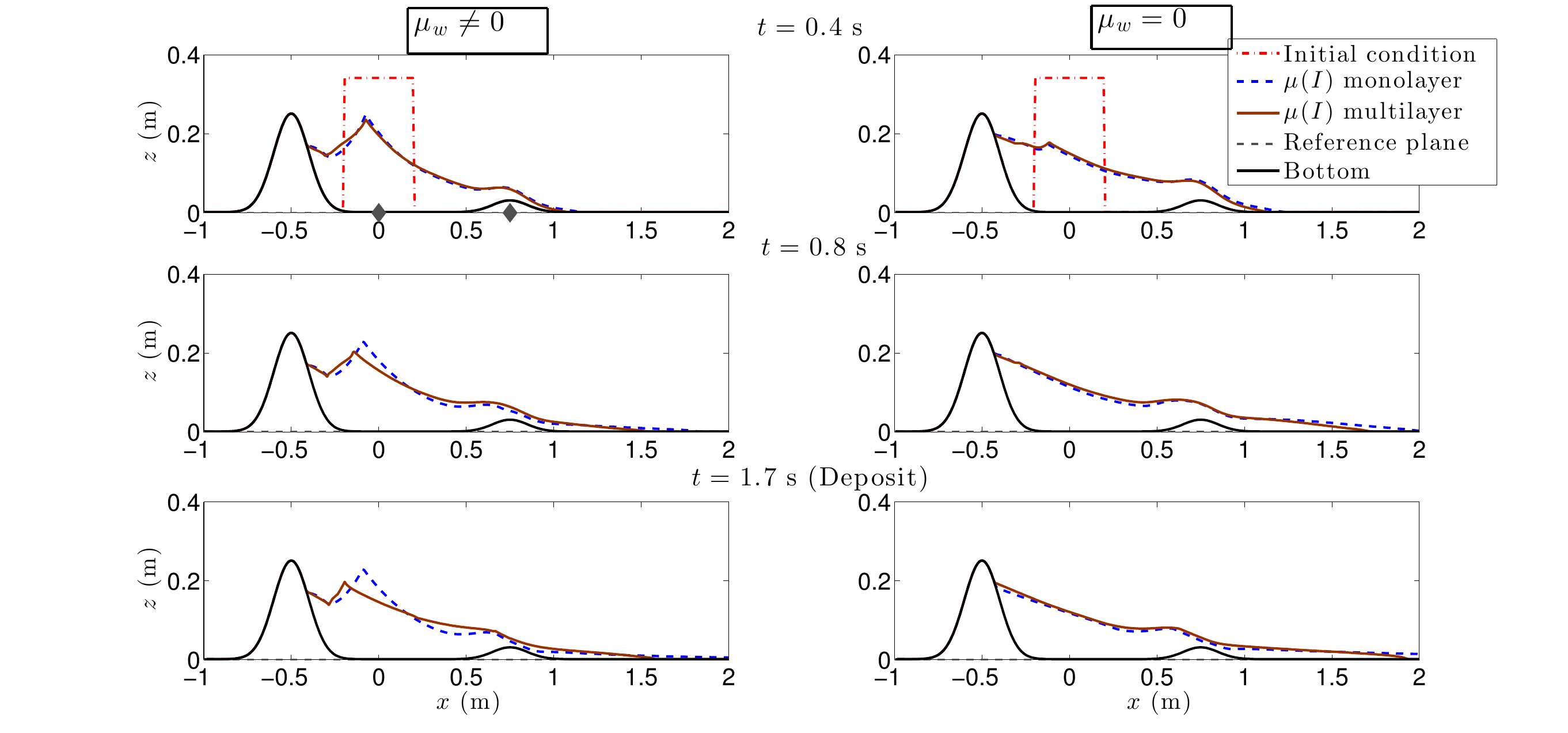}
 \caption{\label{fig:WB_16_muw} \footnotesize \textit{Free surface evolution in local coordinates at different times, computed with the multilayer model (solid brown line) and the monolayer model (dashed blue line), taking into account the side walls friction term (left hand side) and without this term (right hand side)}}

 \end{center}
 \end{figure}

In figure \ref{fig:WB_16_muw} we show the results with and without wall friction for the monolayer and multilayer models. More difference on the shape of the final deposit simulated with the two models is observed when wall friction is considered (left column in figure \ref{fig:WB_16_muw}).  Note that introducing this friction term in the monolayer model adds a constant extra friction over the whole granular layer whereas, in multilayer models, this terms introduces a friction starting from zero at the free surface and increasing with the flow depth.  This will be deeper investigated in subsection \ref{se:h_est}.

\begin{figure}[!h]
 \begin{center}
 \includegraphics[width=0.88\textwidth]{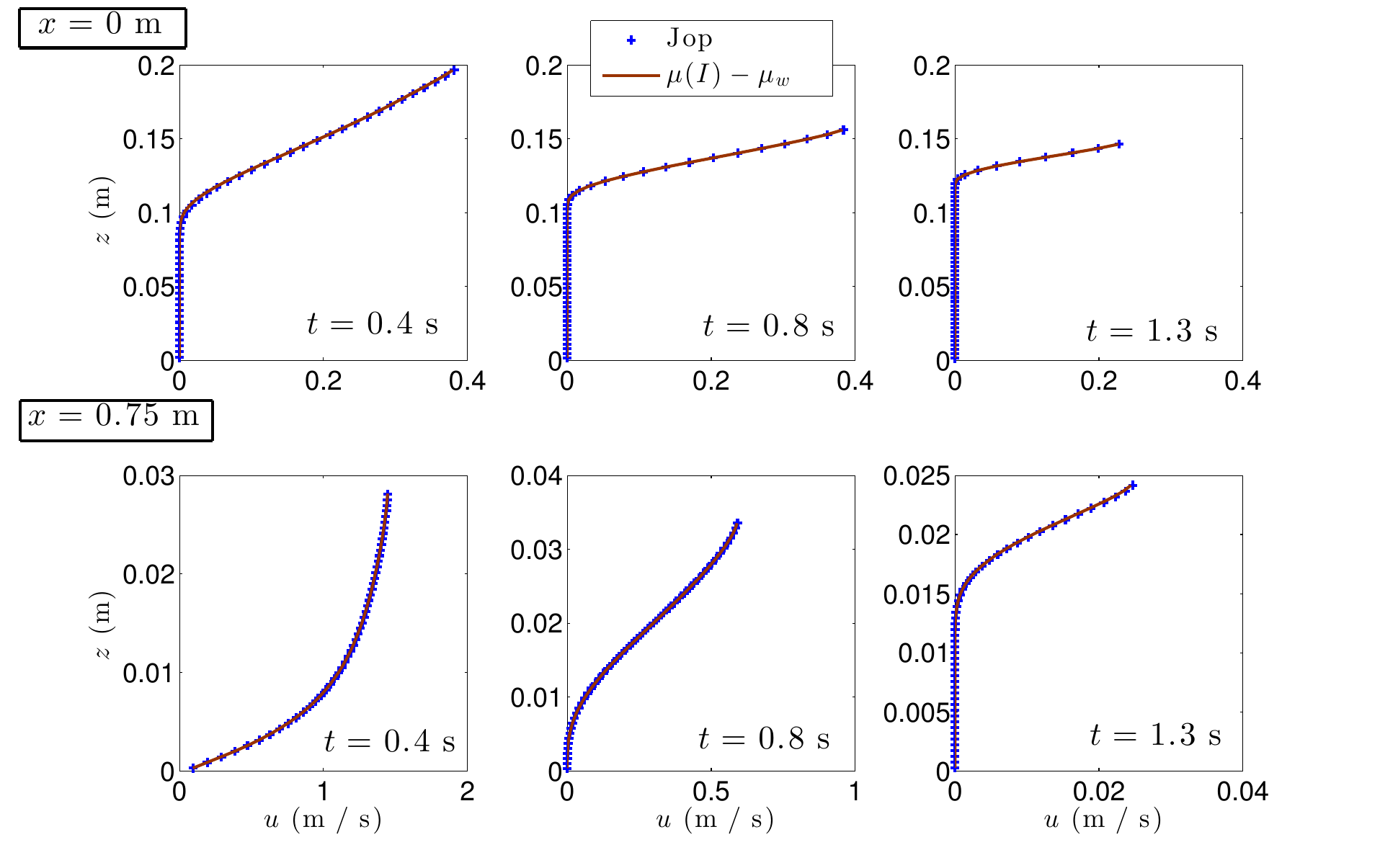}
 \caption{\label{fig:WB_vertical} \footnotesize \textit{Vertical profiles of normal velocity measured at $x = 0$ m and $0.75$ m at times $t = 0.4, 0.8, 1.3$ s. The profiles are computed with the proposed model (solid brown lines) and with the model proposed by Jop et al. \cite{jop:2005} (symbols).}}
 \end{center}
 \end{figure}

The ability of the model to capture the different shapes of the normal profile of the downslope velocity is shown in figure \ref{fig:WB_vertical}. These profiles are shown at different times at two fixed points: the center of the initial released mass ($x=0$ m) and the summit of the second bump of the topography ($x=0.75$ m).
With the proposed multilayer model we can reproduce the Bagnold profile when the flow is accelerating as well as the S-shaped profiles corresponding to the stopping phase. We also show the profiles obtained when including side walls friction in the same way as in Jop et al. \cite{jop:2005}. We see that the results of both models coincide. This is consistent with the remark in section 2, showing that both models match if $sign(u) = sign(\p_z u)$.  \\

\subsection{Solutions at rest with lateral walls friction: multilayer versus monolayer}\label{se:h_est}

With this test we show that an appropriate vertical discretization is essential in order to properly take into account the effect of the side walls friction.

We focus on the steady solutions of system \eqref{eq:swHydro_syst_num}, that is, we assume that $u_\a = 0$.  For the monolayer model ($N=1$) the momentum equation in system \eqref{eq:swHydro_syst_num} give a solution at rest if the following condition is verified:
$$
\left|  \p_{x}\left(\wtd{b}+b+h\right) \right| \leq  \left| \mu_s  + \mu_w \dfrac{h }{W}   �\right|,
$$
where $\wtd{b} = -x\,\tan\theta$. Let us denote by ${\mathcal S} = b+h$ the free surface in local coordinates. Without loss of generality let us suppose that its slope is negative and $\theta\geq 0$.   Then, a solution at rest is defined by ${\mathcal S}$, solution of the following differential equation:
\begin{equation} \label{eq:def_eta_analitique_monolayer}
\p_x {\mathcal S} (x)  = \tan\theta-\mu_s  - \dfrac{\mu_w}{W}\,({\mathcal S}(x)  - b(x)) .
\end{equation}
By setting the initial condition  ${\mathcal S} (x_f)=z_f$, for some constant values $x_f$ and $z_f$, the solution reads
\begin{equation}\label{eq:eta_analitique}
{\mathcal S}  (x) = \dfrac{W}{\mu_w}\left(-\mu_s+|\tan\theta| \right) - \dfrac{\mu_w}{W}\,e^{-\frac{\mu_w}{W}x}\dint_{x}^{x_f}b(s)\,e^{\frac{\mu_w}{W}s}ds + \left( z_f + \dfrac{W}{\mu_w}\left(\mu_s-|\tan\theta|\right)\right)\,e^{-\frac{\mu_w}{W}\left(x-x_f \right)}.
\end{equation}

For the multilayer case, from momentum equation in system \eqref{eq:swHydro_syst_num} we deduce that a  solution at rest is reached if
$$
\left|  g\,\cos\theta\,h\,\p_{x}\left(\wtd{b}+b+h\right) \right| \leq  \left| \dfrac{1}{\r l_{\a}} \Big(K_{\a-\frac{1}{2}} - K_{\a+\frac{1}{2}} \,+\,M_{\a,W}\bigg) \right| \qquad \text{for}\  \a=1,\dots,N.
$$
From the definition of the $K_{\a+1/2}$ and $M_{\a,W}$, previous inequality reads
$$
\left|  \p_{x}\left(\wtd{b}+b+h\right) \right| \leq   \left| \mu_s\ + C_{\a}\, \mu_w \, \dfrac{h }{W}  \, \, \right|,
$$
where
$$
C_{\a}=2  \left( \dsum_{\b=\a+1}^{N} l_\b + \dfrac{l_\a}{2}\right).�
$$
Then, the main difference between the solution at rest of a multilayer model (with $N>1$) and the monolayer model is the coefficient $C_{\a}$ that multiplies $\mu_w$. For the monolayer model this coefficient is $1$.

As a consequence the solution \eqref{eq:eta_analitique} is not a steady solution of the multilayer model. This is because the pressure varies with depth and therefore the friction is smaller for higher layers (that move) and gets bigger for lower layers (they can eventually stop). For example, assuming an odd number of vertical layers, $N = 2n +1$ and $l_\a = 1/N$ for $\a = 1,\dots,N$, then $C_\alpha=\frac2N(N-\alpha+\frac12)$. The only value of $\alpha$ that makes $C_\alpha=1$ is for $\alpha=n+1$, that is, the middle layer. The value of $C_\alpha$ is greater than 1 for lower layers ($\alpha<n+1$), so the friction is bigger and then the material does not move. On the contrary, $C_\alpha<1$ for higher layers ($\alpha>n+1$) which induces a smaller friction and the material moves.  Then, in the multilayer case, the solution ${\mathcal S} $ defined by (\ref{eq:eta_analitique}) is not a steady solution, since the upper part of the granular mass will flow.\\
The solution at rest of the multilayer model converges to the solution defined by the free surface
$$
{\mathcal S} = z_f +(\tan\theta - \mu_s) (x-x_f).
$$

\begin{table}
 \begin{center}
 \begin{tabular}{c|c|c|c|c|c|c}
$\text{Nbr. points in $x$}$ & $L^1$ - Error & $L^1$ - Order & $L^2$ - Error & $L^2$ - Order & $L^{\infty}$ - Error & $L^{\infty}$ - Order\\
\hline & & & & & & \\
50 & 7.02$\times10^{-3}$ & -- & 6.47$\times10^{-3}$ & -- & 9.86$\times10^{-3}$ & --\\
100 & 2.87$\times10^{-3}$ & 1.29 & 2.44$\times10^{-3}$ & 1.41 & 2.06$\times10^{-3}$ & 2.25\\
200 & 1.82$\times10^{-3}$ & 0.65 & 1.55$\times10^{-3}$ & 0.65 & 1.26$\times10^{-3}$ & 0.71\\
400 & 1.06$\times10^{-4}$ & 4.09 & 1.02$\times10^{-4}$ & 3.92 & 2.06$\times10^{-4}$ & 2.61\\
800 & 3.08$\times10^{-5}$ & 1.79 & 2.67$\times10^{-5}$ & 1.94 & 4.65$\times10^{-5}$ & 2.15\\
1600 & 8.01$\times10^{-6}$ & 1.95 & 6.86$\times10^{-6}$ & 1.96 & 1.13$\times10^{-5}$ & 2.03\\
 \end{tabular}
 \caption{\footnotesize \it{Errors and related order for the free surface computed with the monolayer model obtained by varying the number of points in the $x$ direction.}}
\label{tabla_orden}
  \end{center}
 \end{table}

 Let us perform a test showing that the analytical solution defined by (\ref{eq:eta_analitique}) is preserved up to second order by the proposed numerical method when we consider only one layer, $N=1$. On the contrary, when imposing this solution as initial condition in the multilayer model, the mass moves and the new simulated solution at rest is very different. For this test, we assume a flow with the same material and rheological properties as in the previous subsection. We consider the domain $D = [0,2] \times [-0.05,0.05]\times \mathbb{R}$, and a channel width $W=10$ cm. We also consider a bottom topography
$$
b(x) = 0.1\,e^{-100(x-0.5)^2} + 0.35 \,e^{-100x^2},
$$
 over a reference plane of angle $\theta = 16^\circ$. As initial condition the velocities are set to zero and the initial thickness is given by $h={\mathcal S} -b$, where ${\mathcal S} $ is defined by (\ref{eq:eta_analitique}), with $x_f=1$ m and $z_f=b(x_f)$.

\begin{figure}[H]
 \begin{center}
(a) $\theta=16^{\circ}, \quad  b(x) = 0.1\,e^{-100(x-0.5)^2} + 0.35 \,e^{-100x^2}$ \\
   \includegraphics[width=0.42\textwidth]{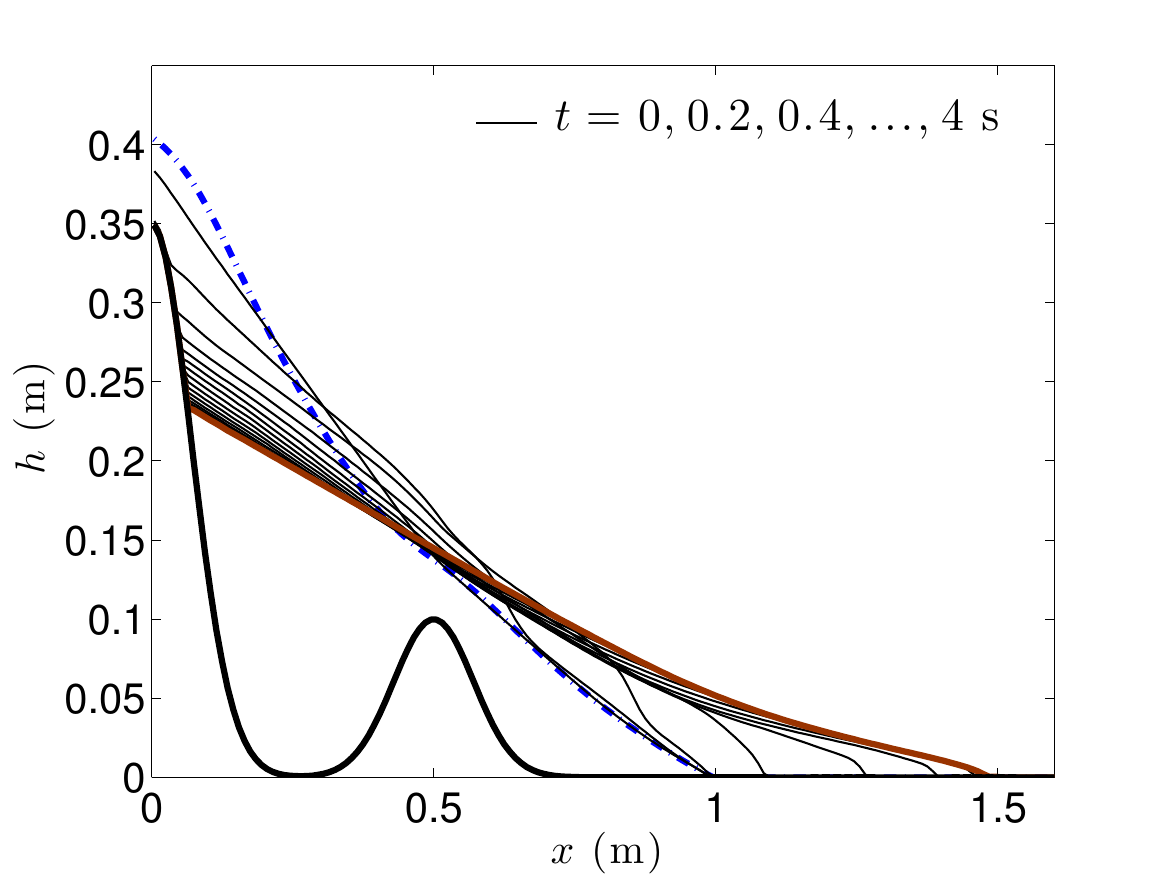}
  \includegraphics[width=0.42\textwidth]{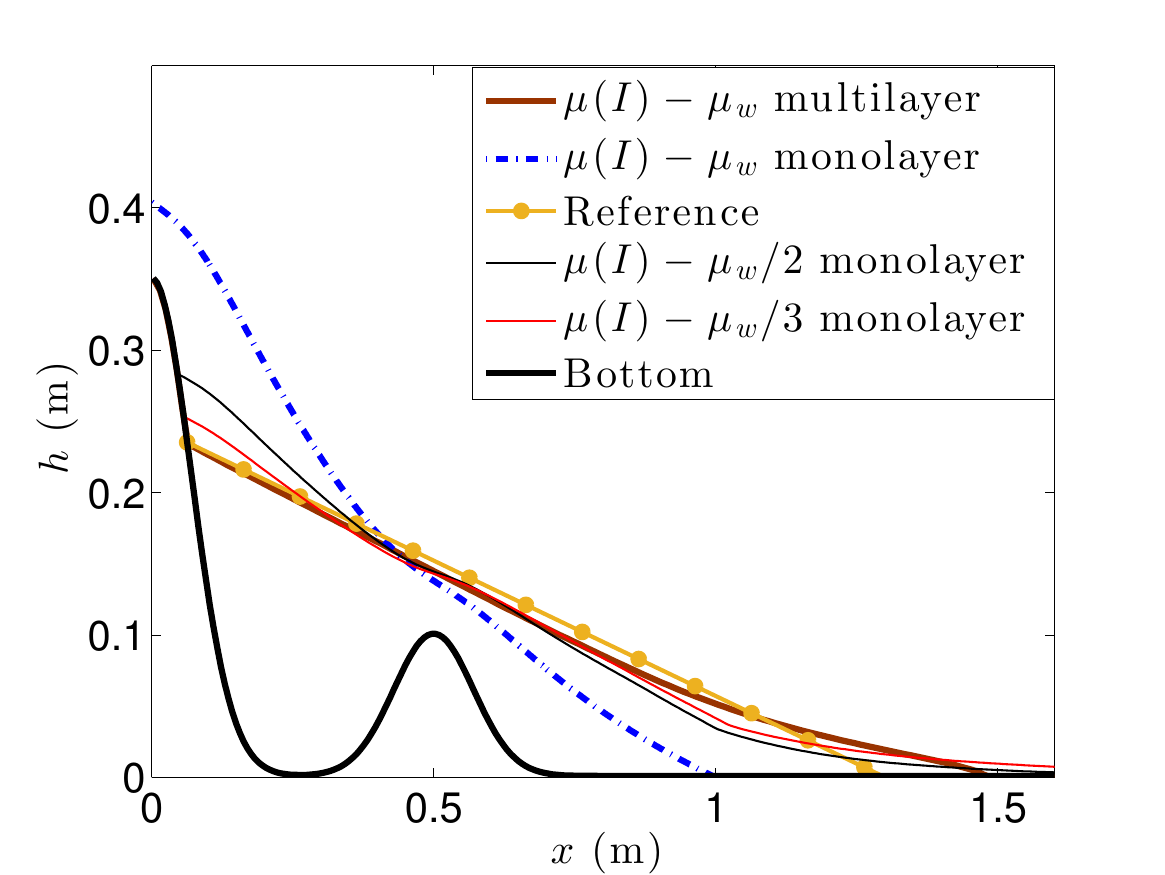} \\
(b) $\theta=16^{\circ}, \quad  b(x) = 0$ \\
   \includegraphics[width=0.42\textwidth]{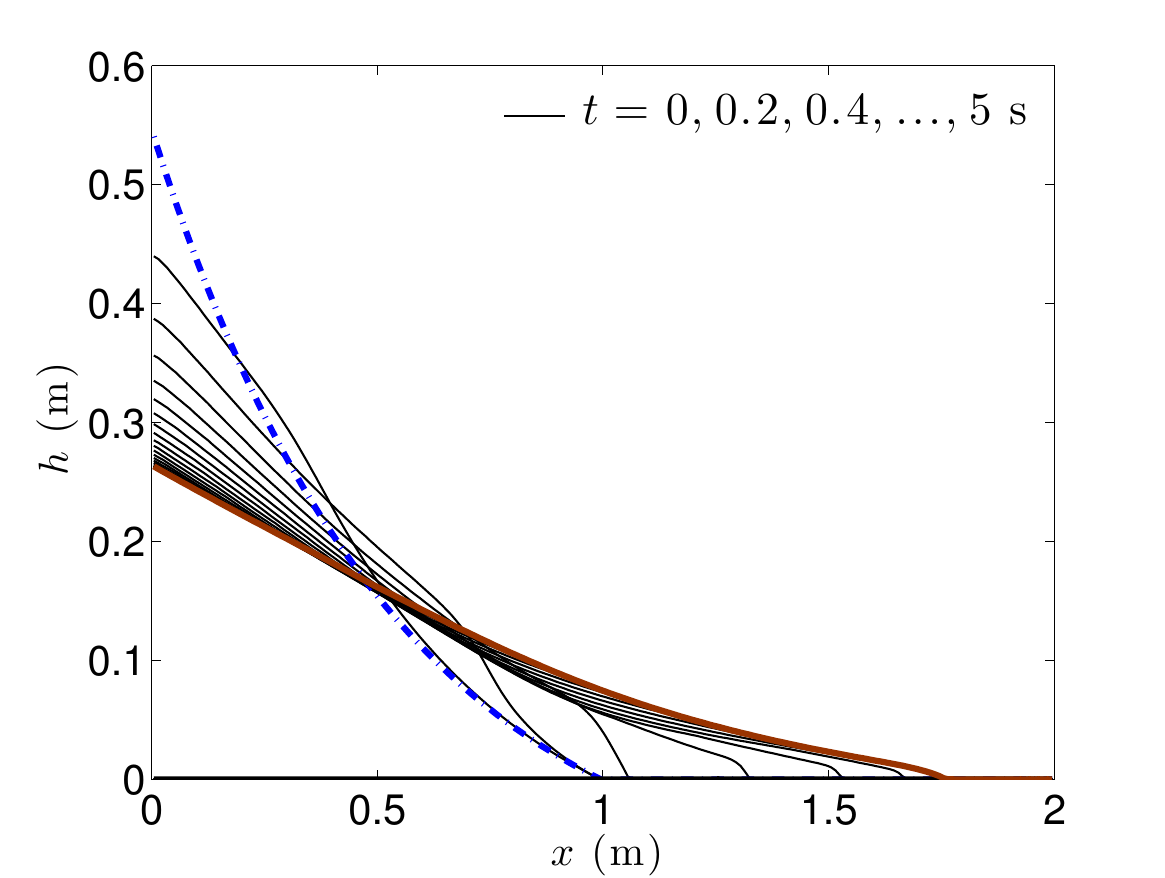}
  \includegraphics[width=0.42\textwidth]{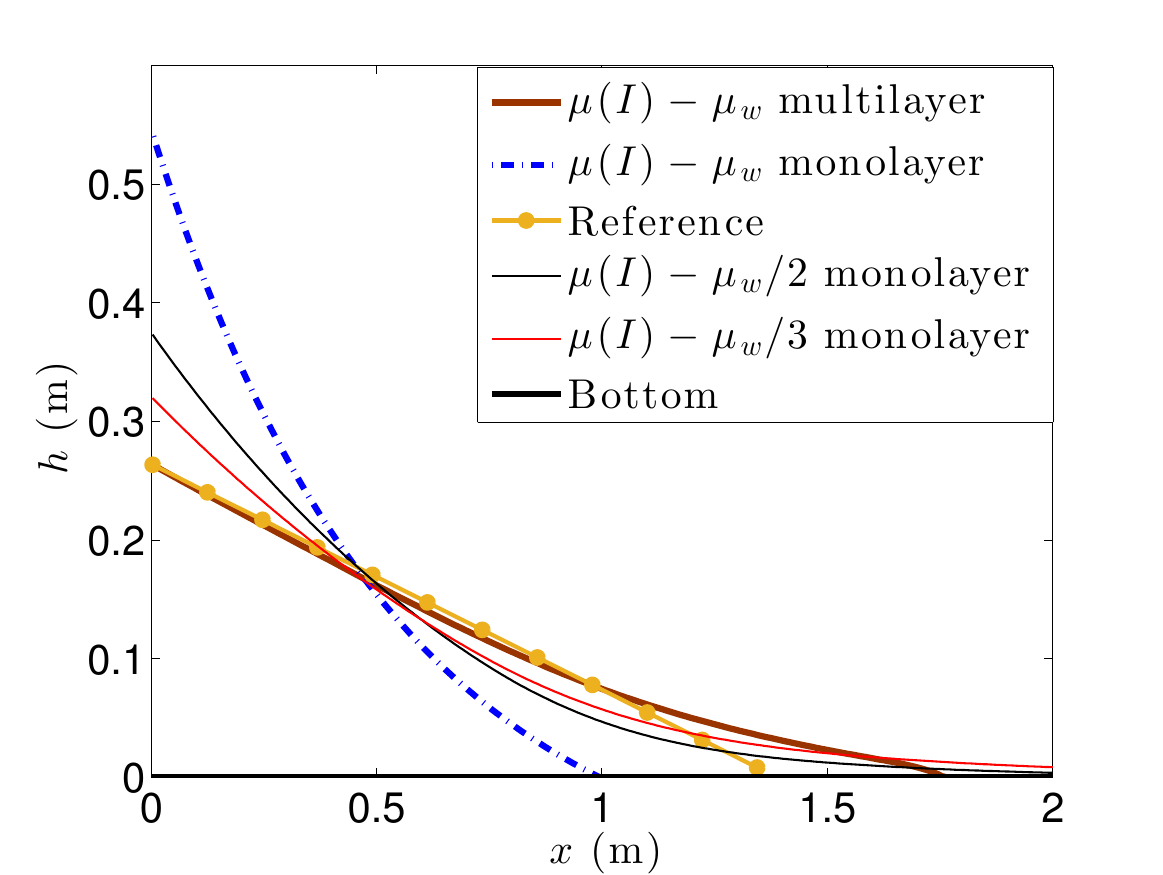} \\
(c) $\theta=0^{\circ}, \quad  b(x) = 0$ \\
   \includegraphics[width=0.42\textwidth]{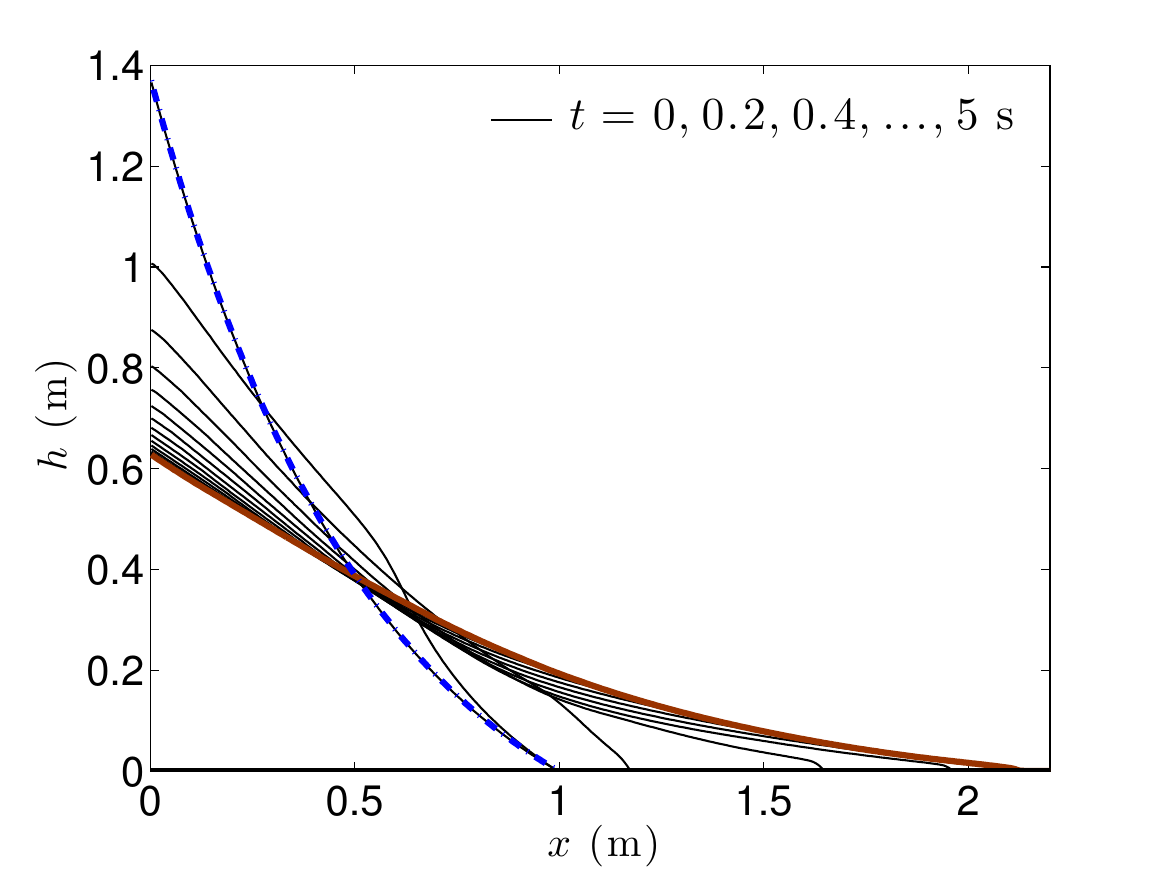}
  \includegraphics[width=0.42\textwidth]{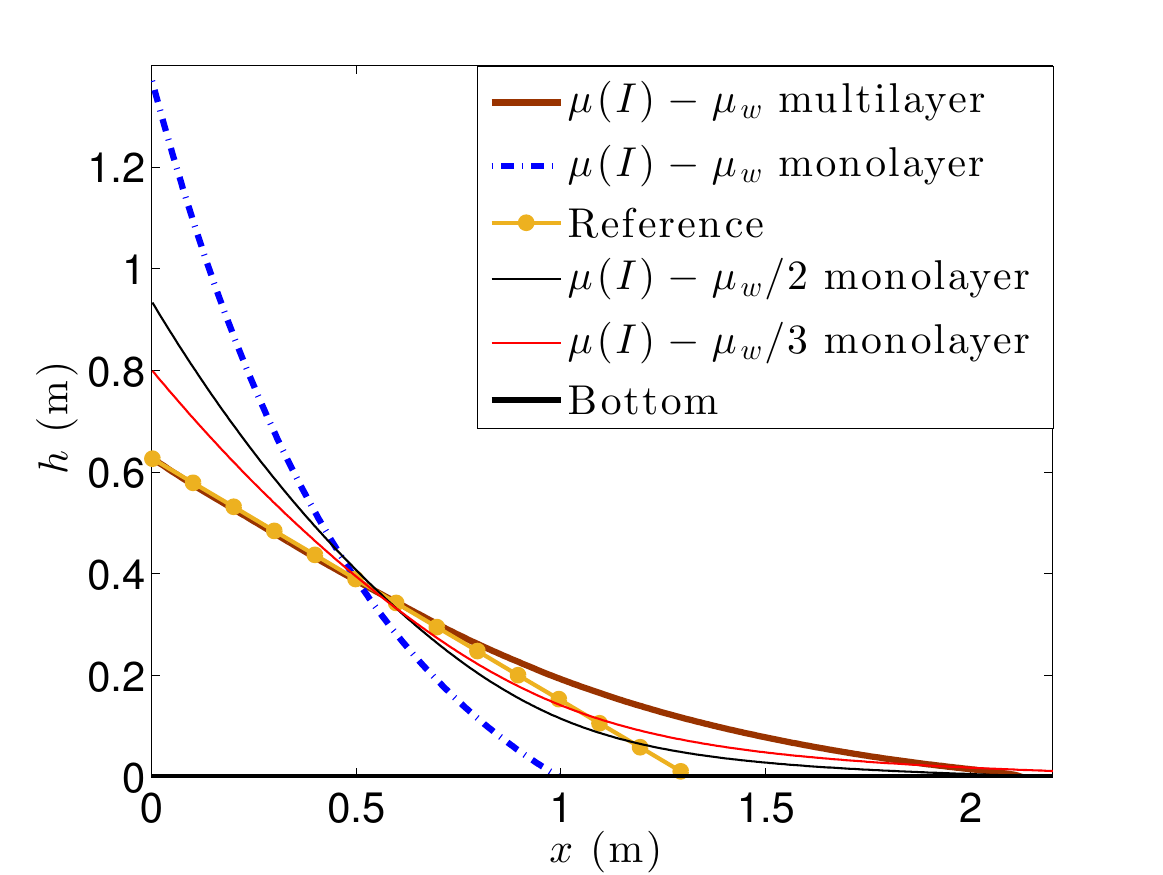}
 \caption{\label{fig:h_est} \footnotesize \textit{Left column: evolution in time of the thickness profile for monolayer (dot-dashed blue line) and multilayer (solid brown line) models with the side walls friction term. Right column: deposit obtained with the monolayer and the multilayer models. Solid black lines (solid red lines) represent the deposit with the monolayer model taking a lower side walls friction coefficient $\mu_w/2$ ($\mu_w/3$). As a reference of the theoretical solution for the multilayer model, we plot a line (point-solid gold line) whose slope is ($\tan \theta -\mu_s$) in local coordinates. }}
 \end{center}
 \end{figure}

In this test we consider 20 layers in the multilayer model and 200 nodes in the horizontal direction. Results are shown in figure \ref{fig:h_est}a for monolayer and multilayer solutions with side walls friction. Table \ref{tabla_orden} shows that the monolayer model keeps the steady solution to second order accuracy whereas the solution for the multilayer model evolves in time to a different deposit (figure \ref{fig:h_est}a). We can observe that the slope of the final deposit obtained with the multilayer model is very close to ($tan(\theta)-\mu_s$) in local coordinates, that is the slope of the solution at rest at which the multilayer model converges. The line with this slope is named {\it Reference} in figure \ref{fig:h_est}. Note that the slope of the computed deposit must always be lower than the slope of this {\it Reference} line, given by the angle of repose of the material. Note that the difference of runout distances predicted by the monolayer and the multilayer models is close to 50$\%$  of the extension of the initial condition.

We also show that we cannot introduce the side walls friction effect by using a monolayer model, even taking a lower friction coefficient $\mu_w/2$, $\mu_w/3$, etc. The deposit widely differs from the solution obtained with the multilayer model in both, the shape and the runout. When the friction coefficient $\mu_w/2$ is considered, the runout are $1.67$, $2.15$, $2.16$ m in cases (a), (b), (c) in figure \ref{fig:h_est} respectively, whereas by using the multilayer model the obtained runout are $1.48$, $1.76$, $2.1$ m respectively. Note also that despite the runout is larger, the height of the material in the initial part of the column is also bigger than the obtained in the multilayer case.

Let us also remark that the bottom topography reduces the exponential shape of the free surface profile (see the influence of $b(x)$ in \eqref{eq:eta_analitique}). Therefore, the solutions with the multilayer and monolayer model are even more different in the case of flat bottom (see figures  \ref{fig:h_est}b and \ref{fig:h_est}c).

\subsection{Laboratory experiments: dam break over rigid and erodible beds } \label{se:labexp}
We compare here our numerical simulation with the laboratory experiments of \cite{mangeney:2010} in the case of a rigid bed (i. e. not covered by a layer of erodible particles). This
configuration was not investigated in our previous work \cite{fernandezNieto:2016} due to the difficulty to deal with dry areas ($h=0$)
from a numerical point of view. When numerical models cannot handle dry areas, a thin layer of material is generally added on these dry zones.
We will investigate here what is the error related to such artificial thin layer.
We also study the time evolution of the flow/no-flow interface with either a variable or a constant friction coefficient.

We release a granular column of height $h=14$ cm and length $20$ cm over an inclined plane of slope $\theta$, confined in a channel of $W=10$ cm.
The granular material in the experiments is made of subspherical glass beads with the material and rheological properties described in previous section \ref{se:WB}. For the numerical simulation, we use 20 layers in the multilayer model.

In this test the friction with the lateral walls is modelled as in our previous work by adding $0.1$ to the friction coefficient $\mu_s$ \cite{fernandezNieto:2016}. As discussed in \cite{fernandezNieto:2016}, hydrostatic models are not able to reproduce the first instants of the granular collapse due to the strong effect of non-hydrostatic pressure. Indeed hydrostatic models spread much faster than experiments at the beginning \cite{mangeney:2005}. As a result,
side walls friction is not well approximated in such models because the flowing layer is overestimated during the first instants.
Despite these limitation, we compare our simulation with laboratory experiments with and without taking into account the extra friction term on the lateral walls.

 \begin{figure}[!h]
 \begin{center}
 (a) $\theta=0^\circ$\\[0.2cm]
  \includegraphics[width=0.9\textwidth]{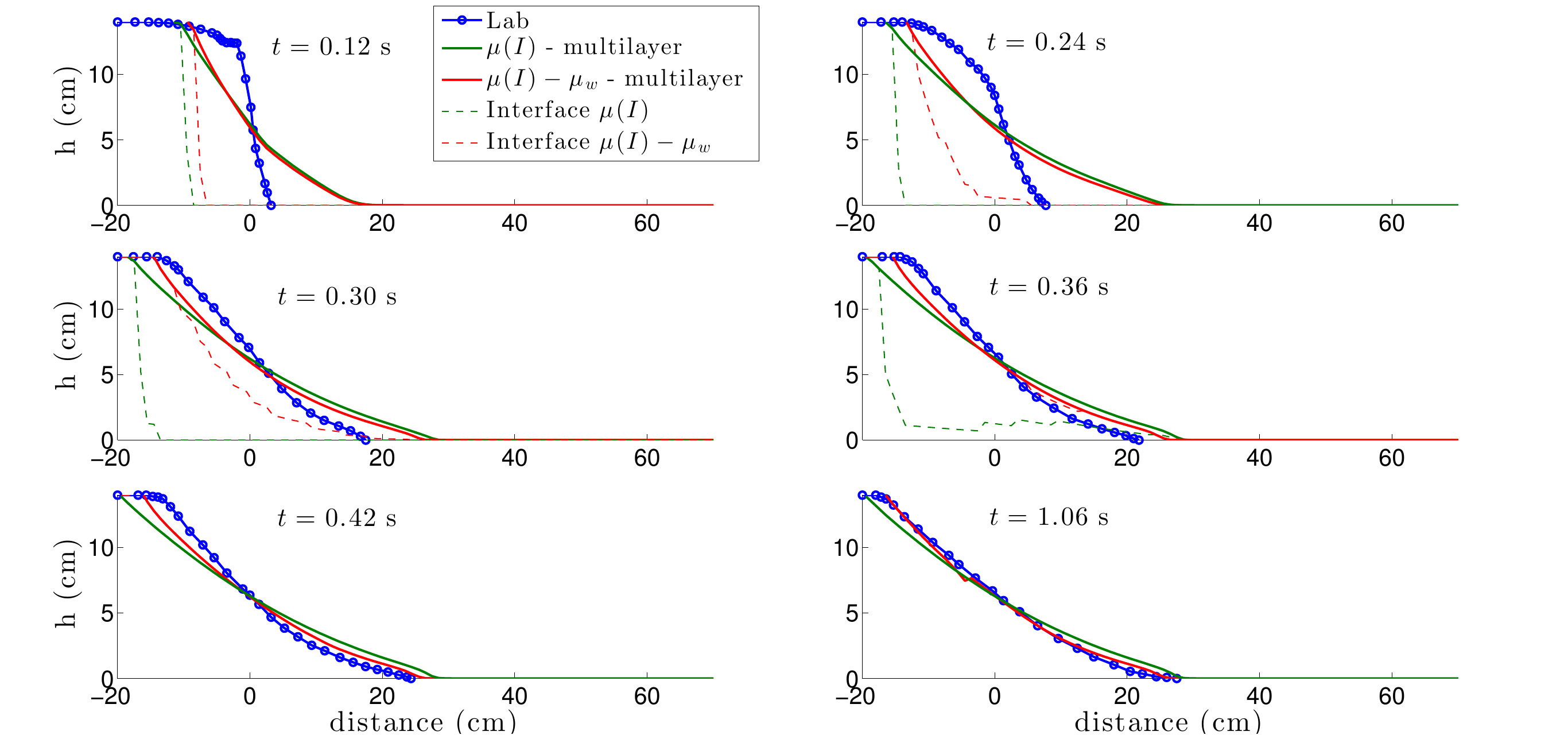}\\[0.2cm]
(b) $\theta=16^\circ$
  \includegraphics[width=0.9\textwidth]{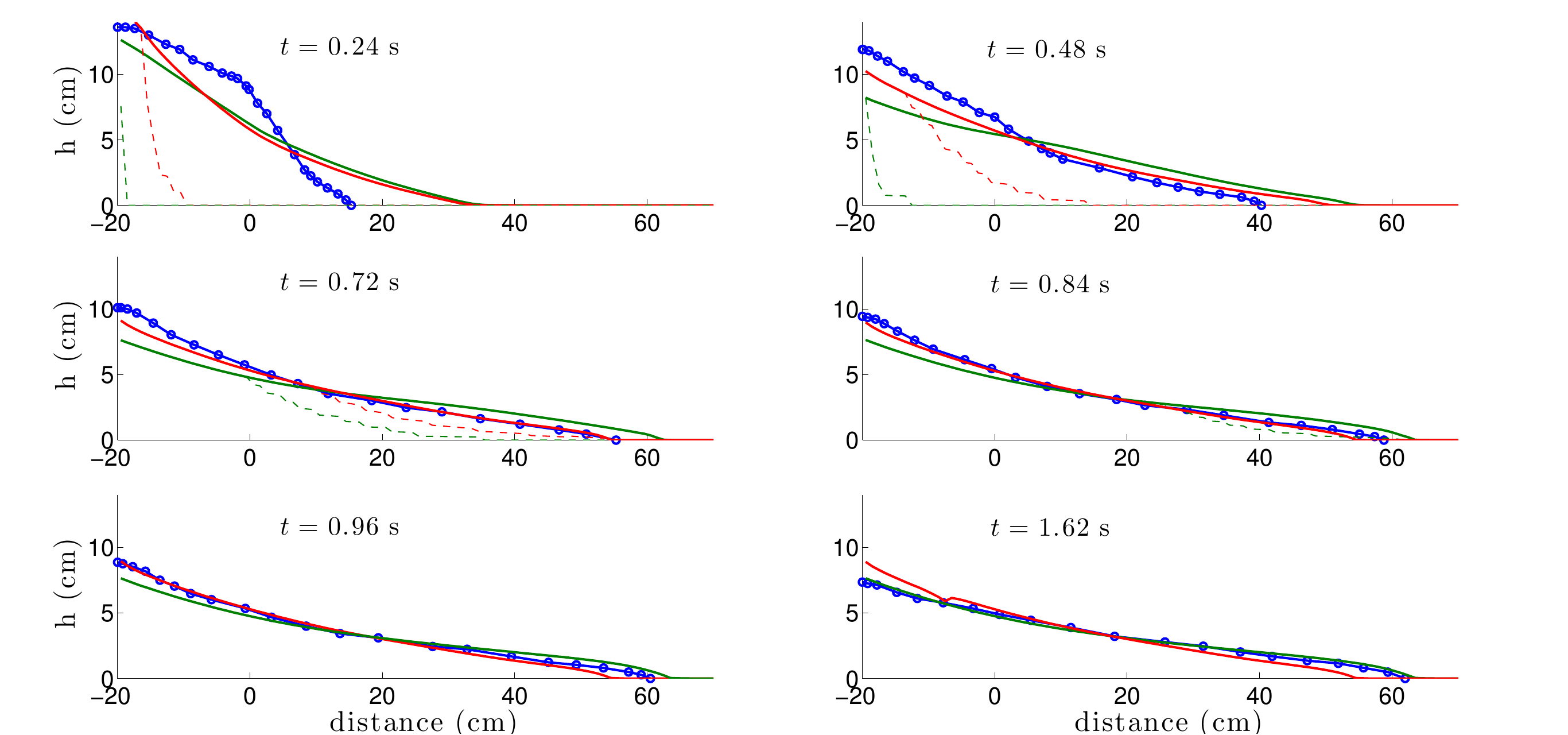}
 \caption{\label{fig:Anne_0}\footnotesize \textit{Thickness of the granular mass as a function of the position along the slope in the laboratory experiments (solid-circle blue line), with the $\mu(I)$-multilayer model (solid green line), and the model adding the side walls friction term ($\mu(I)-\mu_w$-multilayer, solid red line), for a slope (a) $\theta = 0^{\circ}$, (b) $\theta=16^{\circ}$ in the rigid bed case (without an erodible bed over the slope). Thin dashed lines are the flow/no-flow interfaces.
 }}
 \end{center}
 \end{figure}

\begin{figure}[!h]
 \begin{center}
 \includegraphics[width=0.9\textwidth]{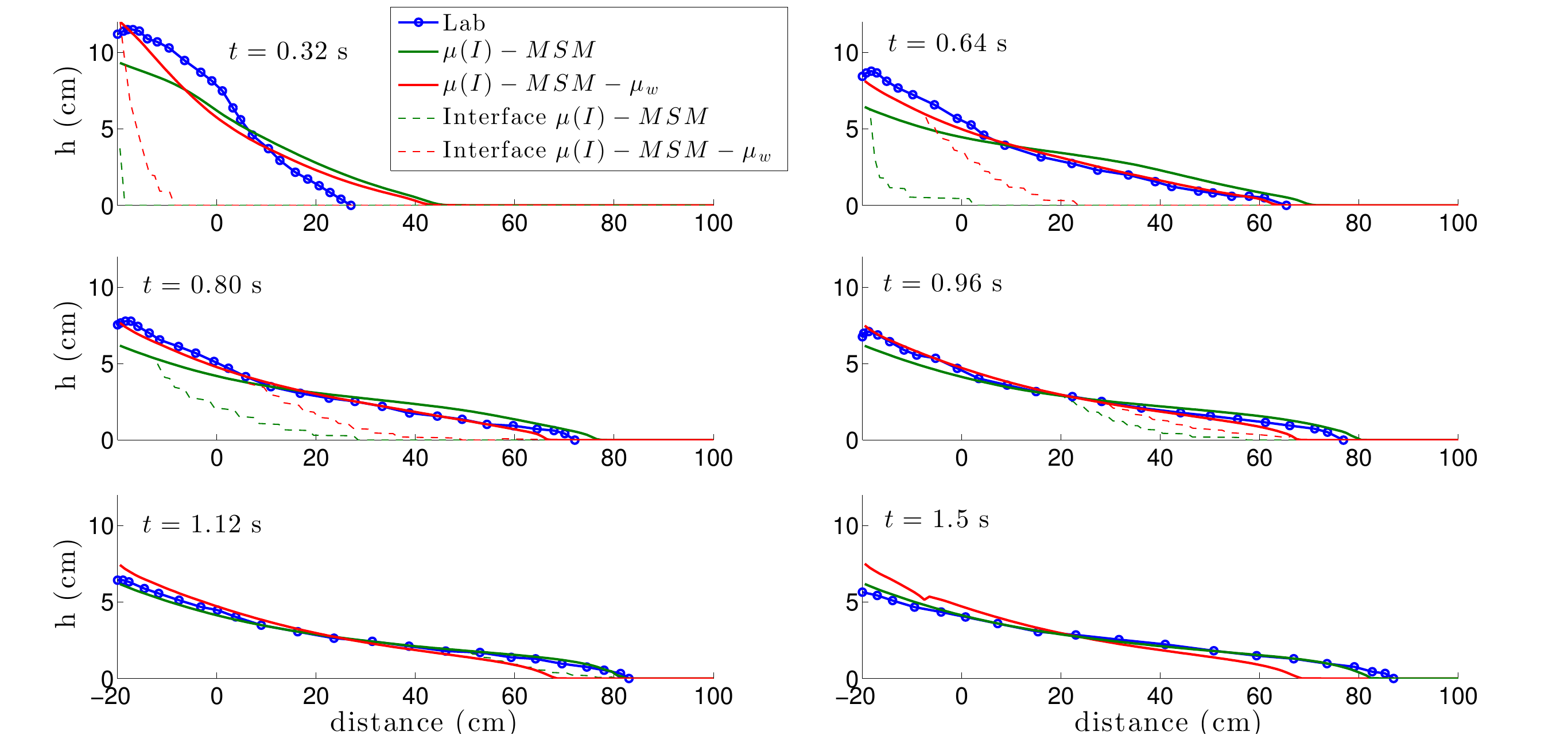}
 \caption{\label{fig:Anne_19}\footnotesize \textit{Thickness of the granular mass as a function of the position along the slope in the laboratory experiments (solid-circle blue line), with the $\mu(I)$-multilayer model (solid green line), and the model adding the side walls friction term ($\mu(I)-\mu_w$-multilayer, solid red line), for a slope $\theta = 19^{\circ}$. Thin dashed lines are the flow/no-flow interfaces.}}
 \end{center}
 \end{figure}
 
 \begin{figure}[!h]
 \begin{center}
  (a) $\theta=22^\circ$,\quad $h_i = 0$ mm\\
 \includegraphics[width=0.9\textwidth]{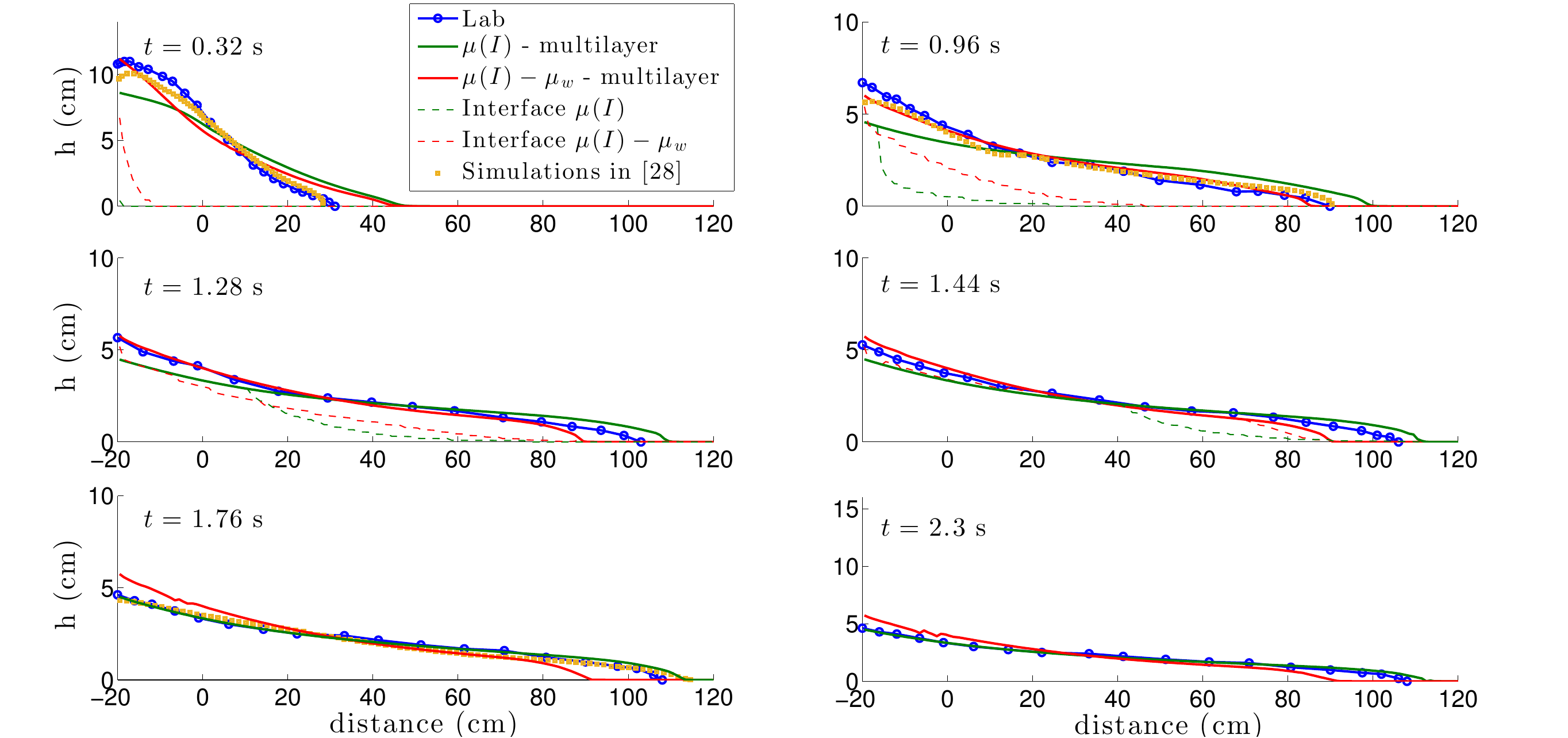}
 (b) $\theta=22^\circ$,\quad $h_i = 4.6$ mm\\
 \includegraphics[width=0.9\textwidth]{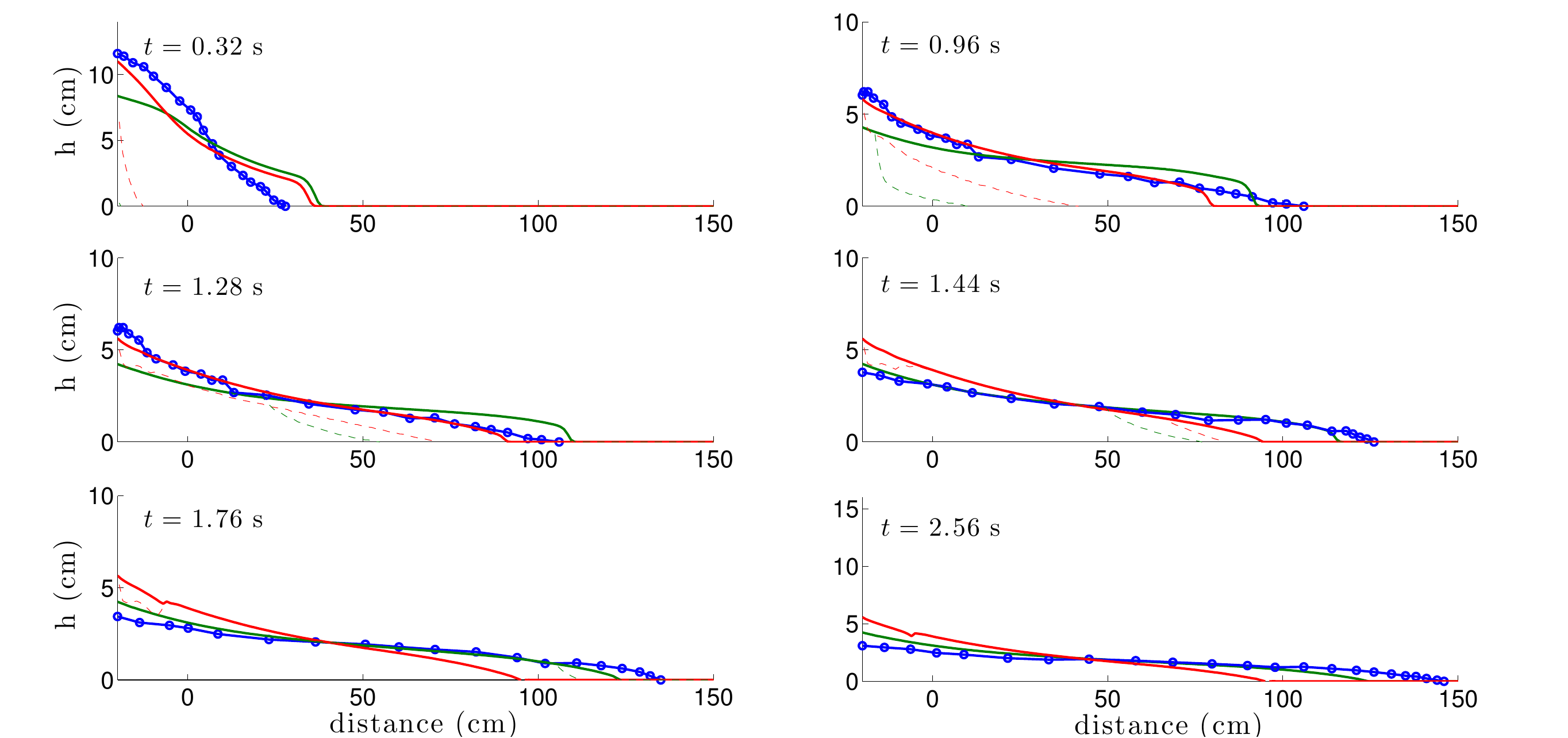}
 \caption{\label{fig:Anne_22}\footnotesize \textit{Thickness of the granular mass as a function of the position along the slope in the laboratory experiments (solid-circle blue line), with the $\mu(I)$-multilayer model (solid green line), and the model adding the side walls friction term ($\mu(I)-\mu_w$-multilayer, solid red line), for a slope $\theta=22^{\circ}$ (a) in the rigid bed case, (b) with an erodible bed $h_i = 4.6$ mm. Thin dashed lines are the flow/no-flow interfaces and gold circles are the simulations of Martin et al. \cite{martin:2017} based on a complete visco-plastic model (i. e. without the shallow approximation).}}
 \end{center}
 \end{figure}
\newpage
Figures \ref{fig:Anne_0}, \ref{fig:Anne_19} and \ref{fig:Anne_22} show the results with and without adding the side walls friction term for different slopes of the inclined plane, $\theta=0^\circ,\, 16^\circ,\, 19^\circ,\, 22^\circ$. We see that the new term makes increase the effective friction, and then the approximation of the free surface improves at short times, while the runout in the final deposit decrease. The comparisons only make sense at final times as consequence of the hydrostatic assumption. As the slope $\theta$ increases, the flow gets thinner and the downslope velocity gets higher compared to the velocity normal to the bottom. As a result, the hydrostatic approximation (i. e. shallow flow approximation) is more correct for higher slopes. Indeed, we can see that the time evolution of the free surface is close to the one obtained with the complete visco-plastic model of Martin et al. \cite{martin:2017} where non-hydrostatic pressure is taken into account (represented by gold circles in figure \ref{fig:Anne_22}a). One of our objective here is to show that multilayer models can be a powerful tool to approximate the flow/no-flow interface position. In order to compute this interface we consider a threshold for flow, i. e. the material is assumed to flow if the velocity is higher than $1$ cm.s$^{-1}$.

   \begin{figure}[!h]
 \begin{center}
 (Left) $\mu(I)$ - multilayer \qquad\qquad\qquad\qquad\qquad (Right) $\mu(I)$ - $\mu_w$ - multilayer\\[0.2cm]
 \includegraphics[width=0.45\textwidth]{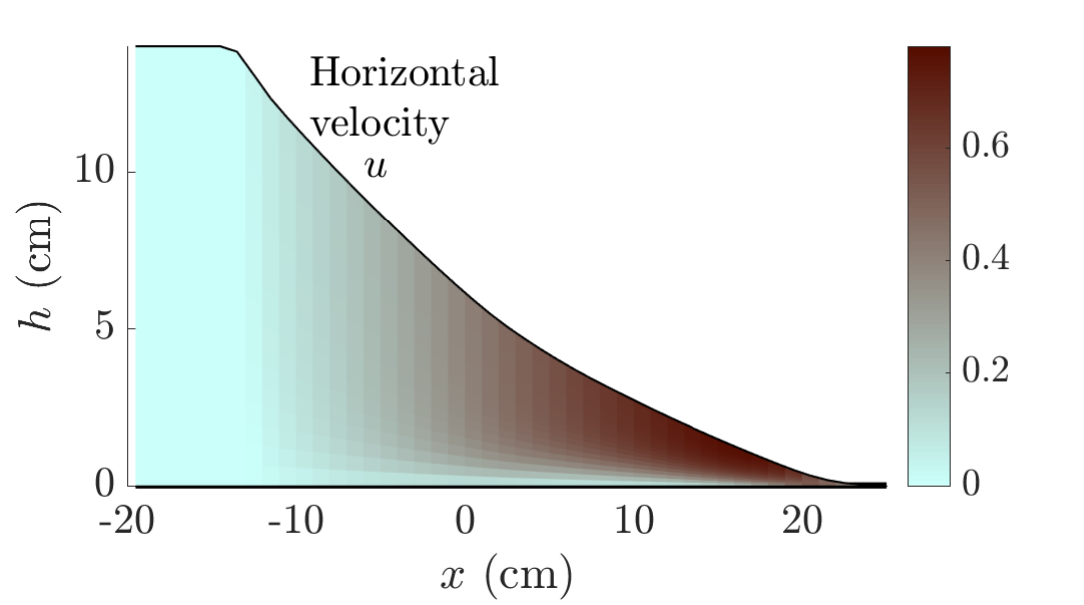}
  \includegraphics[width=0.45\textwidth]{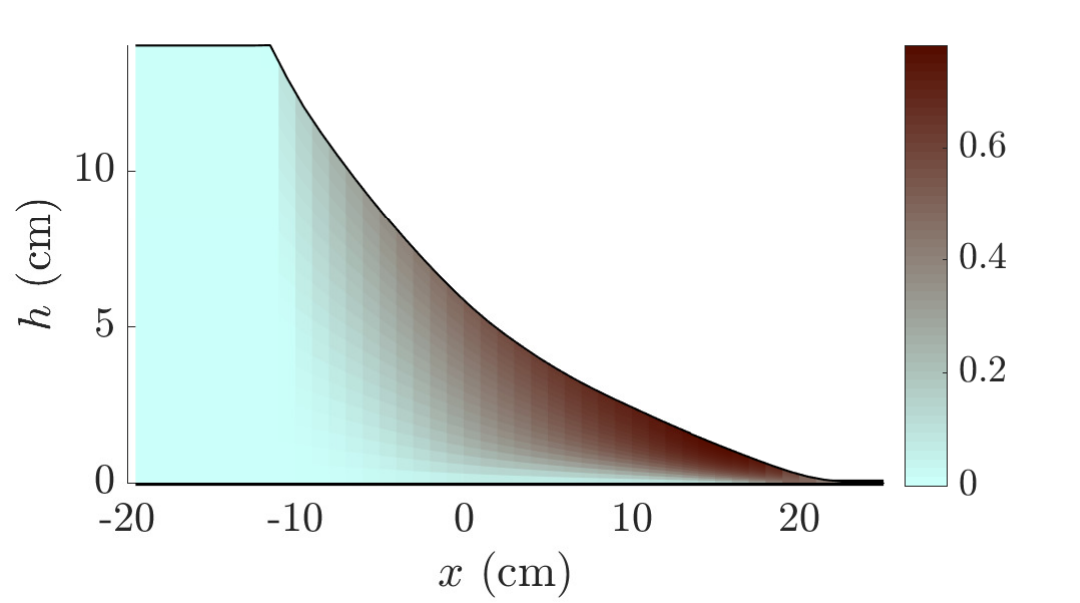}
    \includegraphics[width=0.45\textwidth]{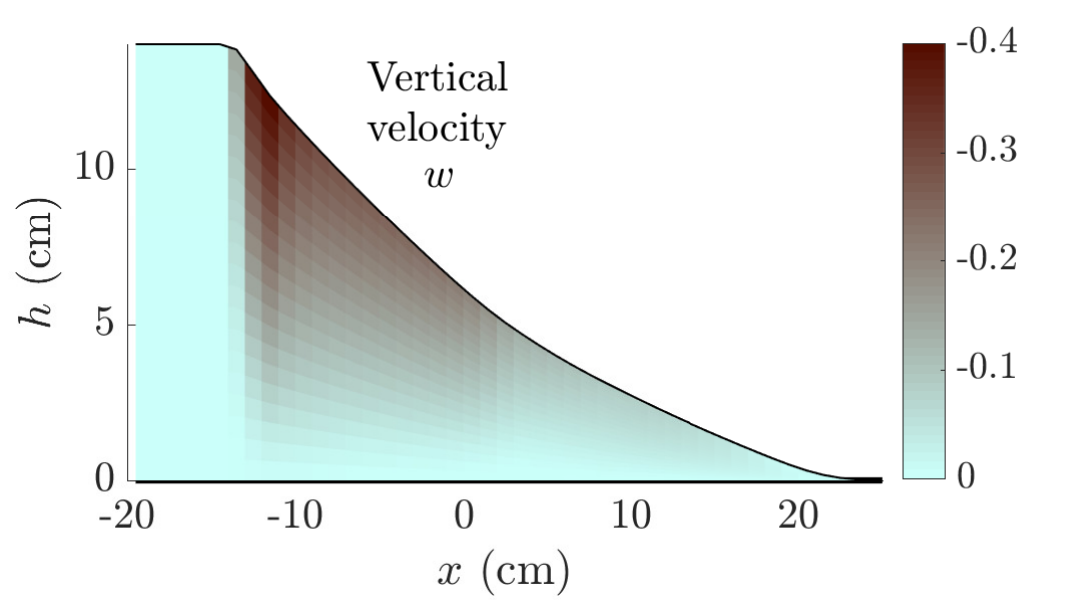}
      \includegraphics[width=0.45\textwidth]{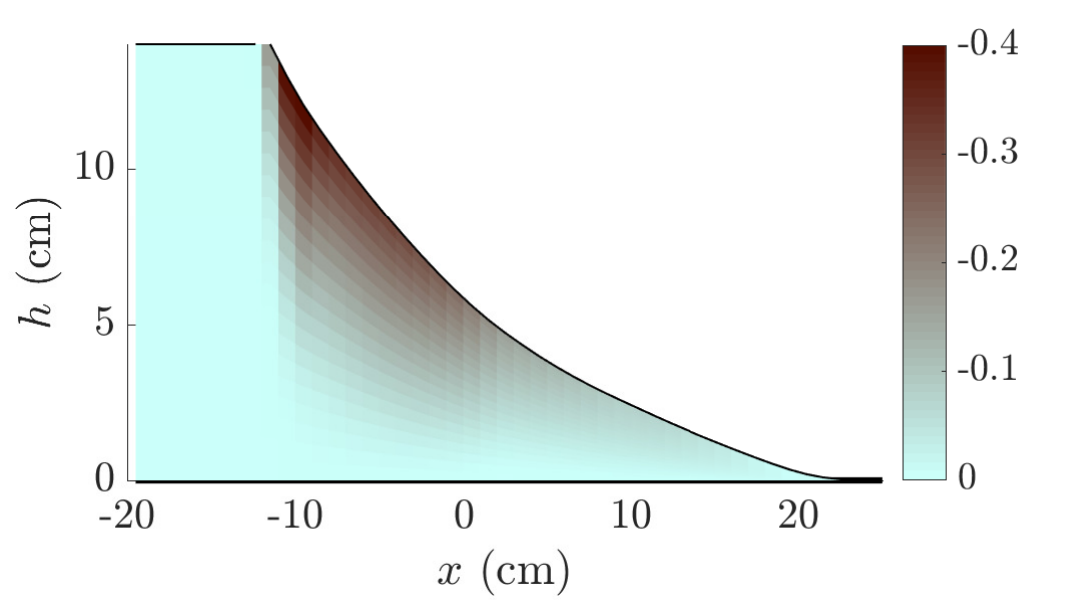}
    \includegraphics[width=0.45\textwidth]{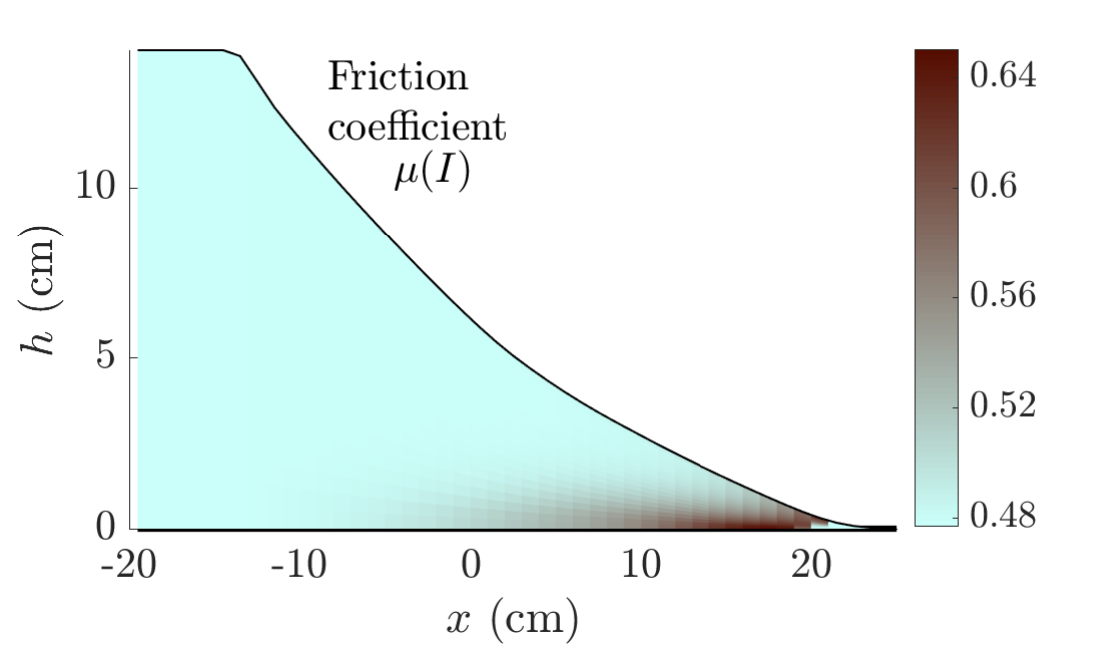}
        \includegraphics[width=0.45\textwidth]{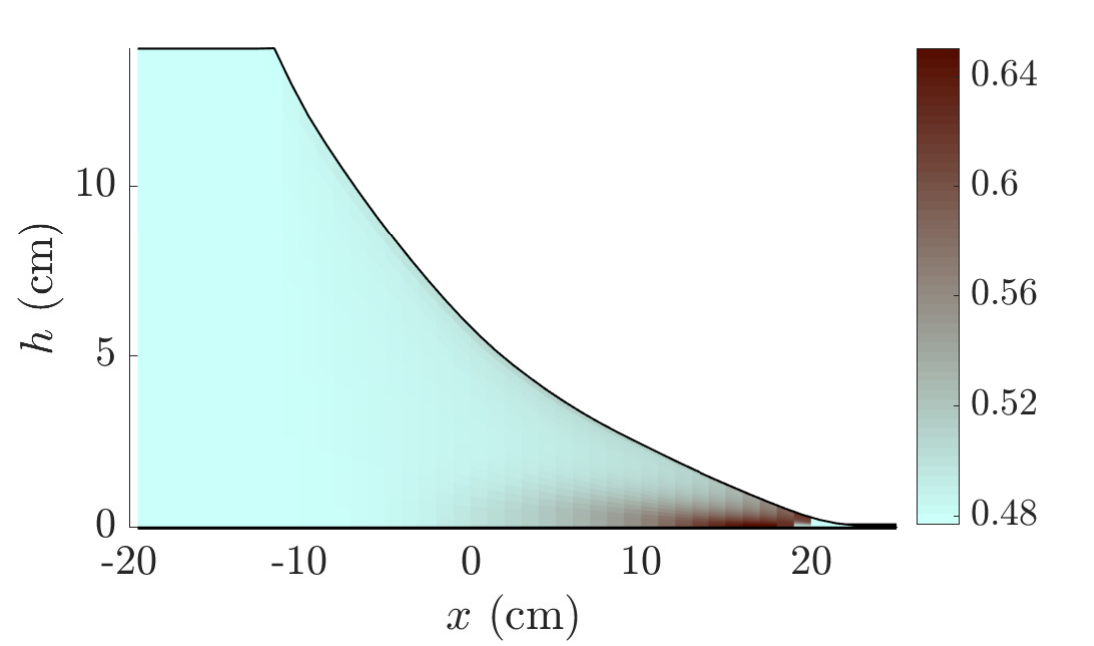}
  \caption{\label{fig:Anne_0_3D}\footnotesize \textit{Free surface in the case $\theta = 0^\circ$ at time $t = 0.18$ s computed with the $\mu(I)$-multilayer model (left hand side) and with the $\mu(I)$-$\mu_w$-multilayer (right hand side). Colors represent the distribution of horizontal velocities $u$ (upper part), vertical velocities $w$ (middle part) and variable coefficient of friction $\mu(I)$ (lower part). }}
 \end{center}
 \end{figure}

   \begin{figure}[!h]
 \begin{center}
  (Left) $\mu(I)$ - multilayer \qquad\qquad\qquad\qquad\qquad (Right) $\mu(I)$ - $\mu_w$ - multilayer\\[0.2cm]
 \includegraphics[width=0.45\textwidth]{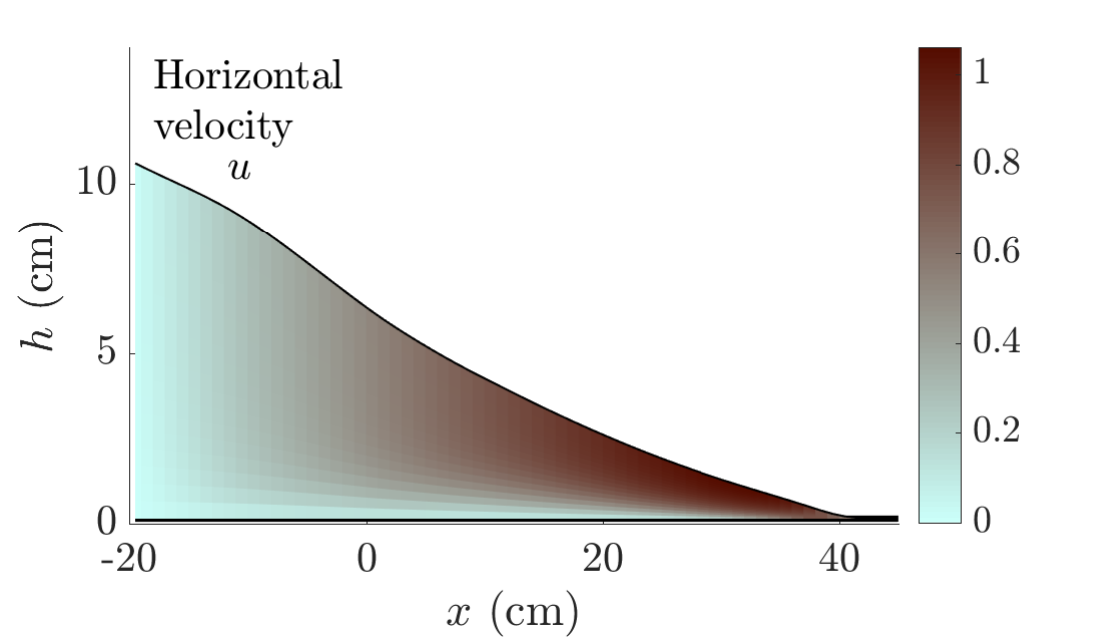}
  \includegraphics[width=0.45\textwidth]{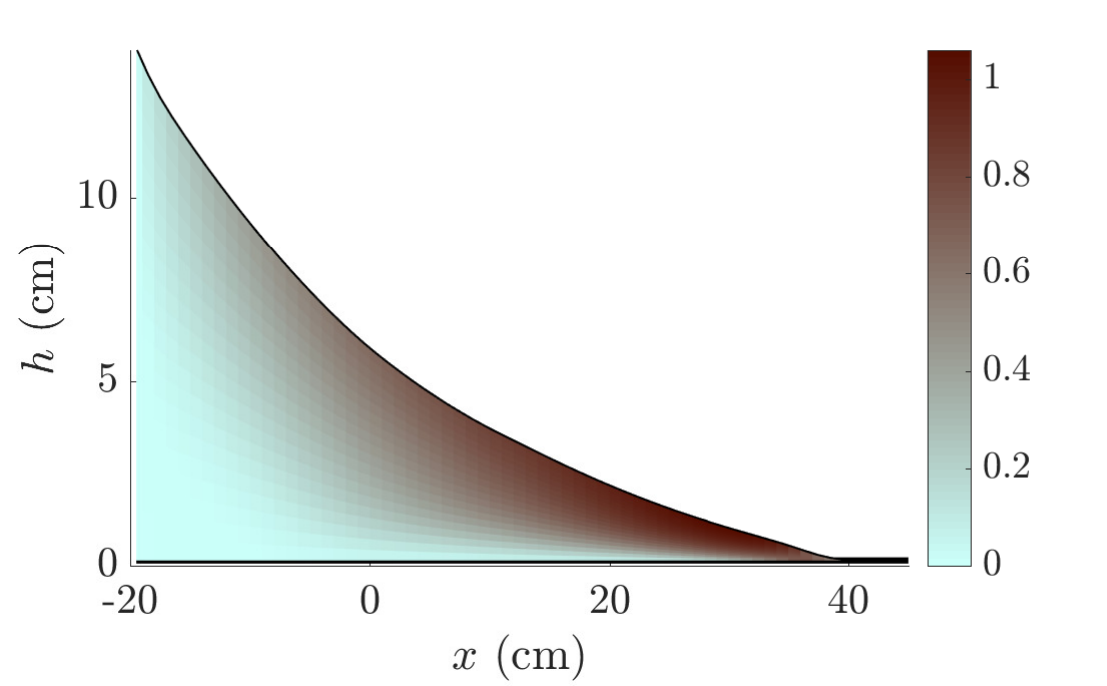}
    \includegraphics[width=0.45\textwidth]{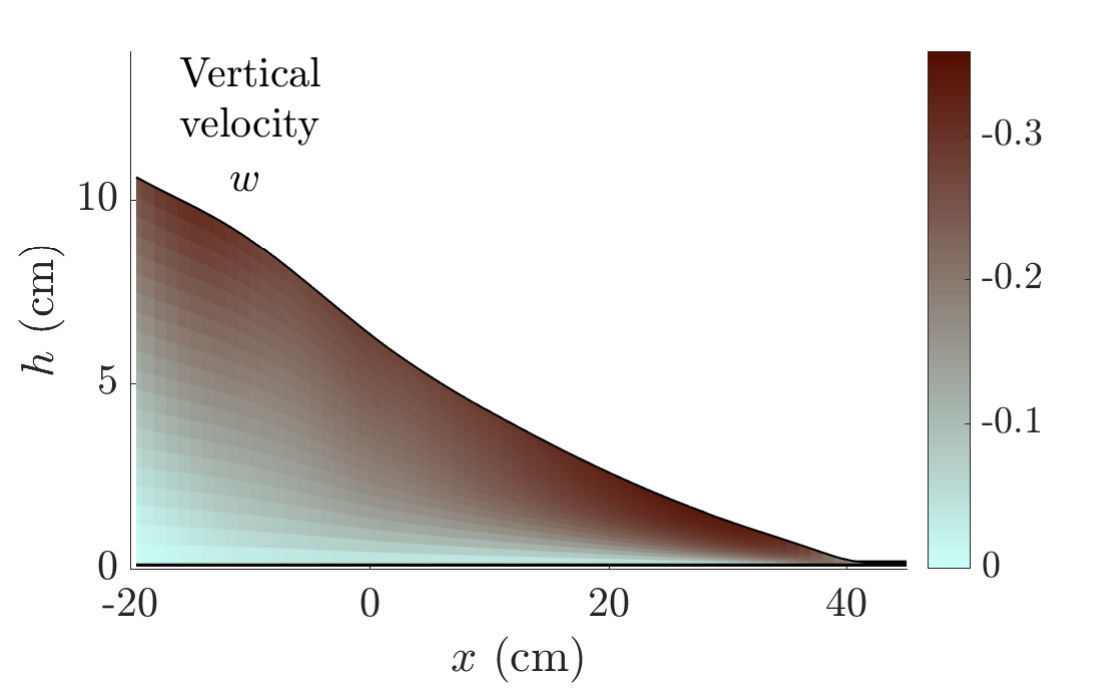}
     \includegraphics[width=0.45\textwidth]{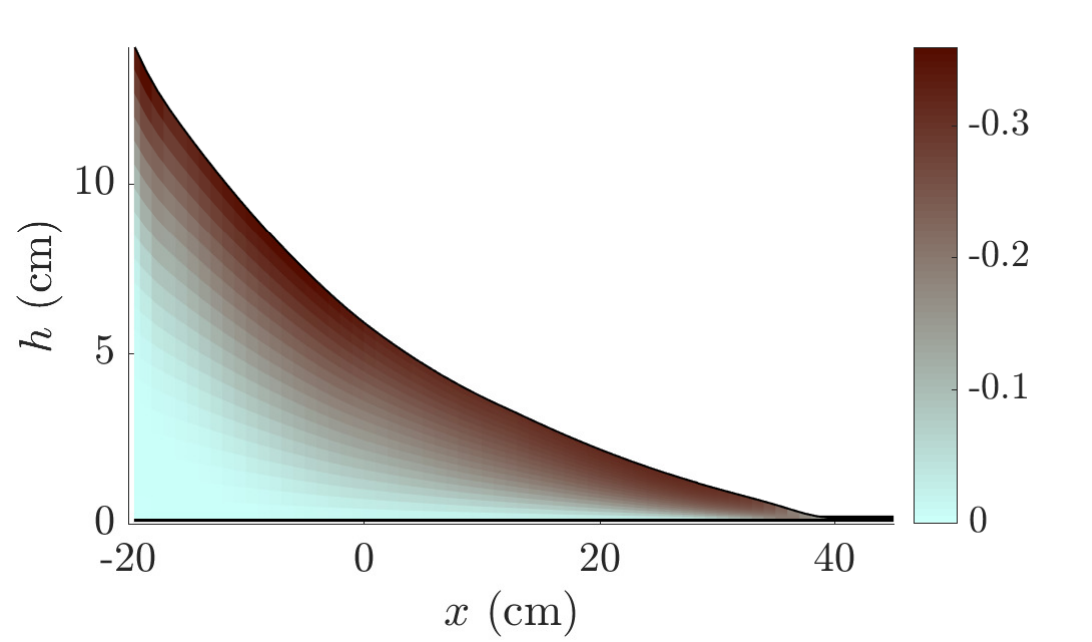}
    \includegraphics[width=0.45\textwidth]{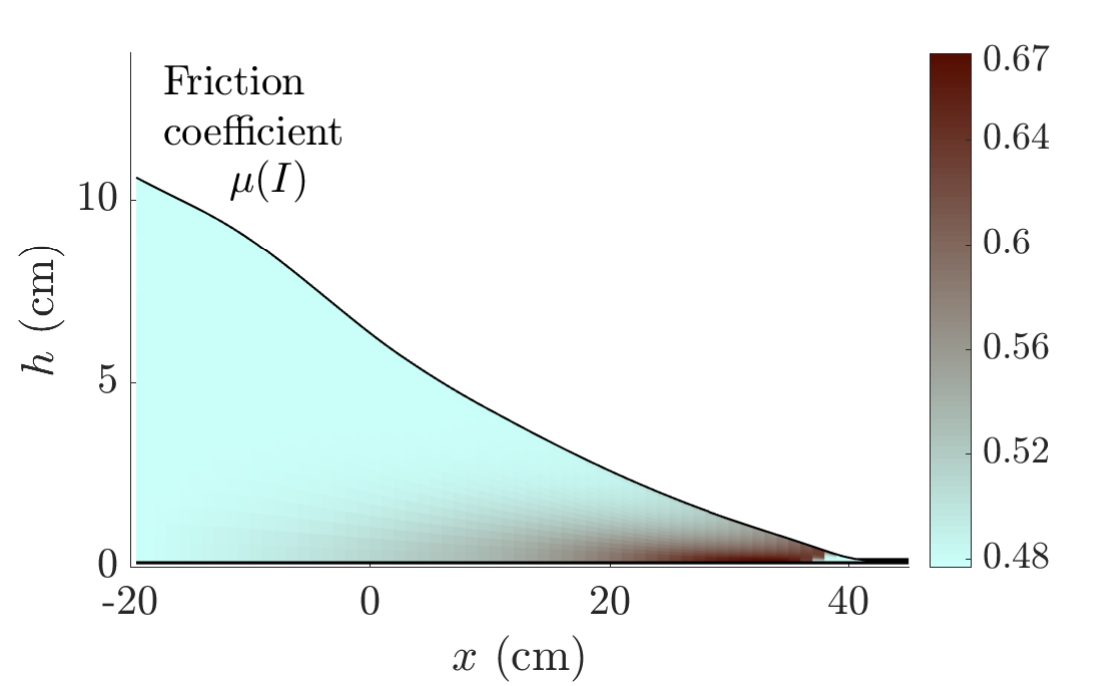}
    \includegraphics[width=0.45\textwidth]{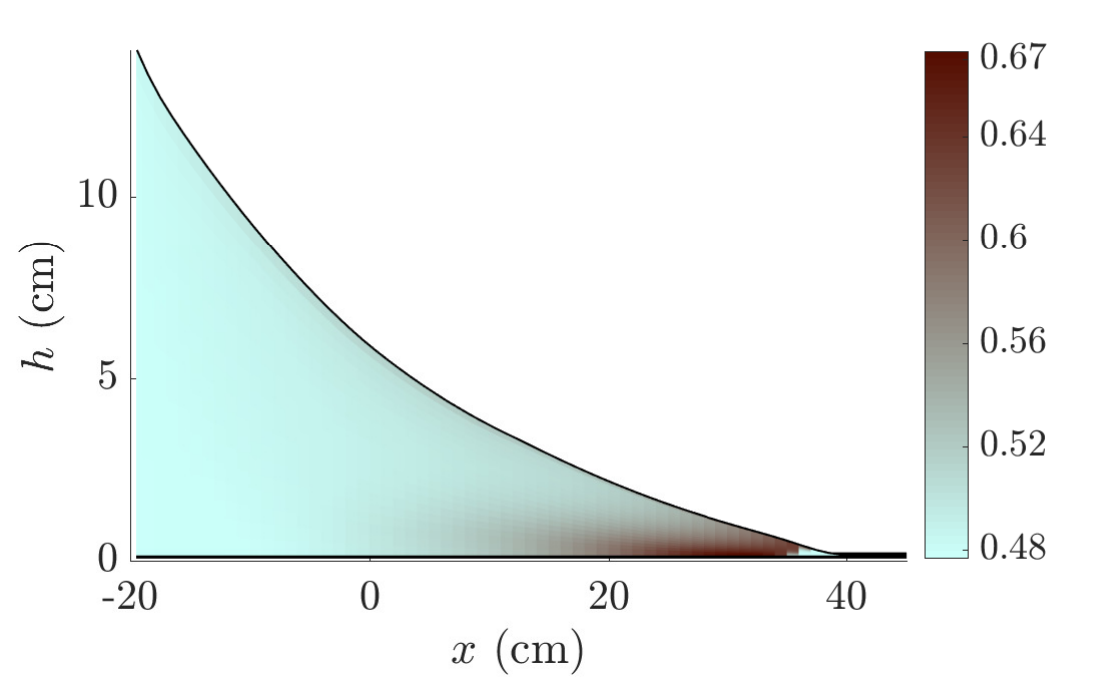}
  \caption{\label{fig:Anne_16_3D}\footnotesize \textit{Free surface in the case $\theta = 16^\circ$ at time $t = 0.3$ s computed with the $\mu(I)$-multilayer model (left hand side) and with the $\mu(I)$-$\mu_w$-multilayer (right hand side). Colors represent the distribution of horizontal velocities $u$ (upper part), vertical velocities $w$ (middle part) and variable coefficient of friction $\mu(I)$ (lower part). }}
 \end{center}
 \end{figure}

Figures \ref{fig:Anne_0_3D} and \ref{fig:Anne_16_3D} show the distribution of the horizontal and vertical velocities, and the variable friction coefficient computed with the $\mu(I)$ - multilayer and the $\mu(I)$ - $\mu_w$ - multilayer model for the slopes $\theta = 0^\circ$ and $\theta = 16^\circ$ at an intermediate time.  We see that the variable friction coefficient is greater close to the front since the strain rates are also greater and the pressure is small leading to high inertial number too.
We see that the absolute value of the velocities (horizontal and vertical) computed with the $\mu(I)$-$\mu_w$ multilayer model are lower close to the bottom due to the fact that the new friction term is greater there.

 \begin{figure}[!h]
 \begin{center}
  \includegraphics[width=0.9\textwidth]{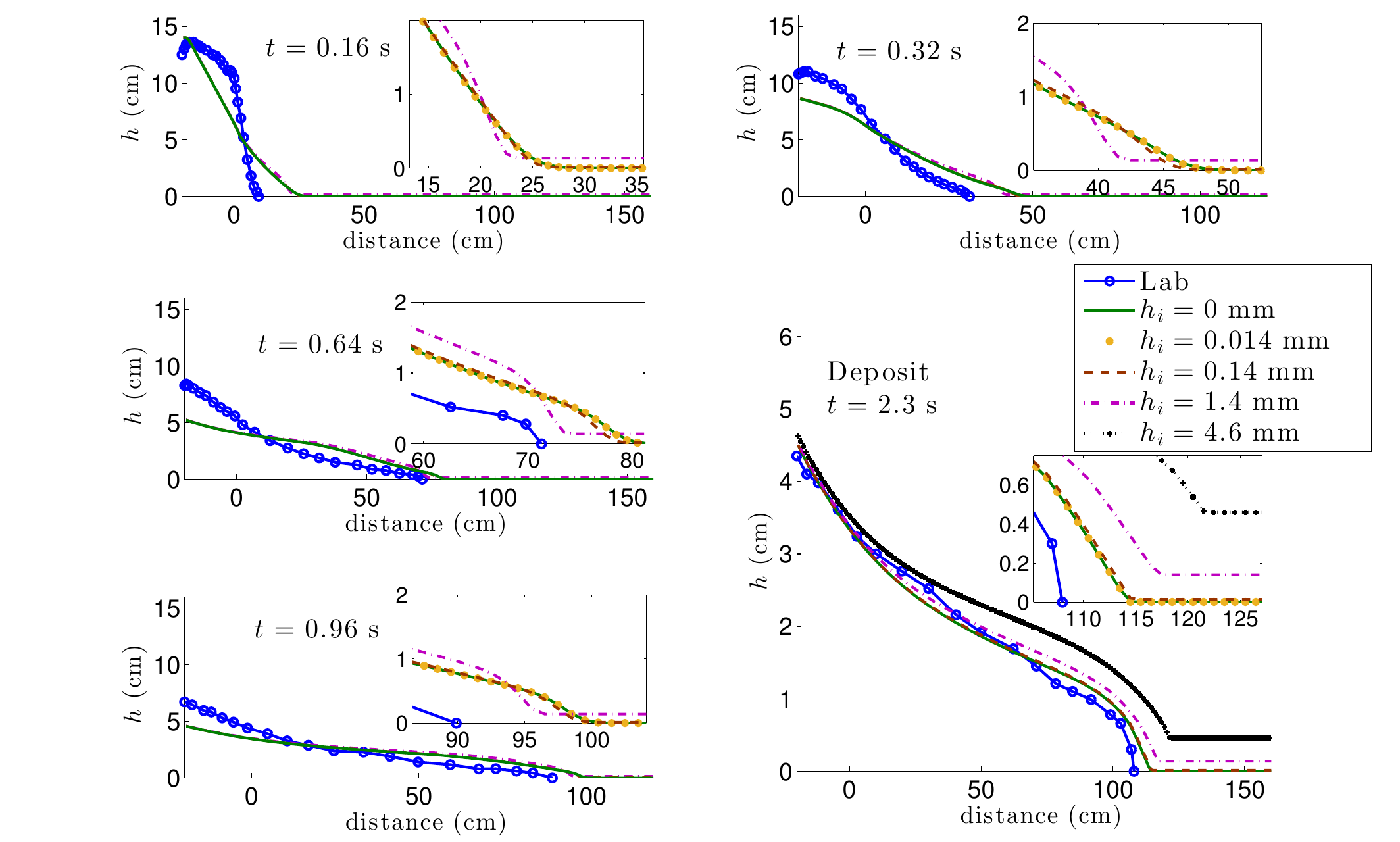}
  \caption{\label{fig:Anne_22_eps}\footnotesize \textit{Deposit obtained in the laboratory experiments (solid-circle blue line), with the $\mu(I)$-multilayer model, for a slope $\theta = 22^{\circ}$ at different times and for thicknesses of the thin layer $h_i = 0.014$ mm (gold points), $0.14$ mm (dashed brown line), $1.4$ mm (dot-dashed magenta line) and $4.6$ mm (dotted-cross black line). The solid green line is rigid bed case.
  }}
 \end{center}
 \end{figure}
Dealing with a rigid bed involving wet/dry fronts is usually hard numerically. Therefore, a thin layer of material is sometimes added on the rigid bed to get rid from numerical issues while expecting to get similar result to the case of true rigid bed.
To quantify the error related to this artificial layer, we simulate here the collapse over a thin layer of material of thickness $h_i$ of a mass with initial thickness: $$h(x,0) = \left\{\begin{matrix}
14 \text{ cm}& \text{if }x \leq 0;\\
h_i & \text{otherwise},
\end{matrix}\right.\qquad\text{with}\quad h_i = 0,\,0.014,\,0.14,\,1.4,\,4.6 \text{ mm.}$$

Figure \ref{fig:Anne_22_eps} shows the collapsing mass profiles and the deposits simulated for a slope $\theta = 22^\circ$. We see that, when the layer is thin enough ($h_i = 0.014$ mm), the simulated mass profiles and deposit are similar to the case when $h_i=0$ (true rigid bed). Slight differences appear at $h_i = 0.14$ mm and get stronger for larger thicknesses ($h_i = 1.4$ mm). In this case, we observe an increase of the runout distance and a different shape of the deposit, in particular near the front as shown in inset zooms in figure \ref{fig:Anne_22_eps} at intermediates times. Note that $h_i = 0.14$ mm represents about $0.1\%$ of the thickness of the initial granular column.\\

  \begin{figure}[!h]
 \begin{center}
 (Left) $h_i = 0$ mm \qquad\qquad\qquad\qquad\qquad\quad (Right) $h_i = 4.6$ mm\\
\includegraphics[width=0.4\textwidth]{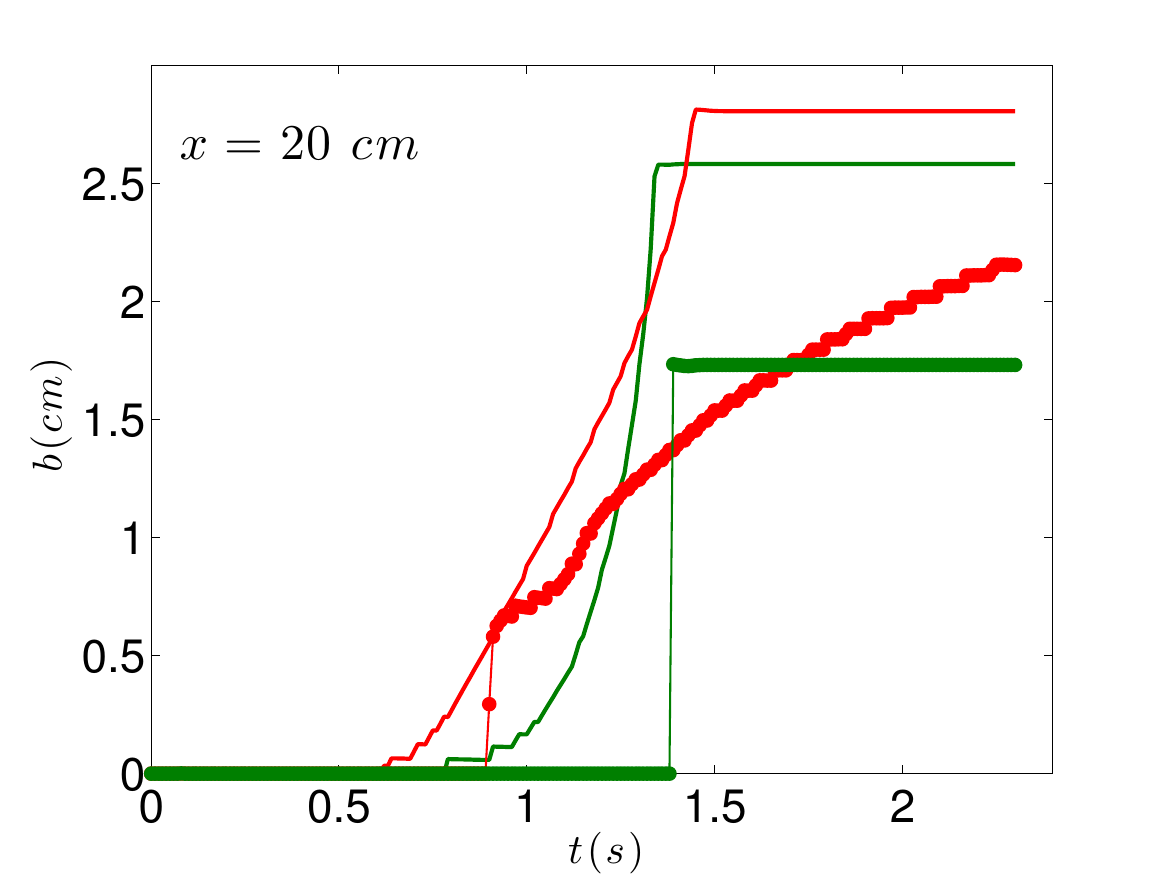}
\includegraphics[width=0.4\textwidth]{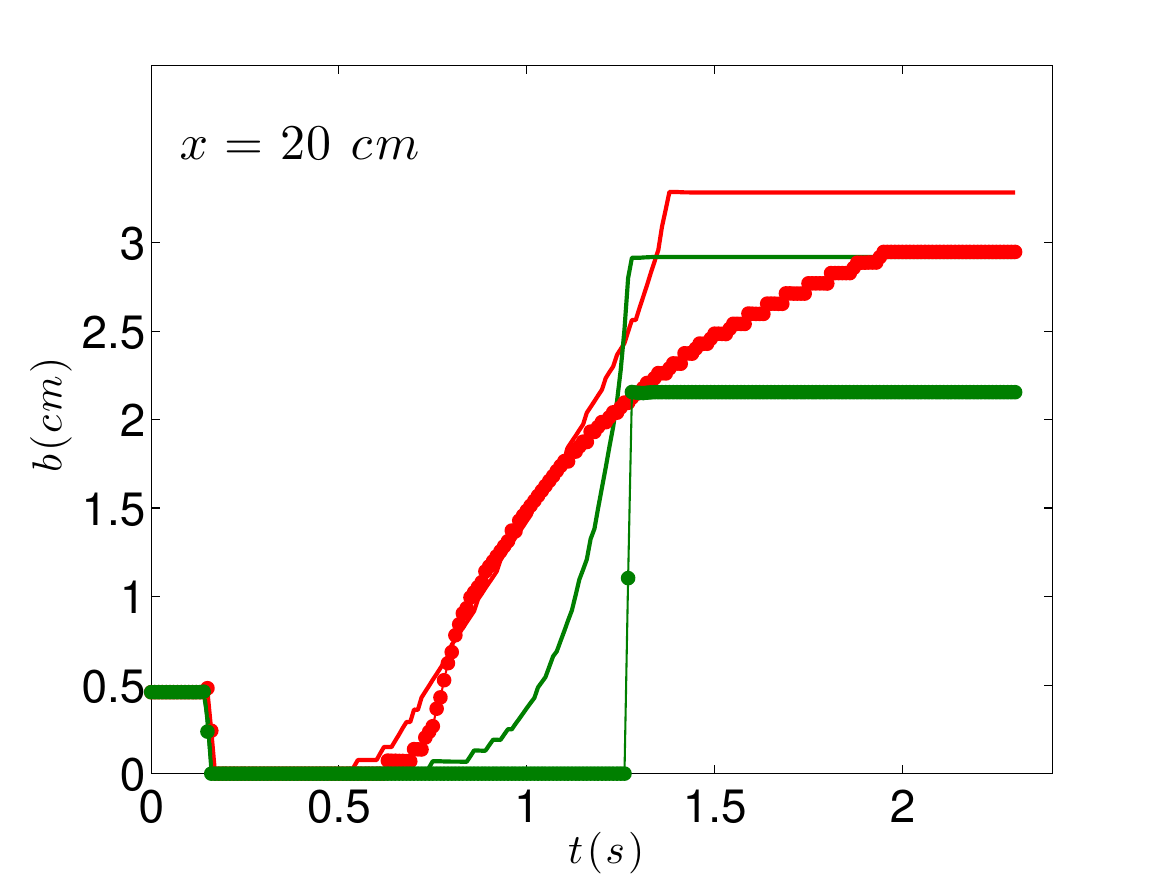}\\
\includegraphics[width=0.4\textwidth]{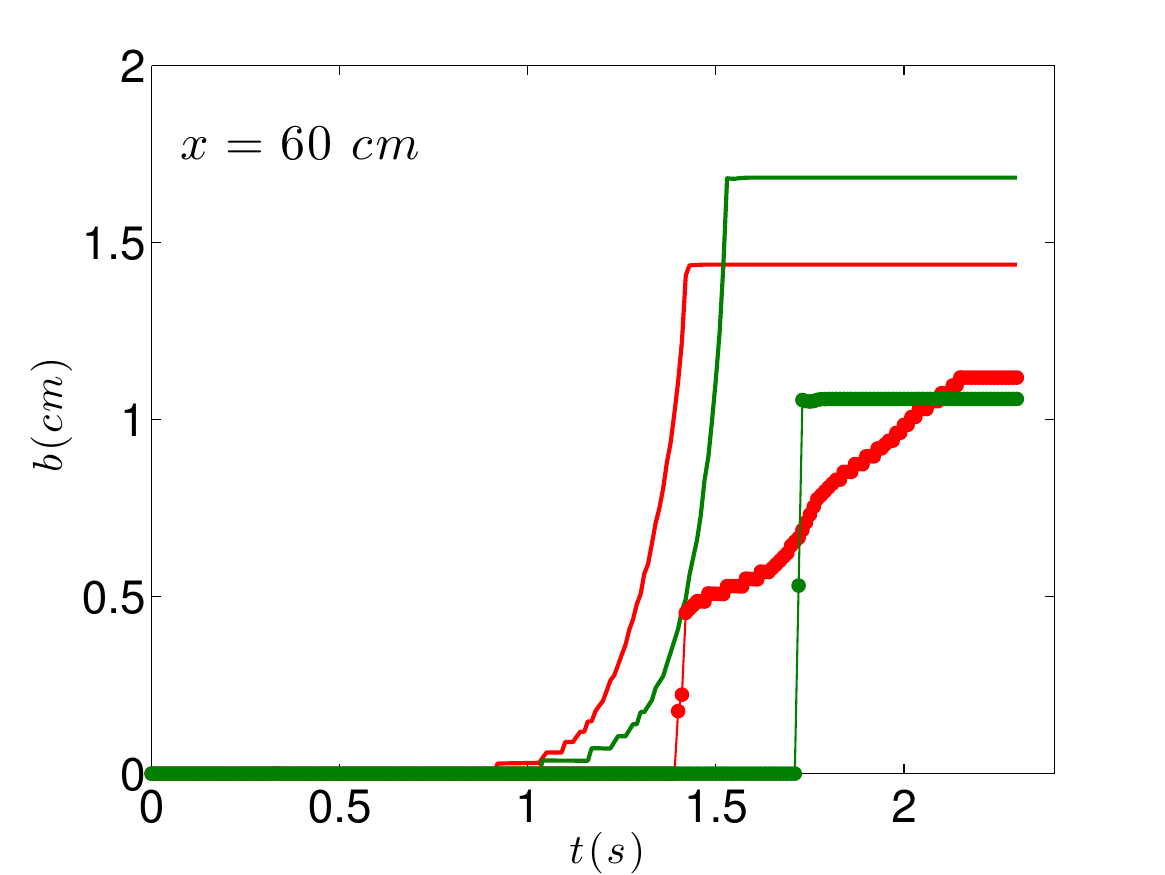}
\includegraphics[width=0.4\textwidth]{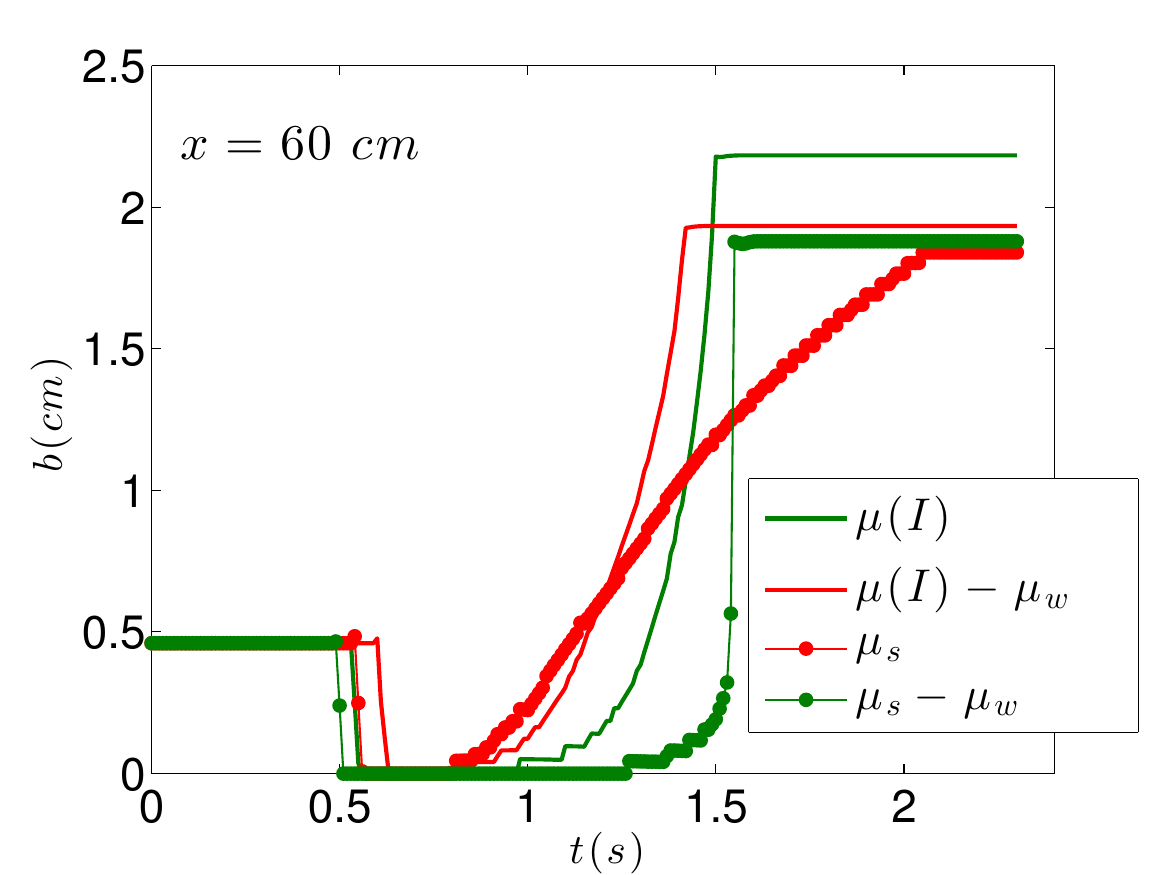}
  \caption{\label{fig:Anne_22_interfaz_hi46}\footnotesize \textit{Time evolution of the flow/no-flow interface computed for the granular collapse ($W=10$ cm) over a slope $\theta = 22^\circ$, covered by a layer of thickness $h_i =0$ mm (left column) and $h_i =4.6$ mm (right column) of the same material, at $x = 20, 60$ cm. The solid green  (green symbols) lines represent the simulations by using the variable friction coefficient $\mu(I)$ (constant coefficient $\mu_s$) without adding the side walls contribution. The solid red (red symbols) lines represent the simulations with the variable friction coefficient $\mu(I)$ (constant coefficient $\mu_s$) adding the side walls friction term.}}
 \end{center}
 \end{figure}

Figure \ref{fig:Anne_22_interfaz_hi46} shows the evolution of the flow/no-flow interface $b(x,t)$ for the granular collapse over a slope $\theta =22^\circ$ at $x = 20$ and $x=60$ cm for flow over a rigid bed (left column) and over an erodible bed of thickness $h_i = 4.6$ mm (right column).The simulations are performed using the multilayer model (with $50$ layers) with a variable friction coefficient $\mu(I)$ or a constant coefficient $\mu_s$, and adding or not the side walls friction term.

When the variable friction coefficient $\mu(I)$ is used (with and without the friction term at lateral walls) to simulate granular collapse over a rigid bed, the flow/no-flow interface goes from the bottom to the top of the granular layer until the whole thickness stops. For granular collapse over erodible bed, the flow/no-flow interface penetrates into the erodible bed very rapidly (i. e. erosion of the granular bed), stays at the bottom for a while (i. e. the whole thickness is flowing) and then goes up to the free surface. This qualitative behaviour is very similar to what is observed in experiments (see e. g. \cite{mangeney:2010,lusso:2017b}). Adding walls friction with the $\mu(I)$ rheology makes the flow/no-flow interface goes up earlier and change the shape of its time evolution up to the free surface. With a constant friction coefficient $\mu_s$ and no wall effects for flows over a rigid bed, the mass moves all over the depth until all the granular thickness suddenly stops, contrary to what is observed experimentally. When adding walls friction, the flow/no-flow interface propagates from the bottom to the top due to increasing friction with depth. For flows over erodible bed with $\mu_s$, the flow/no-flow interface penetrates into the erodible layer as rapidly as with $\mu(I)$ but then, again, goes abruptly up to the free surface. Adding walls friction in this case drastically change the flow/no-flow behaviour that get closer to the results obtained with $\mu(I)$.

   \begin{figure}[!h]
 \begin{center}
 \includegraphics[width=0.55\textwidth]{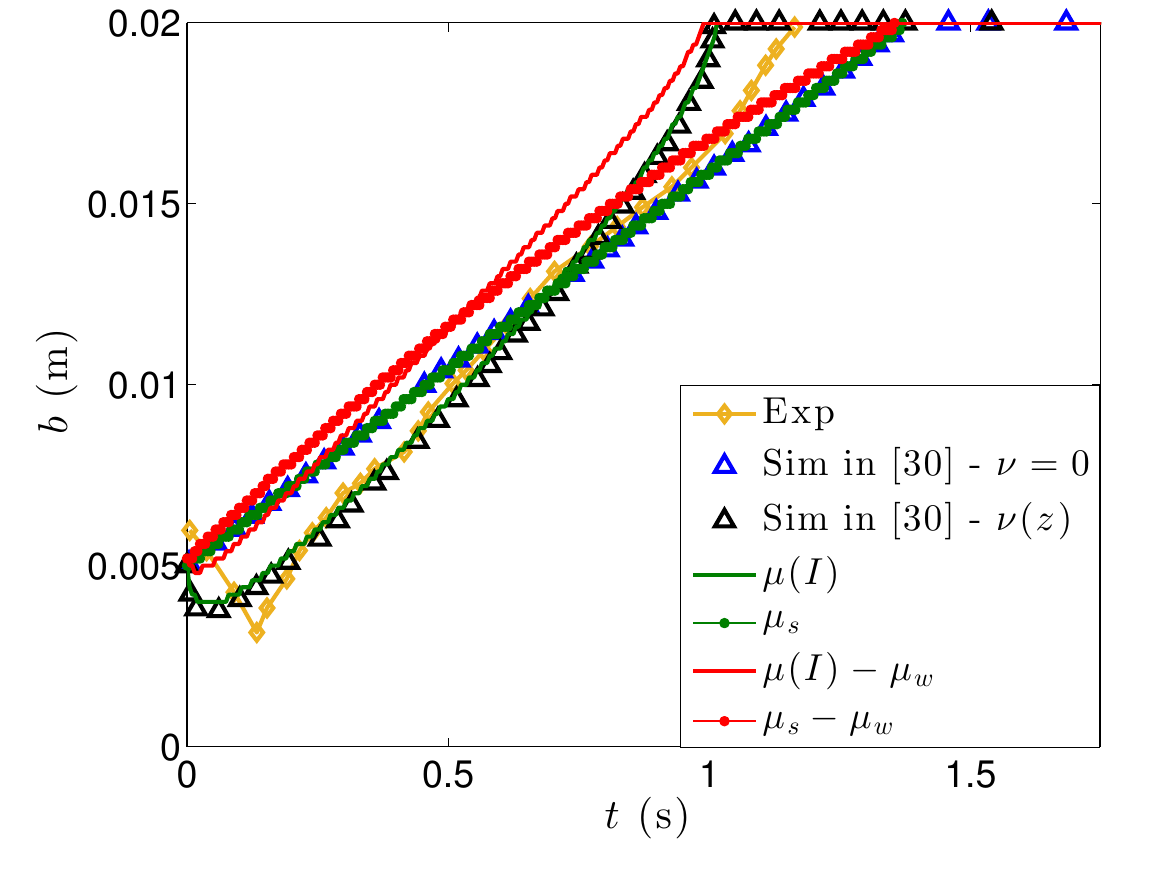}
  \caption{\label{fig:Anne_22_interfaz_unif}\footnotesize \textit{Time evolution of the flow/no-flow interface computed with the multilayer model for a uniform flow, with linear initial velocity profile, over a slope $\theta = 22^\circ$ which is covered by a static layer of thickness $h_i =5$ mm. In this case the channel with is $W=20$ cm. The solid green  (green symbols) lines represent the simulations with the variable friction coefficient $\mu(I)$ (constant coefficient $\mu_s$) without adding the side walls contribution. The solid red (red symbols) lines represent the simulations with the variable friction coefficient $\mu(I)$ (constant coefficient $\mu_s$) adding the side walls friction term. The diamonds-solid gold line is the experiments in \cite{lusso:2017b} for this test.}}
 \end{center}
 \end{figure}

Lusso et al. \cite{lusso:2017b} investigated the evolution of the flow/no-flow interface $b(t)$ through a simplified model which takes into account the variation in the direction normal to the topography but not in the downslope direction. They compare their results to what was measured experimentally in the well-developped shallow flow following granular collapse over a channel of width $W=20$ cm and slope $\theta = 22^\circ$ covered by a static layer of thickness $h_i = 5$ mm. The configuration and material properties of theses experiments are the same as those exposed previously. They compared the position of the flow/no-flow interface at $x =90$ cm with analytical and numerical solution of the non depth-averaged shallow equations for uniform flow in the downslope direction. The parameter and initial condition of the test are:
$$
\mu_s = \tan(26^\circ), \quad \mu_2 = \tan(28^\circ), \quad d_s = 0.7\ \text{mm}, \quad \varphi = 0.62, \quad I_0 = 0.279,
$$
$$
 h_0=h(t=0,x) = 2\ \text{cm},\quad b_0=b(t=0,x) = 5\ \text{mm}.
$$
The linear initial profile of velocity is assumed in the moving layer ($b_0<z<h_0$):
$$
u(t=0,z) = 70\,\left(z-b_0\right)\ \text{m/s} \quad \text{if }z>b_0;\quad u(t=0,z) = 0 \ \text{m/s}\quad \text{if }z<b_0.
$$
This test is simulated here. In order to improve the precision $100$ vertical layers are used. Figure \ref{fig:Anne_22_interfaz_unif} shows the evolution of the flow/no-flow interface $b(t)$ computed with the variable coefficient of friction $\mu(I)$ and the constant coefficient $\mu_s$, including or not the side walls friction term. We see that our result without lateral wall friction agrees almost perfectly with the ones presented in Lusso et al. \cite{lusso:2017b} (see cases $\nu = 0$ and $\nu=\nu(Z)$ in figure 16 in \cite{lusso:2017b}).  By using the friction constant coefficient the profile flow/no-flow interface evolution in time is a straight line, whereas the convex shaped profile observed in experiments is reproduced for the variable friction $\mu(I)$. Based on these results, Lusso et al. \cite{lusso:2017b} suggested that for uniform flows erosion (i. e. penetration within the erodible layer) can only be obtained for a variable friction coefficient (called viscosity in their paper) and not for a constant friction coefficient $\mu_s$. In the case of granular collapse presented above, erosion is also obtained with $\mu_s$, certainly due to the non-uniformity of the flow and in particular to downslope pressure gradients (see section 5 in \cite{lusso:2017b}).

\section{Conclusions}\label{se:conclusions}
 This work provides two main contributions. First, we have introduced a 2D-model that takes into account side walls effect through the viscous term $\p_y\left(\eta\p_y u\right)$ and a Coulomb-type boundary condition. This model follows from a dimensional analysis of Navier-Stokes equations and the hypothesis of a one-dimensional flow, that is, no transversal velocity ($v=0$). We have also shown that this model matches with the one proposed in \cite{jop:2005} under some specific assumptions, which, in particular, are verified for uniform flows. In section \ref{se:test_unif} we show that both, Bagnold and S-shaped vertical profiles of velocity can be automatically recovered by using the multilayer approach. This is not possible for the models proposed by Gray \& Edwards \cite{gray:2014}, Edwards \& Gray \cite{edwards:2015} and Baker et al. \cite{baker:2016} because of the prescribed Bagnold profile neither for the model proposed by Capart et al. \cite{capart:2015}, which only deals with S-shaped profiles. We also quantified the influence of the lateral friction term on the shape of the normal profiles of the downslope velocity and on the maximum velocity as a function of the channel width $W$. In particular we were able to calculate what is the minimum channel width for which the granular mass flows over its all thickness, the minimum width for which Bagnold profiles is obtained instead of S-shaped profiles and the minimum width required to obtain a velocity profile independent of the channel width. This analysis may be helpful when designing and analysing laboratory experiments.\\

Secondly, a multilayer discretization for this model is proposed. We present a numerical scheme with an appropriate treatment of the rheological terms in order to obtain a well balanced scheme. To this aim, we use a hydrostatic reconstruction taking into account the friction term. In section \ref{se:WB} we show that this hydrostatic reconstruction gives the well balance property of the scheme.

Our simulations show that important differences in the final deposit are obtained wether the side walls friction term is approach by a multilayer or by a single-layer model. This is proved in section \ref{se:h_est}, where the monolayer model is able to preserve with second order accuracy steady solution quite different of the ones computed when a vertical discretization is considered. We conclude that including the side walls friction term using single-layer models is not appropriate, since they preserve non-physical solution due to the overestimation of the lateral friction term obtained because of the depth-average hypothesis. It cannot be solve by using a lower friction coefficient $\mu_w/2$, $\mu_w/3$, etc. since the obtained profiles show important differences with the expected ones, for example in the shape, runout and initial height.\\

Finally, we compared our simulation with laboratory data \cite{mangeney:2010} of granular collapse over a rigid bed showing the ability of multilayer models to approximate the flow/no-flow interface. However, the approximation of the lateral walls friction is not good enough and we still need to add $0.1$ to the friction coefficient. Two interesting conclusions can be drawn from our analysis. Firstly, for the dam break problem, similar results can be obtained when a very thin layer of material (about $0.1\%$ of the initial height of the dam) is added to the rigid bed instead of having a true the rigid bed. Secondly, considering a no-slip condition or a friction condition at the bottom in the multilayer approach reduces to multiply by a factor 2 the inertial number in the variable friction coefficient $\mu(I)$ at the bottom. We showed that the convex shape of the time evolution of the flow/no-flow interface is reproduced only with a variable friction coefficient and not with a constant friction coefficient, in agreement with \cite{lusso:2017b}. Our results on granular collapses show that erosion of an underlying erodible bed can occur with both constant and variable friction coefficient which is not the case for uniform flows \cite{lusso:2017b}. This is the result of pressure gradients in the downslope direction.

In conclusion, shallow multilayer models appear to be a very interesting alternative to shallow depth-averaged models by making it possible to describe changes of the velocity profiles, lateral wall effects and erosion processes with still reasonable computational cost.\\

\section*{Acknowledgements}
This research has been partially supported by the Spanish Government and FEDER through the research projects MTM2012-38383-C02-02 and  MTM2015-70490-C2-2-R, by the ANR contract ANR-11-BS01-0016 LANDQUAKES, the USPC PEGES project and the ERC contract ERC-CG-2013-PE10-617472 SLIDEQUAKES. Authors kindly acknowledge useful discussions with F. Bouchut and M. Farin.

\bibliographystyle{plain}

\bibliography{Biblio}
\end{document}